\documentclass[prl,aps,twocolumn,superscriptaddress]{revtex4-1}
\usepackage{bm}
\usepackage[colorlinks=true,linkcolor=blue,citecolor=blue]{hyperref}
\usepackage{times}
\usepackage{amsmath}
\usepackage{amssymb}
\usepackage{amsthm}
\usepackage{amsfonts}
\usepackage{enumerate}
\usepackage{latexsym}
\usepackage{ifpdf}
\usepackage{natbib}
\usepackage{psfrag}
\newcommand{\beq}{\begin{equation}}
\newcommand{\eeq}{\end{equation}}
\usepackage{graphicx}
\usepackage{makeidx}
\usepackage{color}
\usepackage{array}
\usepackage{multirow}
\usepackage{siunitx}
\usepackage{}

\begin{document}
\title{PERSPECTIVE: Emergent phases in rare earth nickelate heterostructures}

\author{J. Chakhalian}
\email{jak.chakhalian@rutgers.edu}
\affiliation{Department of Physics and Astronomy, Rutgers University, Piscataway, New Jersey 08854, USA}
\author {S. Middey}
\email{smiddey@iisc.ac.in}
\affiliation  {Department of Physics, Indian Institute of Science, Bengaluru 560012, India}

\begin{abstract}
The prediction of high $T_c$ superconductivity in layers of LaNiO$_3$ through orbital engineering has led to extensive research efforts over the last fifteen years. During this period, a plethora of thin films and heterostructures based rare-earth nickelate family with perovskite structure has been synthesized and explored. In this short perspective, we briefly review the complexity of bulk $RE$NiO$_3$, spotlighting several recent findings of emergent phenomena in heterostructures containing the interface between $RE$NiO$_3$ and another transition metal oxide. Finally, we outline potentially  interesting future directions linked to time-domain dynamics to harness new Mott and topological phases in artificial structures of $RE$NiO$_3$.
\end{abstract}

\maketitle

\section{Introduction}
\begin{figure}
    \centering
    \includegraphics[width=0.45\textwidth]{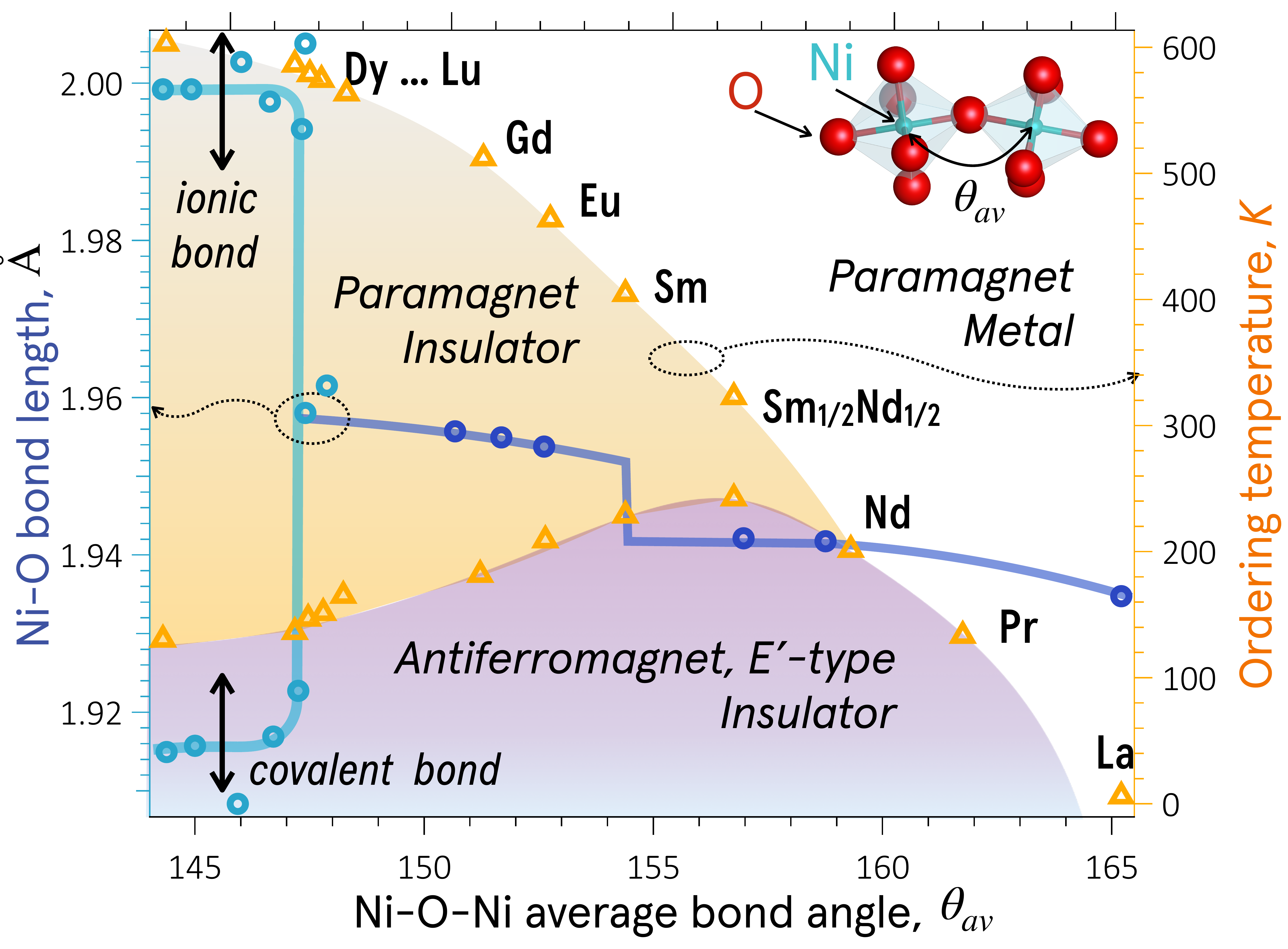}
    \caption{Ni-O bond lengths (left axis) and phase transition temperatures (right axis) have been plotted as a function of Ni-O-Ni bond angle of bulk $RE$NiO$_3$ series. All data have been adapted from Ref.~\onlinecite{Zhou:2004p153105}.}
    \label{Fig1}
\end{figure}

The famous volume titled "Magnetism and the Chemical bond" by John B. Goodenough was released in December 1964 \cite{Goodenough:1963book}
and rapidly earned a reputation of `Atlas Maior' of quantum materials with correlated electrons. Somewhat surprisingly, this encyclopedic compendium contains no mention of rare-earth nickelates despite an early report on the synthesis~\cite{Wold:1957p4911}. However, two years later, in the series of insightful articles on complex oxides with perovskite structure, Goodenough et al., described the majority of structural, electronic, and magnetic properties of these compounds~\cite{Goodenough:1965p1031,Goodenough:1966p1415}. During the 1970s and 80s, the investigation of rare-earth nickelates based on the chemical formula \textit{RE}NiO$_3$, where $RE$ is a rare-earth ion, remained rather dormant until after the discovery of high $T_c$ superconductivity in cuprate oxides with partial perovskite crystal structure~\cite{Torrance:1992p8209}. Moreover, since in  the periodic table Ni is located next to Cu, there was an active research for the nickel-based high $T_c$ superconducting oxides~\cite{Anisimov:1999p7901}, which is still ongoing~\cite{Li:2019p624,Pickett:2021p7}. With the renewed interest in nickelates, a massive high-pressure synthesis effort was put forward, finally yielding the complete $RE$NiO$_3$ family that spans from La to Lu (see Fig.~\ref{Fig1}), thus making the whole family available for the systematic exploration~\cite{Medarde:1997p1679, Catalan:2008p729}.

We begin this perspective article with a brief introduction about the bulk $RE$NiO$_3$ highlighting the complexity of their electronic and magnetic behavior. As shown in the phase diagram of \textit{bulk} $RE$NiO$_3$ (Fig.~1), the first member of the nickelate series, LaNiO$_3$ with rhombohedral $R\bar{3}c$ structure remains metallic down to the lowest  probed temperature. The intermediate members with $RE$ = Pr and Nd undergo a first-order metal-to-insulator transition (MIT)\ from paramagnetic metallic (PMM) state with orthorhombic (\textit{Pbnm}) structure to an antiferromagnetic insulating (AFI) state accompanied with a lowering of the symmetry to monoclinic (\textit{P2$_1$/n})~\cite{Medarde:1997p1679, Catalan:2008p729}.
For even smaller \textit{RE} ions (e.g., Sm...Lu), as the Ni-O-Ni bond angle decreases further away from the maximal value of $\sim 165^o$ in LaNiO$_3$, the MIT\ temperature ($T_{\mathrm{MIT}}$) and magnetic transition ($T_{N}$) separate from each other (see Fig. 1). Within the insulating phase, the entire family has a monoclinic ($P2_1/n$) structure  with two nonequivalent Ni sites in the unit cell. As for the spin structure, neutron diffraction measurements found the  magnetic wave vector to be (1/2, 0, 1/2)$_{\mathrm{ortho}}$ [(1/4, 1/4, 1/4) in pseudo cubic notation]~\cite{Garcia:1994p978}.  The spin arrangement of this $E'$-type antiferromagnetic ($E'$-AFM) phase can be visualized as a sequence of either $\uparrow \uparrow \downarrow \downarrow$ or $\uparrow  \rightarrow \downarrow \leftarrow $ pseudo-cubic (1 1 1) planes. Resonant X-ray scattering (RXS) experiments on single crystalline thin film samples ruled out any orbital ordering~\cite{Scagnoli:2005p155111,Scagnoli:2006p100409} even though the Ni$^{3+}$ ($t_{2g}^6 e_g^1, S =1/2$) is expected to be a Jahn-Teller active in a purely ionic picture. RXS experiments further confirmed the existence of spin non-collinearity within the $E^\prime$-AFM phase~\cite{Scagnoli:2008p115138}.

From the electronic structure viewpoint,  rare-earth nickelates are multi-band materials, as revealed by the Hall effect and thermopower measurements~\cite{Ha:2013p125150,Ojha:2019p235153}. Recent angle-resolved photoemission studies have also supported this multi-band character~\cite{Eguchi:2009p115122,Yoo:2015p8746,Dhaka:2015p035127}. In the metallic phase, nickelates behave as a non Fermi liquid~\cite{Zhou:2005p226602,Liu:2013p2714,Liu:2020p1402}. Tunneling measurement also claimed  a pseudogap phase in metallic LaNiO$_3$~\cite{Allen:2015p062503}. However, the simultaneous nature of the electronic and structural transitions still remains puzzling. The observation of a magnetic transition for NdNiO$_3$ and PrNiO$_3$ adds further complexity in understanding the mechanism of the MIT. For example, optical conductivity measurements find a gradual opening of a bandgap across the transition through the changes in the electrical resistivity. At the same time, lattice constants show a sharp jump in magnitude across the transition~\cite{Katsufuji:1995:p4830}. In addition, the optical band gap ($\sim$ 1 eV) seems much larger than the gap, evaluated from electrical transport data~\cite{Catalan:2000p606}.

Since the insulating  phase with monoclinic symmetry consists of two inequivalent Ni, charge disproportionation (CD) [Ni$^{+3}$+Ni$^{+3}\rightarrow$Ni$^{+3+\delta}$+Ni$^{+3-\delta}$] driven insulating phase would have been a natural explanation as the origin of the MIT~\cite{Alonso:1999p3871,Alonso:1999p4754,Staub:2002p126402,Mazin:2007p176406}. The observation of two distinct magnetic moments (1.4 $\mu_B$ and 0.7 $\mu_B$) in highly distorted YNiO$_3$ supports the CD scenario~\cite{Alonso:1999p3871}. However, this conclusion is in contradiction with the magnetic moments of all Ni sites, which are almost identical across the series of  $RE$NiO$_3$ with $RE$=Sm, Eu, Pr and Nd~\cite{Garcia:1994p978,Vobornik:1999pR8426,Carvajal:1998p456}. Complimentarily, the findings of the similar energy dependence of the pseudocubic (1/4 1/4 1/4) magnetic peak across the  Ni $L_{3,2}$ edges, observed in RXS experiments across the series with different transition temperatures and structural distortions, is also inconsistent with a conventional CD phase~\cite{Bodenthin:2011p036002}. While a spin density wave (SDW) transition can capture the simultaneous MIT and magnetic transition of NdNiO$_3$ and PrNiO$_3$ and the CD can be a simple consequence of a site-centered SDW transition~\cite{Lee:2010p165119, Lee:2011p016405},   the appearance of the simultaneous structural transition can not be accounted for by this mechanism. Though the importance of the electron-electron correlation effect behind the MIT was demonstrated~\cite{Stewart:2011p176401}, a pure Mott transition does not require either magnetic or structural transition.

 The strong O-Ni covalency and the effective negative charge energy lead  to the ground state of Ni ions to be more of 3$d^8\underline{L}$ character ($\underline{L}$  denotes a ligand hole antiferromagnetically coupled with the  Ni $e_g$ electrons), instead of  the ionic 3$d^7$  configuration~\cite{Barman:1994p8475,Bisogni:2016p13017}. This  electronic  configuration can easily explain the absence of any orbital ordering with one electron in each $e_g$ state. Considering the dominant contribution of this 3$d^8\underline{L}$ configuration, a new  CD mechanism has been suggested  where the charge redistribution occurs on the oxygen sub-lattice across the MIT~\cite{Mizokawa:2000p11263}.  This  has been named a bond-disproportionation (BD) transition. Furthermore, this process  3$d^8\underline{L}$ + 3$d^8\underline{L} \rightarrow$ 3$d^8$ (ionic, $S$=1) + 3$d^8\underline{L}^2$ (covalent, $S$=0) became known  as a site-selective Mott transition  ~\cite{Park:2012p156402,Johnston:2014p106404,Subedi:2015p075128,Haule:2017p2045}. Considering the proposed  BD scenario, the spin-configuration of $E^\prime$-AFM phase should be thought as a sequence of $\uparrow$-0-$\downarrow$-0  (1 1 1)$_\mathrm{pc}$ planes. Remarkably, based on the extensive bond-angle vs. bond-length analysis shown in Fig. 1, Goodenough and collaborators proposed an analogous picture of the bond disproportionation into a periodic pattern of ionic and covalent  Ni sites starting around Y-Lu~\cite{Zhou:2004p153105}.

\section{Epitaxial  stabilization of $RE$NiO$_3$}
The continuous advancement in  various physical vapor deposition techniques have enabled stabilization of various complex oxides in a single  crystalline thin film form with atomic precision, ~\cite{Schlom:2008p2429,Martin:2010p89,Zubko:2011p141,Hwang:2012p103,Chakhalian:2014p1189,Bhattacharya:2014p65,Liu:2016p133,Ramesh:2019p257}. This in turn  offers a new set of  control knobs such as quantum confinement, epitaxial strain, interfacial charge transfer, geometrical lattice engineering  etc., to realize a plethora of emergent phenomena~\cite{Schlom:2008p2429,Martin:2010p89,Zubko:2011p141,Hwang:2012p103,Chakhalian:2014p1189,Bhattacharya:2014p65,Liu:2016p133,Ramesh:2019p257}. The stabilization of single crystalline $RE$NiO$_3$ thin films are highly  non-trivial due  to the requirement of the +3 high oxidation state of Ni. Moreover,  the synthesis of stoichiometric $RE$NiO$_3$ in bulk form requires high oxygen pressure at elevated temperature. As a consequence, only LaNiO$_3$ could be grown as a large single crystal~\cite{Alonso:1999p3871,Alonso:1999p4754,Zhang:2017p2730,Guo:2018p43,Klein:2021}. During the initial synthesis work  on  nickelate films, it was demonstrated that even though thin films of LaNiO$_3$ can be grown without high oxygen pressure~\cite{Prasad:1993p1898,Yang:1995p2643},   the bulk-like metal-insulator transition could be  observed only in NdNiO$_3$ films  post annealed under high oxygen pressure~\cite{DeNatale:1995p2992}. The  explanation for the low pressure  film growth was attributed to a coherent film/substrate interface that dramatically lowers the free energy ~\cite{Kaul:2004p861,Gorbenko:2002p4026}. Utilizing this heteroepitaxial stabilization, high quality single crystalline $RE$NiO$_3$ films with $RE$= La, Pr, Nd, Sm, Eu on perovskite substrates have been grown under low oxygen partial pressure by pulsed laser deposition~\cite{Catalan:2000p606,Boris:2011p937,Chakhalian:2011p116805,Liu:2013p2714,Meyers:2013p075116,Bruno:2013p195108,Hepting:2014p227206,Ranjan:2020p071601}, sputtering~\cite{Son:2010p062114,Ha:2012p233,Hauser:2015p092104,Catalano:2014p116110} and oxide molecular beam epitaxy~\cite{Nikolaev:1999p118,May:2010p014110,Feigl:2013p51}.

The prediction of potential high  $T_c$ superconductivity via orbital engineering~\cite{Chaloupka:2008p016404,Hansmann:2009p016401}  has resulted in a massive research drive  for the growth of various artificial structures (e.g., thin-film, heterostructure, superlattice, etc.) of $RE$NiO$_3$, and  have been summarized in recent review articles~\cite{Middey:2016p305,Catalano:2018p046501}. This Perspective article focuses on several recent developments concerning the $RE$NiO$_3$/$A_{n^\prime}B_{n^{\prime\prime}}$O$_{n^{\prime\prime\prime}}$ heterostructure, where  $A_{n^\prime}B_{n^{\prime\prime}}$O$_{n^{\prime\prime\prime}}$ represents another transition metal oxide.

\section{Emergent phases of $RE$NiO$_3$ through octahedral engineering}

\begin{figure*}
    \centering
    \includegraphics[width=0.85\textwidth]{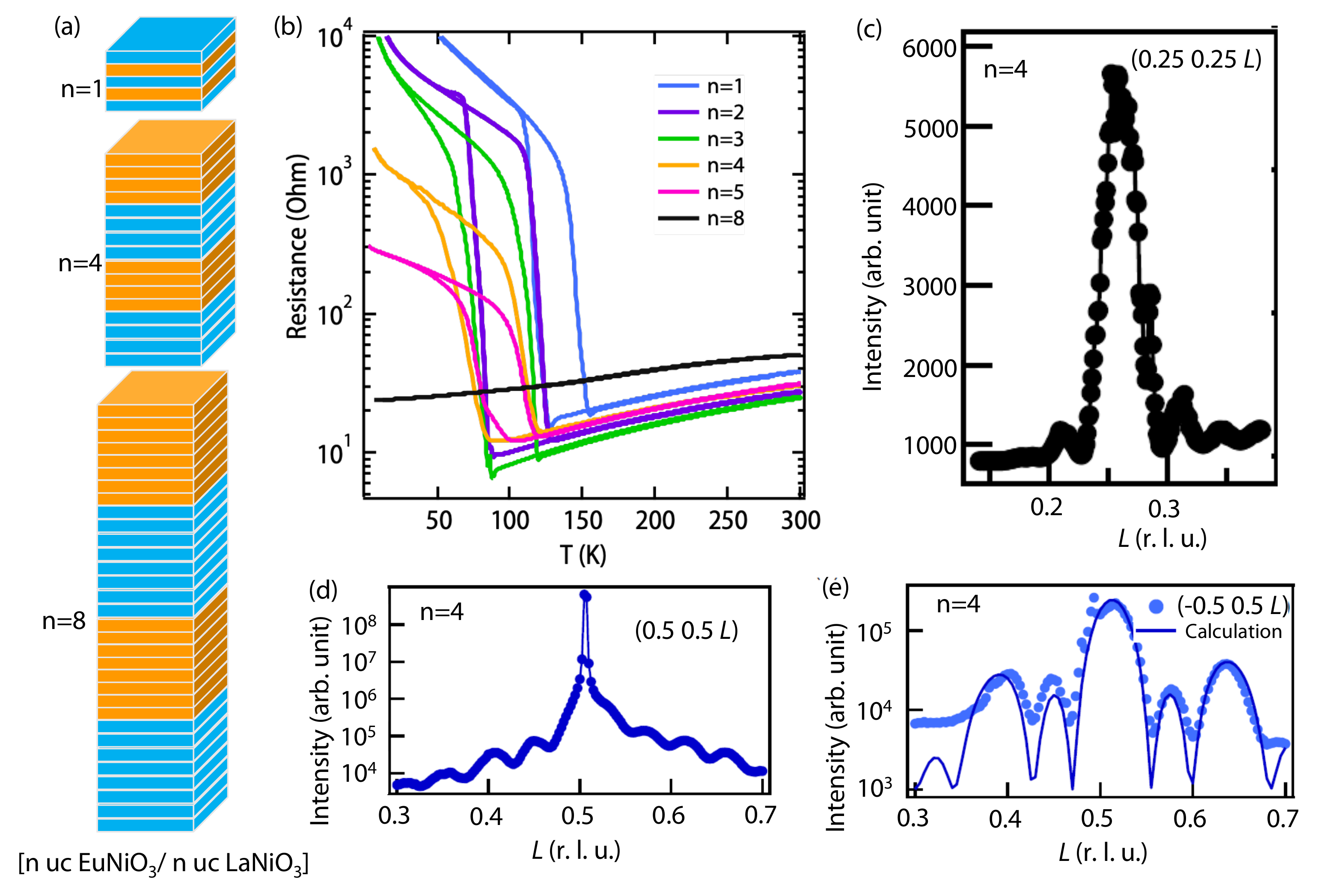}
    \caption{(a) Schematics of deposition sequence for n uc ENO/ n uc LNO superlattice. Cyan and orange color represent ENO and LNO, respectively. (b) Temperature dependent electrical resistivity of n ENO/ n LNO SLs. (c) $L$ scan around (1/4 1/4 1/4) resonant soft x-ray magnetic reflection with the photon energy tuned to the Ni $L_3$ edge for the n=4 SL. The absence of any satellite along $L$ signifies no magnetic contrast between ENO and LNO layers i.e. the $E^\prime$-type antiferromagnetic spin arrangement is present throughout the entire film. $L$ scans for the n=4 sample through (d) (1/2 1/2 1/2) and (e) (-1/2 1/2 1/2) reflections. The (1/2 1/2 1/2) peak is allowed for both orthorhombic and monoclinic symmetry but forbidden for rhombohedral symmetry.  The absence of any satellite established no significant structural contrast between ENO and LNO.   The (-1/2 1/2 1/2) reflection is allowed for monoclinic and forbidden for both orthorhombic \& rhombohedral symmetry. The presence of a satellite peak in this case, signifies contrast in the breathing mode distortion between the ENO and LNO layers. LNO layers have become orthorhombic  without (or much suppressed compared to ENO) breathing mode distortion. The  data shown in panel (b)-(e) have been adapted  from Refs.~\cite{Kim:2020p127601,Patel:2020p041113}). }
    \label{Fig2}
\end{figure*}

When a hetero-interface between two perovskite oxides is formed, there can be a mismatch in lattice constants, octahedral rotational pattern, lattice symmetry, number of electrons on the transition metal site, $d$ orbital configuration, spin ordering pattern, and so on. Collectively,  they may lead to novel emergent phenomena.   The first group of examples that we describe here, consists of an interface between two members of the $RE$NiO$_3$ series to illustrate how pure structural effects can introduce new phases of matter. From a structural viewpoint, LaNiO$_3$ (LNO) is  a   special member of the $RE$NiO$_3$ series as it has a different octahedral rotational pattern ($a^{-}a^{-}a^{-}$) compared to the other members ($a^{-}b^{+}c^{-}$)~\cite{Glazer:1972p09401,May:2010p014110,Tung:2013p205112}. Since $RE$NiO$_3$ members have a strong tendency to retain their bulk-like symmetry even in thin-film form~\cite{Tung:2013p205112}, a strong structural competition is anticipated at the interface between LNO and other members of $RE$NiO$_3$. This effect   has been indeed observed in ultrathin $n$ uc EuNiO$_3$/$m$ uc LaNiO$_3$ [nENO/mLNO] superlattices (here uc is unit cell in pseudocubic  notation, $n$ and $m$ =1, 2)~\cite{Middey:2018p156801,Middey:2018p045115,Middey:2018p081602,Patel:2020p041113}. The 1ENO/2LNO SL having rhombohedral symmetry remains metallic, whereas the superlattices with $n\ge m$ have orthorhombic/monoclinic symmetry  and exhibit first-order MIT as a function of temperature. Moreover, resonant scattering experiments have found that the metal phase of 2ENO/1LNO holds monoclinic symmetry~\cite{Middey:2018p156801}. These results demonstrate that the \textit{structural symmetry change} is not a necessary factor for MIT,   thus solving a long-standing puzzle about the origin of MIT. The competition between interfacial unit cell and bulk-like unit cells were further investigated  by growing nENO/nLNO SLs (Fig.~\ref{Fig2}(a)) ~\cite{Patel:2020p041113,Kim:2020p127601}. The SL with $n$=8 uc remains metallic, similar to bulk LNO, whereas all short periodic SLs exhibit simultaneous MIT and antiferromagnetic transition (Fig.~\ref{Fig2}(b)). Most importantly, thicker bulk-like LaNiO$_3$  layers in these SLs also become antiferromagnetic (Fig.~\ref{Fig2}(c)). Moreover, the bond disproportionation (BD) and charge disproportionation (CD) compete with each other within these SLs~\cite{Patel:2020p041113,Kim:2020p127601}, in contrast to their cooperative nature in  bulk $RE$NiO$_3$~\cite{Staub:2002p126402,Green:2016p195127}. These emergent behaviors are linked to the fact that LNO layers in short periodic SLs are forced to follow the octahedral tilt pattern of highly distorted ENO layers but have zero or greatly reduced BD (see Fig.~\ref{Fig2}(d)-(e))~\cite{Kim:2020p127601}. LaNiO$_3$ layers also follow the octahedral rotational pattern of NdNiO$_3$ in NdNiO$_3$/LaNiO$_3$ superlattices~\cite{Lee:2021p54466}. The interfacial length scale is found to be different in the case of NdNiO$_3$/SmNiO$_3$ SLs, signifying the importance of bulk energetics in artificial quantum materials~\cite{Dominguez:2020p1182}.

 \begin{figure*}
    \centering
    \includegraphics[width=0.85\textwidth]{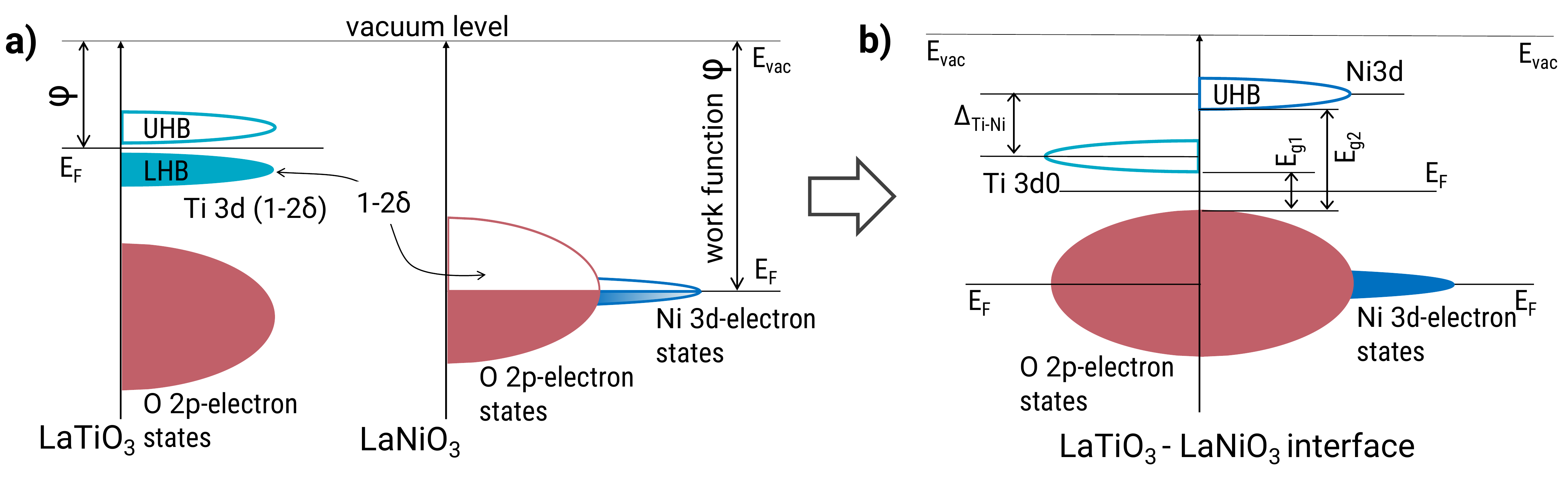}
    \caption{(a) and (b)  Charge transfer between  metallic LaNiO$_3$ and insulating  LaTiO$_3$ resulting in the formation of a new Mott state at the interface. Adapted from Ref.~\onlinecite{Cao:2016p10418}}
    \label{Fig3}
\end{figure*}

\section{Emergent phases of $RE$NiO$_3$ due to interfacial charge transfer}
The phenomena due to charge transfer across the interface between two different semiconducting layers remain at the forefront of condensed matter physics. The interesting effects include a $p-n$ junction, two-dimensional electron gas hosting quantum Hall effect, etc. Similar charge transfer physics is being investigated in the case of complex oxide heterostructures. However, the underlying process is more complex due to electron-phonon coupling, strong electron interaction, strong anisotropy of $d$ orbitals, diverse spin orders, etc. Broadly  speaking, charge transfer in oxide heterostructures is controlled by two factors: equalization of chemical potential across the interface and the stability of a certain redox pair. For example, in $AB^{+n}$O$_3$/$AB_1^{+n_1}$O$_3$  heterostructures, an electron transfer from $AB$O$_3$ to $AB_1$O$_3$   can be expected if the pair $B^{+(n+1)}$/$B_1^{+(n_1-1)}$ is more stable than the $B^{+n}$/$B_1^{+n_1}$. Such redox pair-driven charge transfer is short-range and typically limited within the interface's first one/two unit cells.

Since the most stable oxidation state of Ni is +2 and Ni$^{4+}$ is extremely rare,  electron doping into the $RE$NiO$_3$ layers is possible through a sharp interface with other purposely selected perovskites. As a result, such electron transfer can alter orbital and spin configurations, and structural parameters (Ni-O bond lengths and Ni-O-Ni bond angles), resulting in new electronic and magnetic states. Considering  the interface between LaNiO$_3$ and Mott insulating LaTiO$_3$ (LTO), Chen et al.~\cite{Chen:2013p116403} predicted  electron transfer into LaNiO$_3$ would strongly enhance correlation effects on Ni sites.  The follow-up experimental work on 2LTO/2LNO superlattice confirmed the interfacial electron transfer (Ni$^{+3}$+Ti$^{+3} \rightarrow$ Ni$^{+2}$+Ti$^{+4}$, see Fig.~\ref{Fig3}) with overall insulating ground state~\cite{Cao:2016p10418}. The charge excitation gap ($E_{g1}$ in Fig.~\ref{Fig3}) was found to be around 0.2 eV. X-ray absorption spectroscopy measurement determined a charge gap of 1.3 eV between Ti $3d$ and Ni upper Hubbard band (UHB) state~\cite{Cao:2016p10418}. From these, the correlated gap between UHBs and LHBs (lower Hubbard band) for Ni was deduced to be 1.5 eV. Detailed experimental work on a series of $RE$NiO$_3$/GdTiO$_3$ heterostructures ($RE$=La, Nd, Sm) further revealed that the hybridization effect of $RE$NiO$_3$ has a major impact on the electron transfer process~\cite{Grisolia:2016p484}. Incorporating polar electric fields along with the charge transfer, a large orbital polarization of Ni was reported using LaTiO$_3$/LaNiO$_3$/LaAlO$_3$  heterostructures~\cite{Chen:2013p186402,Disa:2015p026801}. Since  Ti$^{+4}$ with an empty 3$d^0$ shell can not donate any electron, no charge transfer has been found in SrTiO$_3$/$RE$NiO$_3$ heterostructures~\cite{Kaiser:2011p116402,Chen:2021p1295}. Similarly, no charge transfer has been seen in LaFeO$_3$/SmNiO$_3$ heterostructures~\cite{Liao:2018p9515}. This can be readily understood from the fact that  Fe$^{+5}$ (3$d^5$) is in a half-filled state and thus highly stable. In both cases~\cite{Chen:2021p1295,Liao:2018p9515}, the modulation of $T_\mathrm{MIT}$ has been linked with the change in the octahedral rotational pattern of NiO$_6$  octahedra near the interfaces.

Several manganite-nickelate heterostructures have been also extensively investigated. In bulk, CaMnO$_3$ is an insulator and undergoes an antiferromagnetic transition around 125 K~\cite{Zeng:1999p8784}. As argued before, an interfacial charge distribution would be very unlikely in  LaNiO$_3$/CaMnO$_3$ heterostructure since in CaMn$^{+4}$O$_3$, Mn has highly stable half-filled $t_{2g}$ orbitals. This simple consideration is consistent with the experiment~\cite{Grutter:2013p087202}. The most surprising observation in $n$ uc LaNiO$_3$/m uc CaMnO$_3$ is ferromagnetism within the first interfacial unit cell of CaMnO$_3$ when the thickness of LNO layers is thick enough to make it metallic~\cite{Grutter:2013p087202}. In the absence of any charge transfer, the origin of the unexpected ferromagnetism was explained in terms of double exchange among interfacial Mn ions in the presence of an itinerant electron in the adjacent LNO layers.
In contrast, interfacial charge transfer (ICT) is expected in LaNiO$_3$/LaMn$^{+3}$O$_3$ (LNO/LMO) interface as the pair Mn$^{+4}$-Ni$^{+2}$ is more stable than Mn$^{+3}$-Ni$^{+3}$. XAS measurements performed on 2 uc LNO/ 2 uc LMO SL indeed found Ni with +2 and Mn with +4 charge states~\cite{Hoffman:2013p144411}.

LNO/LMO SLs grown along [111] direction exhibit exchange bias, which is absent in [001]-oriented SLs~\cite{Gibert:2012p195}. Follow-up XAS measurements revealed that the charge transfer effect is stronger in [111]-oriented case~\cite{Piamonteze:2015p014426}. Ferromagnetic coupling between Ni and Mn has been confirmed by XMCD (x-ray magnetic circular dichroism) study~\cite{Hoffman:2013p144411,Piamonteze:2015p014426}. ICT has been also reported at the interface of nickelates with mixed-valent manganite~\cite{Hoffman:2016p041038,Fabbris:2018p180401,Xu:2018p30803,Chen:2020p054408}. Among several reports, the most notable observation is the existence of helical spin arrangements within LNO layers, containing atomic-like Ni$^{+2}$ with 3$d^8$ electron configuration~\cite{Hoffman:2016p041038,Fabbris:2018p180401}.

Wrobel et al. investigated a series of La$_2$CuO$_4$/LaO/LaNiO$_3$ heterostructures to separate the dopant and doped layers from each other~\cite{Wrobel:2018p035001}. The doped electrons in LaNiO$_3$ result in interfacial charge disproportionation that strongly influences electrical transport behaviors.

All the compounds discussed so far are 3$d$ transition metal oxides (TMOs), where the effect of spin-orbit coupling (SOC) is usually disregarded in the determination of the ground state as the crystal field (CF) splitting $\Delta_{CF}$ is approximately an order of magnitude larger than the intrinsic  SOC $\lambda$. However, in 4d and 5d TMOs, the enhancement of $\lambda$ makes it comparable to $\Delta_{CF}$. As a result of the competing interactions dominated by SOC, several unusual quantum states, including topological insulators, quantum spin liquids, Weyl semimetals, and Kitaev magnets, have been recently predicted~\cite{Witczak:2014p57}. A prototypical example of this class of compounds is the Ruddlesden-Popper (RP) series (Sr$_{n+1}$Ir$_n$O$_{3n+1}$, $n = 1, 2, ..., \infty$) of iridium oxides (Ir$^{4+}$, 5d$^5$). In the layered perovskite Sr$_2$IrO$_4$ ($n = 1$), the t$_{2g}$ band is split by the strong SOC leading to the formation of $J_{eff} = 1/2$ and $J_{eff} = 3/2$ subbands. A modest value of the Hubbard $U$ further opens a gap and splits the narrow $J_{eff} = 1/2$ band into LHB and UHB, giving rise to a unique spin-orbit entangled Mott-insulating ground state with antiferromagnetic long-range ordering~\cite{Kim:2008p076402}. On the other hand, for the perovskite SrIrO$_3$ ($n = \infty$), the increased Ir 5$d$ bandwidth $W$, together with comparable $\lambda$ and $U$, eventually prevents a Mott gap opening and results in an intriguing correlated semimetallic ground state~\cite{Moon:2008p226402}. To validate this idea, Liu et al. investigated  SrIrO$_3$/LaNiO$_3$ superlattices, where a full electron transfer from Ir to Ni was observed, resulting in $S$=1 spin configurations of interfacial Ni sites~\cite{Liu:2019p19863}. Moreover, the crystal field splitting for the interfacial IrO$_6$ octahedra was dominant over the SOC, which, together with the Hund’s coupling, results in $S$=1 state of Ir sites. Such suppression of SOC effects through ICT warrants careful checks and reinterpretations of SOC-driven effects in thin films and heterostructures of 5$d$ oxides.

 \section{Summary and outlook}

 Starting with a brief description about bulk $RE$NiO$_3$  and principle of epitaxial stabilization, we have discussed emergent behavior of $RE$NiO$_3$ heterostructures, focusing  primarily on octahedral engineering and interfacial charge transfer. In each of these examples, Ni remains in six fold coordination with oxygen. Interestingly, the local symmetry of Ni can be converted to pyramidal environment through interface engineering and topotactic reduction of $RE$NiO$_3$-based heterostructure, leading to the observation of large orbital polarization of Ni~\cite{Liao:2019p589,Ortiz:2021p165137}. It would be very interesting to probe electronic and spin reconstruction of such pyramidal NiO$_5$-units utilizing different members of $RE$NiO$_3$ series.
 Apart from interfacial charge doping, additional carriers (either electron or hole) can be introduced in $RE$NiO$_3$  by chemical substitution~\cite{Cheong:1994p1087,Garcia:1995p13563,Xiang:2010p032114,Shi:2014p4860} and external electric field~\cite{Asanuma:2010p142110,Scherwitzl:2010p5517,Simon:20115p122102,Ha:2013p183102}. While the modulation of MIT of $RE$NiO$_3$ has been demonstrated using both routes, the underlying mechanism and the effect of carrier doping on BD phase is still missing. Another new direction would be exploring high entropy rare-earth nickelate combining at-least five members of the series in equimolar portion~\cite{Rost:2015p8485,Ranjan:2020p071601}.

  \begin{figure}[h]
    \centering
    \includegraphics[width=0.45\textwidth]{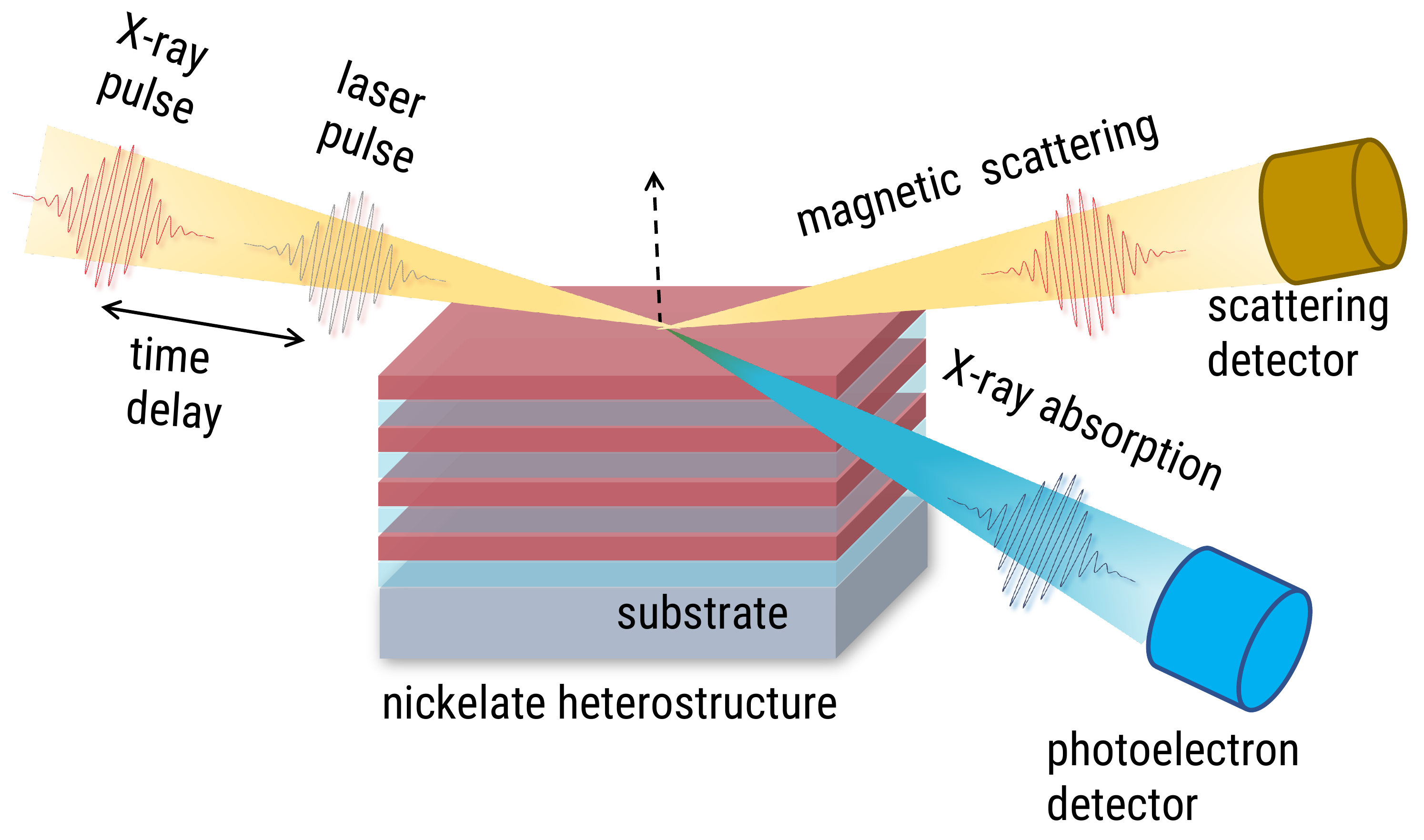}
    \caption{Experimental arrangement of a pump-probe setup, used in Ref.~\onlinecite{Stoica:2020disentangling}.}
    \label{Fig4}
\end{figure}

 Due to the presence of simultaneous transitions in $RE$NiO$_3$,  it is highly non-trivial to discern  a key order-parameter responsible for specific physical properties. This  problem has been  addressed by  the recently developed ultra-fast pump-probe technique that  explores dynamics of these coupled transitions  within a broad time window spanning from femto- to nanoseconds~\cite{Forst:2011p854}.  Taking NdNiO$_3$ as a prototypical system, Stoica et al. optically  excited the spin and charge degrees of freedom away from the equilibrium and then  used femtosecond soft x-ray pulses to simultaneously  probe
the recovery dynamics of magnetic and electronic orders including  their characteristic timescale  (see Fig.~\ref{Fig4})~\cite{Stoica:2020disentangling}. In this experiment the authors observed that after the excitation by an optical pump pulse, the magnetism collapses markedly faster than the time scale over which the insulator-to-metal transition occurs. This  observation allowed to exclude magnetism as the prime driver of the MIT. In an another experiment, Abreu et. al. investigated the THz conductivity dynamics in NdNiO$_3$ following  an optical  thermal pump from the insulating ground state into  a metastable metallic phase~\cite{Abreu:2020p7422}.
The application of the THz probe revealed a remarkable  contrast between the first (in NdNiO$_3$) and second (in EuNiO$_3$) order dynamics. These early experiments clearly demonstrate the importance of  pump-probe experiments to investigate nickelates in the time domain and shed light on controversial or unsolved questions in the equilibrium.   Notably, by tracking the characteristic timescales, one can acquire a definitive picture of how charge, magnetic, orbital, and lattice orders emerge and evolve. Moreover, by recording the time evolution of the signals in multiple detection channels, one can figure out the key progenitor for driving the complex transitions.  In the future, it would be interesting to examine if \textit{dynamic strain engineering} can tune the quantum phases. Furthermore, probing and controlling entwined orders can develop into an alternative venue to control these phases with optical stimulation either above or below the band-gap (i.e., electronic versus phonon excitation). For example, it would be intriguing to verify if coherent phonon coupling to spin order during the photo-induced insulator-to-metal transition can substantiate coherence between lattice and magnetic degrees of freedom on picoseconds timescales and at THz speeds.

Overall, it is truly impressive how this particular materials system highlighted by John Goodenough and collaborators almost  60 years ago went through several incarnations only to continue carrying more fascinating  stories for future generations of physicists. We believe many more surprises remain hidden in the structures based on rare-earth nickelate, only to be unraveled with the advancement of experimental techniques.

\section{Acknowledgement}
This paper is written on the monumental occasion of the centennial birthday of Professor John Goodenough.

JC was supported by the U.S. Department of Energy, Office of Science, Office of Basic Energy Sciences under Award Number DE-SC0022160.
SM acknowledges financial support of a  DST Nanomission grant (Grant No. DST/NM/NS/2018/246) and a SERB Early Career Research Award (Grant No. ECR/2018/001512).
JC thanks John W. Freeland (Argonne National  Laboratory) and Richard Averitt (University of California, San-Diego) for numerous fruitful  discussions on ultra-fast  physics of nickelates. We thank Ranjan Kumar Patel and  Nandana Bhattacharya for their  help  during the preparation of the manuscript.


\begin{thebibliography}{122}%
\makeatletter
\providecommand \@ifxundefined [1]{%
 \@ifx{#1\undefined}
}%
\providecommand \@ifnum [1]{%
 \ifnum #1\expandafter \@firstoftwo
 \else \expandafter \@secondoftwo
 \fi
}%
\providecommand \@ifx [1]{%
 \ifx #1\expandafter \@firstoftwo
 \else \expandafter \@secondoftwo
 \fi
}%
\providecommand \natexlab [1]{#1}%
\providecommand \enquote  [1]{``#1''}%
\providecommand \bibnamefont  [1]{#1}%
\providecommand \bibfnamefont [1]{#1}%
\providecommand \citenamefont [1]{#1}%
\providecommand \href@noop [0]{\@secondoftwo}%
\providecommand \href [0]{\begingroup \@sanitize@url \@href}%
\providecommand \@href[1]{\@@startlink{#1}\@@href}%
\providecommand \@@href[1]{\endgroup#1\@@endlink}%
\providecommand \@sanitize@url [0]{\catcode `\\12\catcode `\$12\catcode
  `\&12\catcode `\#12\catcode `\^12\catcode `\_12\catcode `\%12\relax}%
\providecommand \@@startlink[1]{}%
\providecommand \@@endlink[0]{}%
\providecommand \url  [0]{\begingroup\@sanitize@url \@url }%
\providecommand \@url [1]{\endgroup\@href {#1}{\urlprefix }}%
\providecommand \urlprefix  [0]{URL }%
\providecommand \Eprint [0]{\href }%
\providecommand \doibase [0]{http://dx.doi.org/}%
\providecommand \selectlanguage [0]{\@gobble}%
\providecommand \bibinfo  [0]{\@secondoftwo}%
\providecommand \bibfield  [0]{\@secondoftwo}%
\providecommand \translation [1]{[#1]}%
\providecommand \BibitemOpen [0]{}%
\providecommand \bibitemStop [0]{}%
\providecommand \bibitemNoStop [0]{.\EOS\space}%
\providecommand \EOS [0]{\spacefactor3000\relax}%
\providecommand \BibitemShut  [1]{\csname bibitem#1\endcsname}%
\let\auto@bib@innerbib\@empty
\bibitem [{\citenamefont {Goodenough}(1963)}]{Goodenough:1963book}%
  \BibitemOpen
  \bibfield  {author} {\bibinfo {author} {\bibfnamefont {J.~B.}\ \bibnamefont
  {Goodenough}},\ }\href@noop {} {\emph {\bibinfo {title} {Magnetism and the
  chemical bond}}},\ Vol.~\bibinfo {volume} {1}\ (\bibinfo  {publisher}
  {Interscience publishers},\ \bibinfo {year} {1963})\BibitemShut {NoStop}%
\bibitem [{\citenamefont {Wold}, \citenamefont {Post},\ and\ \citenamefont
  {Banks}(1957)}]{Wold:1957p4911}%
  \BibitemOpen
  \bibfield  {author} {\bibinfo {author} {\bibfnamefont {A.}~\bibnamefont
  {Wold}}, \bibinfo {author} {\bibfnamefont {B.}~\bibnamefont {Post}}, \ and\
  \bibinfo {author} {\bibfnamefont {E.}~\bibnamefont {Banks}},\ }\href@noop {}
  {\bibfield  {journal} {\bibinfo  {journal} {Journal of the American Chemical
  Society}\ }\textbf {\bibinfo {volume} {79}},\ \bibinfo {pages} {4911}
  (\bibinfo {year} {1957})}\BibitemShut {NoStop}%
\bibitem [{\citenamefont {Goodenough}\ and\ \citenamefont
  {Raccah}(1965)}]{Goodenough:1965p1031}%
  \BibitemOpen
  \bibfield  {author} {\bibinfo {author} {\bibfnamefont {J.}~\bibnamefont
  {Goodenough}}\ and\ \bibinfo {author} {\bibfnamefont {P.}~\bibnamefont
  {Raccah}},\ }\href@noop {} {\bibfield  {journal} {\bibinfo  {journal}
  {Journal of Applied Physics}\ }\textbf {\bibinfo {volume} {36}},\ \bibinfo
  {pages} {1031} (\bibinfo {year} {1965})}\BibitemShut {NoStop}%
\bibitem [{\citenamefont {Goodenough}(1966)}]{Goodenough:1966p1415}%
  \BibitemOpen
  \bibfield  {author} {\bibinfo {author} {\bibfnamefont {J.~B.}\ \bibnamefont
  {Goodenough}},\ }\href@noop {} {\bibfield  {journal} {\bibinfo  {journal}
  {Journal of Applied Physics}\ }\textbf {\bibinfo {volume} {37}},\ \bibinfo
  {pages} {1415} (\bibinfo {year} {1966})}\BibitemShut {NoStop}%
\bibitem [{\citenamefont {Torrance}\ \emph {et~al.}(1992)\citenamefont
  {Torrance}, \citenamefont {Lacorre}, \citenamefont {Nazzal}, \citenamefont
  {Ansaldo},\ and\ \citenamefont {Niedermayer}}]{Torrance:1992p8209}%
  \BibitemOpen
  \bibfield  {author} {\bibinfo {author} {\bibfnamefont {J.~B.}\ \bibnamefont
  {Torrance}}, \bibinfo {author} {\bibfnamefont {P.}~\bibnamefont {Lacorre}},
  \bibinfo {author} {\bibfnamefont {A.~I.}\ \bibnamefont {Nazzal}}, \bibinfo
  {author} {\bibfnamefont {E.~J.}\ \bibnamefont {Ansaldo}}, \ and\ \bibinfo
  {author} {\bibfnamefont {C.}~\bibnamefont {Niedermayer}},\ }\href {\doibase
  10.1103/PhysRevB.45.8209} {\bibfield  {journal} {\bibinfo  {journal} {Phys.
  Rev. B}\ }\textbf {\bibinfo {volume} {45}},\ \bibinfo {pages} {8209}
  (\bibinfo {year} {1992})}\BibitemShut {NoStop}%
\bibitem [{\citenamefont {Anisimov}, \citenamefont {Bukhvalov},\ and\
  \citenamefont {Rice}(1999)}]{Anisimov:1999p7901}%
  \BibitemOpen
  \bibfield  {author} {\bibinfo {author} {\bibfnamefont {V.~I.}\ \bibnamefont
  {Anisimov}}, \bibinfo {author} {\bibfnamefont {D.}~\bibnamefont {Bukhvalov}},
  \ and\ \bibinfo {author} {\bibfnamefont {T.~M.}\ \bibnamefont {Rice}},\
  }\href {\doibase 10.1103/PhysRevB.59.7901} {\bibfield  {journal} {\bibinfo
  {journal} {Phys. Rev. B}\ }\textbf {\bibinfo {volume} {59}},\ \bibinfo
  {pages} {7901} (\bibinfo {year} {1999})}\BibitemShut {NoStop}%
\bibitem [{\citenamefont {Li}\ \emph {et~al.}(2019)\citenamefont {Li},
  \citenamefont {Lee}, \citenamefont {Wang}, \citenamefont {Osada},
  \citenamefont {Crossley}, \citenamefont {Lee}, \citenamefont {Cui},
  \citenamefont {Hikita},\ and\ \citenamefont {Hwang}}]{Li:2019p624}%
  \BibitemOpen
  \bibfield  {author} {\bibinfo {author} {\bibfnamefont {D.}~\bibnamefont
  {Li}}, \bibinfo {author} {\bibfnamefont {K.}~\bibnamefont {Lee}}, \bibinfo
  {author} {\bibfnamefont {B.~Y.}\ \bibnamefont {Wang}}, \bibinfo {author}
  {\bibfnamefont {M.}~\bibnamefont {Osada}}, \bibinfo {author} {\bibfnamefont
  {S.}~\bibnamefont {Crossley}}, \bibinfo {author} {\bibfnamefont {H.~R.}\
  \bibnamefont {Lee}}, \bibinfo {author} {\bibfnamefont {Y.}~\bibnamefont
  {Cui}}, \bibinfo {author} {\bibfnamefont {Y.}~\bibnamefont {Hikita}}, \ and\
  \bibinfo {author} {\bibfnamefont {H.~Y.}\ \bibnamefont {Hwang}},\ }\href@noop
  {} {\bibfield  {journal} {\bibinfo  {journal} {Nature}\ }\textbf {\bibinfo
  {volume} {572}},\ \bibinfo {pages} {624} (\bibinfo {year}
  {2019})}\BibitemShut {NoStop}%
\bibitem [{\citenamefont {Pickett}(2021)}]{Pickett:2021p7}%
  \BibitemOpen
  \bibfield  {author} {\bibinfo {author} {\bibfnamefont {W.~E.}\ \bibnamefont
  {Pickett}},\ }\href@noop {} {\bibfield  {journal} {\bibinfo  {journal}
  {Nature Reviews Physics}\ }\textbf {\bibinfo {volume} {3}},\ \bibinfo {pages}
  {7} (\bibinfo {year} {2021})}\BibitemShut {NoStop}%
\bibitem [{\citenamefont {Medarde}(1997)}]{Medarde:1997p1679}%
  \BibitemOpen
  \bibfield  {author} {\bibinfo {author} {\bibfnamefont {M.~L.}\ \bibnamefont
  {Medarde}},\ }\href@noop {} {\bibfield  {journal} {\bibinfo  {journal}
  {Journal of Physics: Condensed Matter}\ }\textbf {\bibinfo {volume} {9}},\
  \bibinfo {pages} {1679} (\bibinfo {year} {1997})}\BibitemShut {NoStop}%
\bibitem [{\citenamefont {Catalan}(2008)}]{Catalan:2008p729}%
  \BibitemOpen
  \bibfield  {author} {\bibinfo {author} {\bibfnamefont {G.}~\bibnamefont
  {Catalan}},\ }\href {\doibase 10.1080/01411590801992463} {\bibfield
  {journal} {\bibinfo  {journal} {Phase Transitions}\ }\textbf {\bibinfo
  {volume} {81}},\ \bibinfo {pages} {729} (\bibinfo {year} {2008})}\BibitemShut
  {NoStop}%
\bibitem [{\citenamefont {Garc\'{i}a-Mu\~noz}, \citenamefont
  {Rodr\'{i}guez-Carvajal},\ and\ \citenamefont
  {Lacorre}(1994)}]{Garcia:1994p978}%
  \BibitemOpen
  \bibfield  {author} {\bibinfo {author} {\bibfnamefont {J.~L.}\ \bibnamefont
  {Garc\'{i}a-Mu\~noz}}, \bibinfo {author} {\bibfnamefont {J.}~\bibnamefont
  {Rodr\'{i}guez-Carvajal}}, \ and\ \bibinfo {author} {\bibfnamefont
  {P.}~\bibnamefont {Lacorre}},\ }\href {\doibase 10.1103/PhysRevB.50.978}
  {\bibfield  {journal} {\bibinfo  {journal} {Phys. Rev. B}\ }\textbf {\bibinfo
  {volume} {50}},\ \bibinfo {pages} {978} (\bibinfo {year} {1994})}\BibitemShut
  {NoStop}%
\bibitem [{\citenamefont {Scagnoli}\ \emph {et~al.}(2005)\citenamefont
  {Scagnoli}, \citenamefont {Staub}, \citenamefont {Janousch}, \citenamefont
  {Mulders}, \citenamefont {Shi}, \citenamefont {Meijer}, \citenamefont
  {Rosenkranz}, \citenamefont {Wilkins}, \citenamefont {Paolasini},
  \citenamefont {Karpinski}, \citenamefont {Kazakov},\ and\ \citenamefont
  {Lovesey}}]{Scagnoli:2005p155111}%
  \BibitemOpen
  \bibfield  {author} {\bibinfo {author} {\bibfnamefont {V.}~\bibnamefont
  {Scagnoli}}, \bibinfo {author} {\bibfnamefont {U.}~\bibnamefont {Staub}},
  \bibinfo {author} {\bibfnamefont {M.}~\bibnamefont {Janousch}}, \bibinfo
  {author} {\bibfnamefont {A.~M.}\ \bibnamefont {Mulders}}, \bibinfo {author}
  {\bibfnamefont {M.}~\bibnamefont {Shi}}, \bibinfo {author} {\bibfnamefont
  {G.~I.}\ \bibnamefont {Meijer}}, \bibinfo {author} {\bibfnamefont
  {S.}~\bibnamefont {Rosenkranz}}, \bibinfo {author} {\bibfnamefont {S.~B.}\
  \bibnamefont {Wilkins}}, \bibinfo {author} {\bibfnamefont {L.}~\bibnamefont
  {Paolasini}}, \bibinfo {author} {\bibfnamefont {J.}~\bibnamefont
  {Karpinski}}, \bibinfo {author} {\bibfnamefont {S.~M.}\ \bibnamefont
  {Kazakov}}, \ and\ \bibinfo {author} {\bibfnamefont {S.~W.}\ \bibnamefont
  {Lovesey}},\ }\href {\doibase 10.1103/PhysRevB.72.155111} {\bibfield
  {journal} {\bibinfo  {journal} {Phys. Rev. B}\ }\textbf {\bibinfo {volume}
  {72}},\ \bibinfo {pages} {155111} (\bibinfo {year} {2005})}\BibitemShut
  {NoStop}%
\bibitem [{\citenamefont {Scagnoli}\ \emph {et~al.}(2006)\citenamefont
  {Scagnoli}, \citenamefont {Staub}, \citenamefont {Mulders}, \citenamefont
  {Janousch}, \citenamefont {Meijer}, \citenamefont {Hammerl}, \citenamefont
  {Tonnerre},\ and\ \citenamefont {Stojic}}]{Scagnoli:2006p100409}%
  \BibitemOpen
  \bibfield  {author} {\bibinfo {author} {\bibfnamefont {V.}~\bibnamefont
  {Scagnoli}}, \bibinfo {author} {\bibfnamefont {U.}~\bibnamefont {Staub}},
  \bibinfo {author} {\bibfnamefont {A.~M.}\ \bibnamefont {Mulders}}, \bibinfo
  {author} {\bibfnamefont {M.}~\bibnamefont {Janousch}}, \bibinfo {author}
  {\bibfnamefont {G.~I.}\ \bibnamefont {Meijer}}, \bibinfo {author}
  {\bibfnamefont {G.}~\bibnamefont {Hammerl}}, \bibinfo {author} {\bibfnamefont
  {J.~M.}\ \bibnamefont {Tonnerre}}, \ and\ \bibinfo {author} {\bibfnamefont
  {N.}~\bibnamefont {Stojic}},\ }\href {\doibase 10.1103/PhysRevB.73.100409}
  {\bibfield  {journal} {\bibinfo  {journal} {Phys. Rev. B}\ }\textbf {\bibinfo
  {volume} {73}},\ \bibinfo {pages} {100409} (\bibinfo {year}
  {2006})}\BibitemShut {NoStop}%
\bibitem [{\citenamefont {Scagnoli}\ \emph {et~al.}(2008)\citenamefont
  {Scagnoli}, \citenamefont {Staub}, \citenamefont {Bodenthin}, \citenamefont
  {Garc\'{i}a-Fern\'andez}, \citenamefont {Mulders}, \citenamefont {Meijer},\
  and\ \citenamefont {Hammerl}}]{Scagnoli:2008p115138}%
  \BibitemOpen
  \bibfield  {author} {\bibinfo {author} {\bibfnamefont {V.}~\bibnamefont
  {Scagnoli}}, \bibinfo {author} {\bibfnamefont {U.}~\bibnamefont {Staub}},
  \bibinfo {author} {\bibfnamefont {Y.}~\bibnamefont {Bodenthin}}, \bibinfo
  {author} {\bibfnamefont {M.}~\bibnamefont {Garc\'{i}a-Fern\'andez}}, \bibinfo
  {author} {\bibfnamefont {A.~M.}\ \bibnamefont {Mulders}}, \bibinfo {author}
  {\bibfnamefont {G.~I.}\ \bibnamefont {Meijer}}, \ and\ \bibinfo {author}
  {\bibfnamefont {G.}~\bibnamefont {Hammerl}},\ }\href {\doibase
  10.1103/PhysRevB.77.115138} {\bibfield  {journal} {\bibinfo  {journal} {Phys.
  Rev. B}\ }\textbf {\bibinfo {volume} {77}},\ \bibinfo {pages} {115138}
  (\bibinfo {year} {2008})}\BibitemShut {NoStop}%
\bibitem [{\citenamefont {Zhou}\ and\ \citenamefont
  {Goodenough}(2004)}]{Zhou:2004p153105}%
  \BibitemOpen
  \bibfield  {author} {\bibinfo {author} {\bibfnamefont {J.-S.}\ \bibnamefont
  {Zhou}}\ and\ \bibinfo {author} {\bibfnamefont {J.~B.}\ \bibnamefont
  {Goodenough}},\ }\href {\doibase 10.1103/PhysRevB.69.153105} {\bibfield
  {journal} {\bibinfo  {journal} {Phys. Rev. B}\ }\textbf {\bibinfo {volume}
  {69}},\ \bibinfo {pages} {153105} (\bibinfo {year} {2004})}\BibitemShut
  {NoStop}%
\bibitem [{\citenamefont {Ha}\ \emph {et~al.}(2013{\natexlab{a}})\citenamefont
  {Ha}, \citenamefont {Jaramillo}, \citenamefont {Silevitch}, \citenamefont
  {Schoofs}, \citenamefont {Kerman}, \citenamefont {Baniecki},\ and\
  \citenamefont {Ramanathan}}]{Ha:2013p125150}%
  \BibitemOpen
  \bibfield  {author} {\bibinfo {author} {\bibfnamefont {S.~D.}\ \bibnamefont
  {Ha}}, \bibinfo {author} {\bibfnamefont {R.}~\bibnamefont {Jaramillo}},
  \bibinfo {author} {\bibfnamefont {D.~M.}\ \bibnamefont {Silevitch}}, \bibinfo
  {author} {\bibfnamefont {F.}~\bibnamefont {Schoofs}}, \bibinfo {author}
  {\bibfnamefont {K.}~\bibnamefont {Kerman}}, \bibinfo {author} {\bibfnamefont
  {J.~D.}\ \bibnamefont {Baniecki}}, \ and\ \bibinfo {author} {\bibfnamefont
  {S.}~\bibnamefont {Ramanathan}},\ }\href {\doibase
  10.1103/PhysRevB.87.125150} {\bibfield  {journal} {\bibinfo  {journal} {Phys.
  Rev. B}\ }\textbf {\bibinfo {volume} {87}},\ \bibinfo {pages} {125150}
  (\bibinfo {year} {2013}{\natexlab{a}})}\BibitemShut {NoStop}%
\bibitem [{\citenamefont {Ojha}\ \emph {et~al.}(2019)\citenamefont {Ojha},
  \citenamefont {Ray}, \citenamefont {Das}, \citenamefont {Middey},
  \citenamefont {Sarkar}, \citenamefont {Mahadevan}, \citenamefont {Wang},
  \citenamefont {Zhu}, \citenamefont {Liu}, \citenamefont {Kareev},\ and\
  \citenamefont {Chakhalian}}]{Ojha:2019p235153}%
  \BibitemOpen
  \bibfield  {author} {\bibinfo {author} {\bibfnamefont {S.~K.}\ \bibnamefont
  {Ojha}}, \bibinfo {author} {\bibfnamefont {S.}~\bibnamefont {Ray}}, \bibinfo
  {author} {\bibfnamefont {T.}~\bibnamefont {Das}}, \bibinfo {author}
  {\bibfnamefont {S.}~\bibnamefont {Middey}}, \bibinfo {author} {\bibfnamefont
  {S.}~\bibnamefont {Sarkar}}, \bibinfo {author} {\bibfnamefont
  {P.}~\bibnamefont {Mahadevan}}, \bibinfo {author} {\bibfnamefont
  {Z.}~\bibnamefont {Wang}}, \bibinfo {author} {\bibfnamefont {Y.}~\bibnamefont
  {Zhu}}, \bibinfo {author} {\bibfnamefont {X.}~\bibnamefont {Liu}}, \bibinfo
  {author} {\bibfnamefont {M.}~\bibnamefont {Kareev}}, \ and\ \bibinfo {author}
  {\bibfnamefont {J.}~\bibnamefont {Chakhalian}},\ }\href {\doibase
  10.1103/PhysRevB.99.235153} {\bibfield  {journal} {\bibinfo  {journal} {Phys.
  Rev. B}\ }\textbf {\bibinfo {volume} {99}},\ \bibinfo {pages} {235153}
  (\bibinfo {year} {2019})}\BibitemShut {NoStop}%
\bibitem [{\citenamefont {Eguchi}\ \emph {et~al.}(2009)\citenamefont {Eguchi},
  \citenamefont {Chainani}, \citenamefont {Taguchi}, \citenamefont {Matsunami},
  \citenamefont {Ishida}, \citenamefont {Horiba}, \citenamefont {Senba},
  \citenamefont {Ohashi},\ and\ \citenamefont {Shin}}]{Eguchi:2009p115122}%
  \BibitemOpen
  \bibfield  {author} {\bibinfo {author} {\bibfnamefont {R.}~\bibnamefont
  {Eguchi}}, \bibinfo {author} {\bibfnamefont {A.}~\bibnamefont {Chainani}},
  \bibinfo {author} {\bibfnamefont {M.}~\bibnamefont {Taguchi}}, \bibinfo
  {author} {\bibfnamefont {M.}~\bibnamefont {Matsunami}}, \bibinfo {author}
  {\bibfnamefont {Y.}~\bibnamefont {Ishida}}, \bibinfo {author} {\bibfnamefont
  {K.}~\bibnamefont {Horiba}}, \bibinfo {author} {\bibfnamefont
  {Y.}~\bibnamefont {Senba}}, \bibinfo {author} {\bibfnamefont
  {H.}~\bibnamefont {Ohashi}}, \ and\ \bibinfo {author} {\bibfnamefont
  {S.}~\bibnamefont {Shin}},\ }\href {\doibase 10.1103/PhysRevB.79.115122}
  {\bibfield  {journal} {\bibinfo  {journal} {Phys. Rev. B}\ }\textbf {\bibinfo
  {volume} {79}},\ \bibinfo {pages} {115122} (\bibinfo {year}
  {2009})}\BibitemShut {NoStop}%
\bibitem [{\citenamefont {Yoo}\ \emph {et~al.}(2015)\citenamefont {Yoo},
  \citenamefont {Hyun}, \citenamefont {Moreschini}, \citenamefont {Kim},
  \citenamefont {Chang}, \citenamefont {Sohn}, \citenamefont {Jeong},
  \citenamefont {Sinn}, \citenamefont {Kim}, \citenamefont {Bostwick},
  \citenamefont {Rotenberg}, \citenamefont {Shim},\ and\ \citenamefont
  {Noh}}]{Yoo:2015p8746}%
  \BibitemOpen
  \bibfield  {author} {\bibinfo {author} {\bibfnamefont {H.~K.}\ \bibnamefont
  {Yoo}}, \bibinfo {author} {\bibfnamefont {S.~I.}\ \bibnamefont {Hyun}},
  \bibinfo {author} {\bibfnamefont {L.}~\bibnamefont {Moreschini}}, \bibinfo
  {author} {\bibfnamefont {H.-D.}\ \bibnamefont {Kim}}, \bibinfo {author}
  {\bibfnamefont {Y.~J.}\ \bibnamefont {Chang}}, \bibinfo {author}
  {\bibfnamefont {C.~H.}\ \bibnamefont {Sohn}}, \bibinfo {author}
  {\bibfnamefont {D.~W.}\ \bibnamefont {Jeong}}, \bibinfo {author}
  {\bibfnamefont {S.}~\bibnamefont {Sinn}}, \bibinfo {author} {\bibfnamefont
  {Y.~S.}\ \bibnamefont {Kim}}, \bibinfo {author} {\bibfnamefont
  {A.}~\bibnamefont {Bostwick}}, \bibinfo {author} {\bibfnamefont
  {E.}~\bibnamefont {Rotenberg}}, \bibinfo {author} {\bibfnamefont {J.~H.}\
  \bibnamefont {Shim}}, \ and\ \bibinfo {author} {\bibfnamefont {T.~W.}\
  \bibnamefont {Noh}},\ }\href {http://dx.doi.org/10.1038/srep08746} {\bibfield
   {journal} {\bibinfo  {journal} {Sci. Rep.}\ }\textbf {\bibinfo {volume}
  {5}},\ \bibinfo {pages} {8746} (\bibinfo {year} {2015})}\BibitemShut
  {NoStop}%
\bibitem [{\citenamefont {Dhaka}\ \emph {et~al.}(2015)\citenamefont {Dhaka},
  \citenamefont {Das}, \citenamefont {Plumb}, \citenamefont {Ristic},
  \citenamefont {Kong}, \citenamefont {Matt}, \citenamefont {Xu}, \citenamefont
  {Dolui}, \citenamefont {Razzoli}, \citenamefont {Medarde}, \citenamefont
  {Patthey}, \citenamefont {Shi}, \citenamefont {Radovi\ifmmode~\acute{c}\else
  \'{c}\fi{}},\ and\ \citenamefont {Mesot}}]{Dhaka:2015p035127}%
  \BibitemOpen
  \bibfield  {author} {\bibinfo {author} {\bibfnamefont {R.~S.}\ \bibnamefont
  {Dhaka}}, \bibinfo {author} {\bibfnamefont {T.}~\bibnamefont {Das}}, \bibinfo
  {author} {\bibfnamefont {N.~C.}\ \bibnamefont {Plumb}}, \bibinfo {author}
  {\bibfnamefont {Z.}~\bibnamefont {Ristic}}, \bibinfo {author} {\bibfnamefont
  {W.}~\bibnamefont {Kong}}, \bibinfo {author} {\bibfnamefont {C.~E.}\
  \bibnamefont {Matt}}, \bibinfo {author} {\bibfnamefont {N.}~\bibnamefont
  {Xu}}, \bibinfo {author} {\bibfnamefont {K.}~\bibnamefont {Dolui}}, \bibinfo
  {author} {\bibfnamefont {E.}~\bibnamefont {Razzoli}}, \bibinfo {author}
  {\bibfnamefont {M.}~\bibnamefont {Medarde}}, \bibinfo {author} {\bibfnamefont
  {L.}~\bibnamefont {Patthey}}, \bibinfo {author} {\bibfnamefont
  {M.}~\bibnamefont {Shi}}, \bibinfo {author} {\bibfnamefont {M.}~\bibnamefont
  {Radovi\ifmmode~\acute{c}\else \'{c}\fi{}}}, \ and\ \bibinfo {author}
  {\bibfnamefont {J.}~\bibnamefont {Mesot}},\ }\href {\doibase
  10.1103/PhysRevB.92.035127} {\bibfield  {journal} {\bibinfo  {journal} {Phys.
  Rev. B}\ }\textbf {\bibinfo {volume} {92}},\ \bibinfo {pages} {035127}
  (\bibinfo {year} {2015})}\BibitemShut {NoStop}%
\bibitem [{\citenamefont {Zhou}, \citenamefont {Goodenough},\ and\
  \citenamefont {Dabrowski}(2005)}]{Zhou:2005p226602}%
  \BibitemOpen
  \bibfield  {author} {\bibinfo {author} {\bibfnamefont {J.-S.}\ \bibnamefont
  {Zhou}}, \bibinfo {author} {\bibfnamefont {J.~B.}\ \bibnamefont
  {Goodenough}}, \ and\ \bibinfo {author} {\bibfnamefont {B.}~\bibnamefont
  {Dabrowski}},\ }\href {\doibase 10.1103/PhysRevLett.94.226602} {\bibfield
  {journal} {\bibinfo  {journal} {Phys. Rev. Lett.}\ }\textbf {\bibinfo
  {volume} {94}},\ \bibinfo {pages} {226602} (\bibinfo {year}
  {2005})}\BibitemShut {NoStop}%
\bibitem [{\citenamefont {Liu}\ \emph {et~al.}(2013)\citenamefont {Liu},
  \citenamefont {Kargarian}, \citenamefont {Kareev}, \citenamefont {Gray},
  \citenamefont {Ryan}, \citenamefont {Cruz}, \citenamefont {Tahir},
  \citenamefont {Chuang}, \citenamefont {Guo}, \citenamefont {Rondinelli},
  \citenamefont {Freeland}, \citenamefont {Fiete},\ and\ \citenamefont
  {Chakhalian}}]{Liu:2013p2714}%
  \BibitemOpen
  \bibfield  {author} {\bibinfo {author} {\bibfnamefont {J.}~\bibnamefont
  {Liu}}, \bibinfo {author} {\bibfnamefont {M.}~\bibnamefont {Kargarian}},
  \bibinfo {author} {\bibfnamefont {M.}~\bibnamefont {Kareev}}, \bibinfo
  {author} {\bibfnamefont {B.}~\bibnamefont {Gray}}, \bibinfo {author}
  {\bibfnamefont {P.~J.}\ \bibnamefont {Ryan}}, \bibinfo {author}
  {\bibfnamefont {A.}~\bibnamefont {Cruz}}, \bibinfo {author} {\bibfnamefont
  {N.}~\bibnamefont {Tahir}}, \bibinfo {author} {\bibfnamefont {Y.-D.}\
  \bibnamefont {Chuang}}, \bibinfo {author} {\bibfnamefont {J.}~\bibnamefont
  {Guo}}, \bibinfo {author} {\bibfnamefont {J.~M.}\ \bibnamefont {Rondinelli}},
  \bibinfo {author} {\bibfnamefont {J.~W.}\ \bibnamefont {Freeland}}, \bibinfo
  {author} {\bibfnamefont {G.~A.}\ \bibnamefont {Fiete}}, \ and\ \bibinfo
  {author} {\bibfnamefont {J.}~\bibnamefont {Chakhalian}},\ }\href
  {http://dx.doi.org/10.1038/ncomms3714} {\bibfield  {journal} {\bibinfo
  {journal} {Nat Commun}\ }\textbf {\bibinfo {volume} {4}},\ \bibinfo {pages}
  {2714} (\bibinfo {year} {2013})}\BibitemShut {NoStop}%
\bibitem [{\citenamefont {Liu}\ \emph {et~al.}(2020)\citenamefont {Liu},
  \citenamefont {Humbert}, \citenamefont {Bretz-Sullivan}, \citenamefont
  {Wang}, \citenamefont {Hong}, \citenamefont {Wrobel}, \citenamefont {Zhang},
  \citenamefont {Hoffman}, \citenamefont {Pearson}, \citenamefont {Jiang} \emph
  {et~al.}}]{Liu:2020p1402}%
  \BibitemOpen
  \bibfield  {author} {\bibinfo {author} {\bibfnamefont {C.}~\bibnamefont
  {Liu}}, \bibinfo {author} {\bibfnamefont {V.~F.}\ \bibnamefont {Humbert}},
  \bibinfo {author} {\bibfnamefont {T.~M.}\ \bibnamefont {Bretz-Sullivan}},
  \bibinfo {author} {\bibfnamefont {G.}~\bibnamefont {Wang}}, \bibinfo {author}
  {\bibfnamefont {D.}~\bibnamefont {Hong}}, \bibinfo {author} {\bibfnamefont
  {F.}~\bibnamefont {Wrobel}}, \bibinfo {author} {\bibfnamefont
  {J.}~\bibnamefont {Zhang}}, \bibinfo {author} {\bibfnamefont {J.~D.}\
  \bibnamefont {Hoffman}}, \bibinfo {author} {\bibfnamefont {J.~E.}\
  \bibnamefont {Pearson}}, \bibinfo {author} {\bibfnamefont {J.~S.}\
  \bibnamefont {Jiang}},  \emph {et~al.},\ }\href@noop {} {\bibfield  {journal}
  {\bibinfo  {journal} {Nature communications}\ }\textbf {\bibinfo {volume}
  {11}},\ \bibinfo {pages} {1402} (\bibinfo {year} {2020})}\BibitemShut
  {NoStop}%
\bibitem [{\citenamefont {Allen}\ \emph {et~al.}(2015)\citenamefont {Allen},
  \citenamefont {Hauser}, \citenamefont {Mikheev}, \citenamefont {Zhang},
  \citenamefont {Moreno}, \citenamefont {Son}, \citenamefont {Ouellette},
  \citenamefont {Kally}, \citenamefont {Kozhanov}, \citenamefont {Balents},\
  and\ \citenamefont {Stemmer}}]{Allen:2015p062503}%
  \BibitemOpen
  \bibfield  {author} {\bibinfo {author} {\bibfnamefont {S.~J.}\ \bibnamefont
  {Allen}}, \bibinfo {author} {\bibfnamefont {A.~J.}\ \bibnamefont {Hauser}},
  \bibinfo {author} {\bibfnamefont {E.}~\bibnamefont {Mikheev}}, \bibinfo
  {author} {\bibfnamefont {J.~Y.}\ \bibnamefont {Zhang}}, \bibinfo {author}
  {\bibfnamefont {N.~E.}\ \bibnamefont {Moreno}}, \bibinfo {author}
  {\bibfnamefont {J.}~\bibnamefont {Son}}, \bibinfo {author} {\bibfnamefont
  {D.~G.}\ \bibnamefont {Ouellette}}, \bibinfo {author} {\bibfnamefont
  {J.}~\bibnamefont {Kally}}, \bibinfo {author} {\bibfnamefont
  {A.}~\bibnamefont {Kozhanov}}, \bibinfo {author} {\bibfnamefont
  {L.}~\bibnamefont {Balents}}, \ and\ \bibinfo {author} {\bibfnamefont
  {S.}~\bibnamefont {Stemmer}},\ }\href {\doibase
  http://dx.doi.org/10.1063/1.4907771} {\bibfield  {journal} {\bibinfo
  {journal} {APL Materials}\ }\textbf {\bibinfo {volume} {3}},\ \bibinfo
  {pages} {062503} (\bibinfo {year} {2015})}\BibitemShut {NoStop}%
\bibitem [{\citenamefont {Katsufuji}\ \emph {et~al.}(1995)\citenamefont
  {Katsufuji}, \citenamefont {Okimoto}, \citenamefont {Arima}, \citenamefont
  {Tokura},\ and\ \citenamefont {Torrance}}]{Katsufuji:1995:p4830}%
  \BibitemOpen
  \bibfield  {author} {\bibinfo {author} {\bibfnamefont {T.}~\bibnamefont
  {Katsufuji}}, \bibinfo {author} {\bibfnamefont {Y.}~\bibnamefont {Okimoto}},
  \bibinfo {author} {\bibfnamefont {T.}~\bibnamefont {Arima}}, \bibinfo
  {author} {\bibfnamefont {Y.}~\bibnamefont {Tokura}}, \ and\ \bibinfo {author}
  {\bibfnamefont {J.~B.}\ \bibnamefont {Torrance}},\ }\href {\doibase
  10.1103/PhysRevB.51.4830} {\bibfield  {journal} {\bibinfo  {journal} {Phys.
  Rev. B}\ }\textbf {\bibinfo {volume} {51}},\ \bibinfo {pages} {4830}
  (\bibinfo {year} {1995})}\BibitemShut {NoStop}%
\bibitem [{\citenamefont {Catalan}, \citenamefont {Bowman},\ and\ \citenamefont
  {Gregg}(2000)}]{Catalan:2000p606}%
  \BibitemOpen
  \bibfield  {author} {\bibinfo {author} {\bibfnamefont {G.}~\bibnamefont
  {Catalan}}, \bibinfo {author} {\bibfnamefont {R.~M.}\ \bibnamefont {Bowman}},
  \ and\ \bibinfo {author} {\bibfnamefont {J.~M.}\ \bibnamefont {Gregg}},\
  }\href@noop {} {\bibfield  {journal} {\bibinfo  {journal} {Journal of Applied
  Physics}\ }\textbf {\bibinfo {volume} {87}},\ \bibinfo {pages} {606}
  (\bibinfo {year} {2000})}\BibitemShut {NoStop}%
\bibitem [{\citenamefont {Alonso}\ \emph
  {et~al.}(1999{\natexlab{a}})\citenamefont {Alonso}, \citenamefont
  {Garc\'{i}a-Mu\~noz}, \citenamefont {Fern\'andez-D\'{i}az}, \citenamefont
  {Aranda}, \citenamefont {Mart\'{i}nez-Lope},\ and\ \citenamefont
  {Casais}}]{Alonso:1999p3871}%
  \BibitemOpen
  \bibfield  {author} {\bibinfo {author} {\bibfnamefont {J.~A.}\ \bibnamefont
  {Alonso}}, \bibinfo {author} {\bibfnamefont {J.~L.}\ \bibnamefont
  {Garc\'{i}a-Mu\~noz}}, \bibinfo {author} {\bibfnamefont {M.~T.}\ \bibnamefont
  {Fern\'andez-D\'{i}az}}, \bibinfo {author} {\bibfnamefont {M.~A.~G.}\
  \bibnamefont {Aranda}}, \bibinfo {author} {\bibfnamefont {M.~J.}\
  \bibnamefont {Mart\'{i}nez-Lope}}, \ and\ \bibinfo {author} {\bibfnamefont
  {M.~T.}\ \bibnamefont {Casais}},\ }\href {\doibase
  10.1103/PhysRevLett.82.3871} {\bibfield  {journal} {\bibinfo  {journal}
  {Phys. Rev. Lett.}\ }\textbf {\bibinfo {volume} {82}},\ \bibinfo {pages}
  {3871} (\bibinfo {year} {1999}{\natexlab{a}})}\BibitemShut {NoStop}%
\bibitem [{\citenamefont {Alonso}\ \emph
  {et~al.}(1999{\natexlab{b}})\citenamefont {Alonso}, \citenamefont
  {Mart{\'\i}nez-Lope}, \citenamefont {Casais}, \citenamefont {Aranda},\ and\
  \citenamefont {Fern{\'a}ndez-D{\'\i}az}}]{Alonso:1999p4754}%
  \BibitemOpen
  \bibfield  {author} {\bibinfo {author} {\bibfnamefont {J.~A.}\ \bibnamefont
  {Alonso}}, \bibinfo {author} {\bibfnamefont {M.~J.}\ \bibnamefont
  {Mart{\'\i}nez-Lope}}, \bibinfo {author} {\bibfnamefont {M.~T.}\ \bibnamefont
  {Casais}}, \bibinfo {author} {\bibfnamefont {M.~A.~G.}\ \bibnamefont
  {Aranda}}, \ and\ \bibinfo {author} {\bibfnamefont {M.~T.}\ \bibnamefont
  {Fern{\'a}ndez-D{\'\i}az}},\ }\href {\doibase 10.1021/ja984015x} {\bibfield
  {journal} {\bibinfo  {journal} {Journal of the American Chemical Society}\
  }\textbf {\bibinfo {volume} {121}},\ \bibinfo {pages} {4754} (\bibinfo {year}
  {1999}{\natexlab{b}})},\ \Eprint
  {http://arxiv.org/abs/http://dx.doi.org/10.1021/ja984015x}
  {http://dx.doi.org/10.1021/ja984015x} \BibitemShut {NoStop}%
\bibitem [{\citenamefont {Staub}\ \emph {et~al.}(2002)\citenamefont {Staub},
  \citenamefont {Meijer}, \citenamefont {Fauth}, \citenamefont {Allenspach},
  \citenamefont {Bednorz}, \citenamefont {Karpinski}, \citenamefont {Kazakov},
  \citenamefont {Paolasini},\ and\ \citenamefont
  {d'Acapito}}]{Staub:2002p126402}%
  \BibitemOpen
  \bibfield  {author} {\bibinfo {author} {\bibfnamefont {U.}~\bibnamefont
  {Staub}}, \bibinfo {author} {\bibfnamefont {G.~I.}\ \bibnamefont {Meijer}},
  \bibinfo {author} {\bibfnamefont {F.}~\bibnamefont {Fauth}}, \bibinfo
  {author} {\bibfnamefont {R.}~\bibnamefont {Allenspach}}, \bibinfo {author}
  {\bibfnamefont {J.~G.}\ \bibnamefont {Bednorz}}, \bibinfo {author}
  {\bibfnamefont {J.}~\bibnamefont {Karpinski}}, \bibinfo {author}
  {\bibfnamefont {S.~M.}\ \bibnamefont {Kazakov}}, \bibinfo {author}
  {\bibfnamefont {L.}~\bibnamefont {Paolasini}}, \ and\ \bibinfo {author}
  {\bibfnamefont {F.}~\bibnamefont {d'Acapito}},\ }\href {\doibase
  10.1103/PhysRevLett.88.126402} {\bibfield  {journal} {\bibinfo  {journal}
  {Phys. Rev. Lett.}\ }\textbf {\bibinfo {volume} {88}},\ \bibinfo {pages}
  {126402} (\bibinfo {year} {2002})}\BibitemShut {NoStop}%
\bibitem [{\citenamefont {Mazin}\ \emph {et~al.}(2007)\citenamefont {Mazin},
  \citenamefont {Khomskii}, \citenamefont {Lengsdorf}, \citenamefont {Alonso},
  \citenamefont {Marshall}, \citenamefont {Ibberson}, \citenamefont
  {Podlesnyak}, \citenamefont {Mart\'{\i}nez-Lope},\ and\ \citenamefont
  {Abd-Elmeguid}}]{Mazin:2007p176406}%
  \BibitemOpen
  \bibfield  {author} {\bibinfo {author} {\bibfnamefont {I.~I.}\ \bibnamefont
  {Mazin}}, \bibinfo {author} {\bibfnamefont {D.~I.}\ \bibnamefont {Khomskii}},
  \bibinfo {author} {\bibfnamefont {R.}~\bibnamefont {Lengsdorf}}, \bibinfo
  {author} {\bibfnamefont {J.~A.}\ \bibnamefont {Alonso}}, \bibinfo {author}
  {\bibfnamefont {W.~G.}\ \bibnamefont {Marshall}}, \bibinfo {author}
  {\bibfnamefont {R.~M.}\ \bibnamefont {Ibberson}}, \bibinfo {author}
  {\bibfnamefont {A.}~\bibnamefont {Podlesnyak}}, \bibinfo {author}
  {\bibfnamefont {M.~J.}\ \bibnamefont {Mart\'{\i}nez-Lope}}, \ and\ \bibinfo
  {author} {\bibfnamefont {M.~M.}\ \bibnamefont {Abd-Elmeguid}},\ }\href
  {\doibase 10.1103/PhysRevLett.98.176406} {\bibfield  {journal} {\bibinfo
  {journal} {Phys. Rev. Lett.}\ }\textbf {\bibinfo {volume} {98}},\ \bibinfo
  {pages} {176406} (\bibinfo {year} {2007})}\BibitemShut {NoStop}%
\bibitem [{\citenamefont {Vobornik}\ \emph {et~al.}(1999)\citenamefont
  {Vobornik}, \citenamefont {Perfetti}, \citenamefont {Zacchigna},
  \citenamefont {Grioni}, \citenamefont {Margaritondo}, \citenamefont {Mesot},
  \citenamefont {Medarde},\ and\ \citenamefont
  {Lacorre}}]{Vobornik:1999pR8426}%
  \BibitemOpen
  \bibfield  {author} {\bibinfo {author} {\bibfnamefont {I.}~\bibnamefont
  {Vobornik}}, \bibinfo {author} {\bibfnamefont {L.}~\bibnamefont {Perfetti}},
  \bibinfo {author} {\bibfnamefont {M.}~\bibnamefont {Zacchigna}}, \bibinfo
  {author} {\bibfnamefont {M.}~\bibnamefont {Grioni}}, \bibinfo {author}
  {\bibfnamefont {G.}~\bibnamefont {Margaritondo}}, \bibinfo {author}
  {\bibfnamefont {J.}~\bibnamefont {Mesot}}, \bibinfo {author} {\bibfnamefont
  {M.}~\bibnamefont {Medarde}}, \ and\ \bibinfo {author} {\bibfnamefont
  {P.}~\bibnamefont {Lacorre}},\ }\href {\doibase 10.1103/PhysRevB.60.R8426}
  {\bibfield  {journal} {\bibinfo  {journal} {Phys. Rev. B}\ }\textbf {\bibinfo
  {volume} {60}},\ \bibinfo {pages} {R8426} (\bibinfo {year}
  {1999})}\BibitemShut {NoStop}%
\bibitem [{\citenamefont {Rodr\'{i}guez-Carvajal}\ \emph
  {et~al.}(1998)\citenamefont {Rodr\'{i}guez-Carvajal}, \citenamefont
  {Rosenkranz}, \citenamefont {Medarde}, \citenamefont {Lacorre}, \citenamefont
  {Fernandez-D\'{i}az}, \citenamefont {Fauth},\ and\ \citenamefont
  {Trounov}}]{Carvajal:1998p456}%
  \BibitemOpen
  \bibfield  {author} {\bibinfo {author} {\bibfnamefont {J.}~\bibnamefont
  {Rodr\'{i}guez-Carvajal}}, \bibinfo {author} {\bibfnamefont {S.}~\bibnamefont
  {Rosenkranz}}, \bibinfo {author} {\bibfnamefont {M.}~\bibnamefont {Medarde}},
  \bibinfo {author} {\bibfnamefont {P.}~\bibnamefont {Lacorre}}, \bibinfo
  {author} {\bibfnamefont {M.~T.}\ \bibnamefont {Fernandez-D\'{i}az}}, \bibinfo
  {author} {\bibfnamefont {F.}~\bibnamefont {Fauth}}, \ and\ \bibinfo {author}
  {\bibfnamefont {V.}~\bibnamefont {Trounov}},\ }\href {\doibase
  10.1103/PhysRevB.57.456} {\bibfield  {journal} {\bibinfo  {journal} {Phys.
  Rev. B}\ }\textbf {\bibinfo {volume} {57}},\ \bibinfo {pages} {456} (\bibinfo
  {year} {1998})}\BibitemShut {NoStop}%
\bibitem [{\citenamefont {Bodenthin}\ \emph {et~al.}(2011)\citenamefont
  {Bodenthin}, \citenamefont {Staub}, \citenamefont {Piamonteze}, \citenamefont
  {Garcia-Fernndez}, \citenamefont {Martnez-Lope},\ and\ \citenamefont
  {Alonso}}]{Bodenthin:2011p036002}%
  \BibitemOpen
  \bibfield  {author} {\bibinfo {author} {\bibfnamefont {Y.}~\bibnamefont
  {Bodenthin}}, \bibinfo {author} {\bibfnamefont {U.}~\bibnamefont {Staub}},
  \bibinfo {author} {\bibfnamefont {C.}~\bibnamefont {Piamonteze}}, \bibinfo
  {author} {\bibfnamefont {M.}~\bibnamefont {Garcia-Fernndez}}, \bibinfo
  {author} {\bibfnamefont {M.~J.}\ \bibnamefont {Martnez-Lope}}, \ and\
  \bibinfo {author} {\bibfnamefont {J.~A.}\ \bibnamefont {Alonso}},\
  }\href@noop {} {\bibfield  {journal} {\bibinfo  {journal} {Journal of
  Physics: Condensed Matter}\ }\textbf {\bibinfo {volume} {23}},\ \bibinfo
  {pages} {036002} (\bibinfo {year} {2011})}\BibitemShut {NoStop}%
\bibitem [{\citenamefont {Lee}, \citenamefont {Chen},\ and\ \citenamefont
  {Balents}(2011{\natexlab{a}})}]{Lee:2010p165119}%
  \BibitemOpen
  \bibfield  {author} {\bibinfo {author} {\bibfnamefont {S.}~\bibnamefont
  {Lee}}, \bibinfo {author} {\bibfnamefont {R.}~\bibnamefont {Chen}}, \ and\
  \bibinfo {author} {\bibfnamefont {L.}~\bibnamefont {Balents}},\ }\href
  {\doibase 10.1103/PhysRevB.84.165119} {\bibfield  {journal} {\bibinfo
  {journal} {Phys. Rev. B}\ }\textbf {\bibinfo {volume} {84}},\ \bibinfo
  {pages} {165119} (\bibinfo {year} {2011}{\natexlab{a}})}\BibitemShut
  {NoStop}%
\bibitem [{\citenamefont {Lee}, \citenamefont {Chen},\ and\ \citenamefont
  {Balents}(2011{\natexlab{b}})}]{Lee:2011p016405}%
  \BibitemOpen
  \bibfield  {author} {\bibinfo {author} {\bibfnamefont {S.}~\bibnamefont
  {Lee}}, \bibinfo {author} {\bibfnamefont {R.}~\bibnamefont {Chen}}, \ and\
  \bibinfo {author} {\bibfnamefont {L.}~\bibnamefont {Balents}},\ }\href
  {\doibase 10.1103/PhysRevLett.106.016405} {\bibfield  {journal} {\bibinfo
  {journal} {Phys. Rev. Lett.}\ }\textbf {\bibinfo {volume} {106}},\ \bibinfo
  {pages} {016405} (\bibinfo {year} {2011}{\natexlab{b}})}\BibitemShut
  {NoStop}%
\bibitem [{\citenamefont {Stewart}\ \emph {et~al.}(2011)\citenamefont
  {Stewart}, \citenamefont {Liu}, \citenamefont {Kareev}, \citenamefont
  {Chakhalian},\ and\ \citenamefont {Basov}}]{Stewart:2011p176401}%
  \BibitemOpen
  \bibfield  {author} {\bibinfo {author} {\bibfnamefont {M.~K.}\ \bibnamefont
  {Stewart}}, \bibinfo {author} {\bibfnamefont {J.}~\bibnamefont {Liu}},
  \bibinfo {author} {\bibfnamefont {M.}~\bibnamefont {Kareev}}, \bibinfo
  {author} {\bibfnamefont {J.}~\bibnamefont {Chakhalian}}, \ and\ \bibinfo
  {author} {\bibfnamefont {D.~N.}\ \bibnamefont {Basov}},\ }\href {\doibase
  10.1103/PhysRevLett.107.176401} {\bibfield  {journal} {\bibinfo  {journal}
  {Phys. Rev. Lett.}\ }\textbf {\bibinfo {volume} {107}},\ \bibinfo {pages}
  {176401} (\bibinfo {year} {2011})}\BibitemShut {NoStop}%
\bibitem [{\citenamefont {Barman}, \citenamefont {Chainani},\ and\
  \citenamefont {Sarma}(1994)}]{Barman:1994p8475}%
  \BibitemOpen
  \bibfield  {author} {\bibinfo {author} {\bibfnamefont {S.~R.}\ \bibnamefont
  {Barman}}, \bibinfo {author} {\bibfnamefont {A.}~\bibnamefont {Chainani}}, \
  and\ \bibinfo {author} {\bibfnamefont {D.~D.}\ \bibnamefont {Sarma}},\ }\href
  {\doibase 10.1103/PhysRevB.49.8475} {\bibfield  {journal} {\bibinfo
  {journal} {Phys. Rev. B}\ }\textbf {\bibinfo {volume} {49}},\ \bibinfo
  {pages} {8475} (\bibinfo {year} {1994})}\BibitemShut {NoStop}%
\bibitem [{\citenamefont {Bisogni}\ \emph {et~al.}(2016)\citenamefont
  {Bisogni}, \citenamefont {Catalano}, \citenamefont {Green}, \citenamefont
  {Gibert}, \citenamefont {Scherwitzl}, \citenamefont {Huang}, \citenamefont
  {Strocov}, \citenamefont {Zubko}, \citenamefont {Balandeh}, \citenamefont
  {Triscone} \emph {et~al.}}]{Bisogni:2016p13017}%
  \BibitemOpen
  \bibfield  {author} {\bibinfo {author} {\bibfnamefont {V.}~\bibnamefont
  {Bisogni}}, \bibinfo {author} {\bibfnamefont {S.}~\bibnamefont {Catalano}},
  \bibinfo {author} {\bibfnamefont {R.~J.}\ \bibnamefont {Green}}, \bibinfo
  {author} {\bibfnamefont {M.}~\bibnamefont {Gibert}}, \bibinfo {author}
  {\bibfnamefont {R.}~\bibnamefont {Scherwitzl}}, \bibinfo {author}
  {\bibfnamefont {Y.}~\bibnamefont {Huang}}, \bibinfo {author} {\bibfnamefont
  {V.~N.}\ \bibnamefont {Strocov}}, \bibinfo {author} {\bibfnamefont
  {P.}~\bibnamefont {Zubko}}, \bibinfo {author} {\bibfnamefont
  {S.}~\bibnamefont {Balandeh}}, \bibinfo {author} {\bibfnamefont {J.-M.}\
  \bibnamefont {Triscone}},  \emph {et~al.},\ }\href@noop {} {\bibfield
  {journal} {\bibinfo  {journal} {Nature Communications}\ }\textbf {\bibinfo
  {volume} {7}},\ \bibinfo {pages} {13017} (\bibinfo {year}
  {2016})}\BibitemShut {NoStop}%
\bibitem [{\citenamefont {Mizokawa}, \citenamefont {Khomskii},\ and\
  \citenamefont {Sawatzky}(2000)}]{Mizokawa:2000p11263}%
  \BibitemOpen
  \bibfield  {author} {\bibinfo {author} {\bibfnamefont {T.}~\bibnamefont
  {Mizokawa}}, \bibinfo {author} {\bibfnamefont {D.~I.}\ \bibnamefont
  {Khomskii}}, \ and\ \bibinfo {author} {\bibfnamefont {G.~A.}\ \bibnamefont
  {Sawatzky}},\ }\href {\doibase 10.1103/PhysRevB.61.11263} {\bibfield
  {journal} {\bibinfo  {journal} {Phys. Rev. B}\ }\textbf {\bibinfo {volume}
  {61}},\ \bibinfo {pages} {11263} (\bibinfo {year} {2000})}\BibitemShut
  {NoStop}%
\bibitem [{\citenamefont {Park}, \citenamefont {Millis},\ and\ \citenamefont
  {Marianetti}(2012)}]{Park:2012p156402}%
  \BibitemOpen
  \bibfield  {author} {\bibinfo {author} {\bibfnamefont {H.}~\bibnamefont
  {Park}}, \bibinfo {author} {\bibfnamefont {A.~J.}\ \bibnamefont {Millis}}, \
  and\ \bibinfo {author} {\bibfnamefont {C.~A.}\ \bibnamefont {Marianetti}},\
  }\href {\doibase 10.1103/PhysRevLett.109.156402} {\bibfield  {journal}
  {\bibinfo  {journal} {Phys. Rev. Lett.}\ }\textbf {\bibinfo {volume} {109}},\
  \bibinfo {pages} {156402} (\bibinfo {year} {2012})}\BibitemShut {NoStop}%
\bibitem [{\citenamefont {Johnston}\ \emph {et~al.}(2014)\citenamefont
  {Johnston}, \citenamefont {Mukherjee}, \citenamefont {Elfimov}, \citenamefont
  {Berciu},\ and\ \citenamefont {Sawatzky}}]{Johnston:2014p106404}%
  \BibitemOpen
  \bibfield  {author} {\bibinfo {author} {\bibfnamefont {S.}~\bibnamefont
  {Johnston}}, \bibinfo {author} {\bibfnamefont {A.}~\bibnamefont {Mukherjee}},
  \bibinfo {author} {\bibfnamefont {I.}~\bibnamefont {Elfimov}}, \bibinfo
  {author} {\bibfnamefont {M.}~\bibnamefont {Berciu}}, \ and\ \bibinfo {author}
  {\bibfnamefont {G.~A.}\ \bibnamefont {Sawatzky}},\ }\href {\doibase
  10.1103/PhysRevLett.112.106404} {\bibfield  {journal} {\bibinfo  {journal}
  {Phys. Rev. Lett.}\ }\textbf {\bibinfo {volume} {112}},\ \bibinfo {pages}
  {106404} (\bibinfo {year} {2014})}\BibitemShut {NoStop}%
\bibitem [{\citenamefont {Subedi}, \citenamefont {Peil},\ and\ \citenamefont
  {Georges}(2015)}]{Subedi:2015p075128}%
  \BibitemOpen
  \bibfield  {author} {\bibinfo {author} {\bibfnamefont {A.}~\bibnamefont
  {Subedi}}, \bibinfo {author} {\bibfnamefont {O.~E.}\ \bibnamefont {Peil}}, \
  and\ \bibinfo {author} {\bibfnamefont {A.}~\bibnamefont {Georges}},\ }\href
  {\doibase 10.1103/PhysRevB.91.075128} {\bibfield  {journal} {\bibinfo
  {journal} {Phys. Rev. B}\ }\textbf {\bibinfo {volume} {91}},\ \bibinfo
  {pages} {075128} (\bibinfo {year} {2015})}\BibitemShut {NoStop}%
\bibitem [{\citenamefont {Haule}\ and\ \citenamefont
  {Pascut}(2017)}]{Haule:2017p2045}%
  \BibitemOpen
  \bibfield  {author} {\bibinfo {author} {\bibfnamefont {K.}~\bibnamefont
  {Haule}}\ and\ \bibinfo {author} {\bibfnamefont {G.~L.}\ \bibnamefont
  {Pascut}},\ }\href {\doibase 10.1038/s41598-017-10374-2} {\bibfield
  {journal} {\bibinfo  {journal} {Scientific Reports}\ }\textbf {\bibinfo
  {volume} {7}},\ \bibinfo {pages} {2045} (\bibinfo {year} {2017})}\BibitemShut
  {NoStop}%
\bibitem [{\citenamefont {Schlom}\ \emph {et~al.}(2008)\citenamefont {Schlom},
  \citenamefont {Chen}, \citenamefont {Pan}, \citenamefont {Schmehl},\ and\
  \citenamefont {Zurbuchen}}]{Schlom:2008p2429}%
  \BibitemOpen
  \bibfield  {author} {\bibinfo {author} {\bibfnamefont {D.~G.}\ \bibnamefont
  {Schlom}}, \bibinfo {author} {\bibfnamefont {L.-Q.}\ \bibnamefont {Chen}},
  \bibinfo {author} {\bibfnamefont {X.}~\bibnamefont {Pan}}, \bibinfo {author}
  {\bibfnamefont {A.}~\bibnamefont {Schmehl}}, \ and\ \bibinfo {author}
  {\bibfnamefont {M.~A.}\ \bibnamefont {Zurbuchen}},\ }\href {\doibase
  10.1111/j.1551-2916.2008.02556.x} {\bibfield  {journal} {\bibinfo  {journal}
  {Journal of the American Ceramic Society}\ }\textbf {\bibinfo {volume}
  {91}},\ \bibinfo {pages} {2429} (\bibinfo {year} {2008})}\BibitemShut
  {NoStop}%
\bibitem [{\citenamefont {Martin}, \citenamefont {Chu},\ and\ \citenamefont
  {Ramesh}(2010)}]{Martin:2010p89}%
  \BibitemOpen
  \bibfield  {author} {\bibinfo {author} {\bibfnamefont {L.}~\bibnamefont
  {Martin}}, \bibinfo {author} {\bibfnamefont {Y.-H.}\ \bibnamefont {Chu}}, \
  and\ \bibinfo {author} {\bibfnamefont {R.}~\bibnamefont {Ramesh}},\
  }\href@noop {} {\bibfield  {journal} {\bibinfo  {journal} {Materials Science
  and Engineering: R: Reports}\ }\textbf {\bibinfo {volume} {68}},\ \bibinfo
  {pages} {89} (\bibinfo {year} {2010})}\BibitemShut {NoStop}%
\bibitem [{\citenamefont {Zubko}\ \emph {et~al.}(2011)\citenamefont {Zubko},
  \citenamefont {Gariglio}, \citenamefont {Gabay}, \citenamefont {Ghosez},\
  and\ \citenamefont {Triscone}}]{Zubko:2011p141}%
  \BibitemOpen
  \bibfield  {author} {\bibinfo {author} {\bibfnamefont {P.}~\bibnamefont
  {Zubko}}, \bibinfo {author} {\bibfnamefont {S.}~\bibnamefont {Gariglio}},
  \bibinfo {author} {\bibfnamefont {M.}~\bibnamefont {Gabay}}, \bibinfo
  {author} {\bibfnamefont {P.}~\bibnamefont {Ghosez}}, \ and\ \bibinfo {author}
  {\bibfnamefont {J.-M.}\ \bibnamefont {Triscone}},\ }\href@noop {} {\bibfield
  {journal} {\bibinfo  {journal} {Annual Review of Condensed Matter Physics}\
  }\textbf {\bibinfo {volume} {2}},\ \bibinfo {pages} {141} (\bibinfo {year}
  {2011})}\BibitemShut {NoStop}%
\bibitem [{\citenamefont {Hwang}\ \emph {et~al.}(2012)\citenamefont {Hwang},
  \citenamefont {Iwasa}, \citenamefont {Kawasaki}, \citenamefont {Keimer},
  \citenamefont {Nagaosa},\ and\ \citenamefont {Tokura}}]{Hwang:2012p103}%
  \BibitemOpen
  \bibfield  {author} {\bibinfo {author} {\bibfnamefont {H.~Y.}\ \bibnamefont
  {Hwang}}, \bibinfo {author} {\bibfnamefont {Y.}~\bibnamefont {Iwasa}},
  \bibinfo {author} {\bibfnamefont {M.}~\bibnamefont {Kawasaki}}, \bibinfo
  {author} {\bibfnamefont {B.}~\bibnamefont {Keimer}}, \bibinfo {author}
  {\bibfnamefont {N.}~\bibnamefont {Nagaosa}}, \ and\ \bibinfo {author}
  {\bibfnamefont {Y.}~\bibnamefont {Tokura}},\ }\href {\doibase
  10.1038/nmat3223} {\bibfield  {journal} {\bibinfo  {journal} {Nature Mater.}\
  }\textbf {\bibinfo {volume} {11}},\ \bibinfo {pages} {103} (\bibinfo {year}
  {2012})}\BibitemShut {NoStop}%
\bibitem [{\citenamefont {Chakhalian}\ \emph {et~al.}(2014)\citenamefont
  {Chakhalian}, \citenamefont {Freeland}, \citenamefont {Millis}, \citenamefont
  {Panagopoulos},\ and\ \citenamefont {Rondinelli}}]{Chakhalian:2014p1189}%
  \BibitemOpen
  \bibfield  {author} {\bibinfo {author} {\bibfnamefont {J.}~\bibnamefont
  {Chakhalian}}, \bibinfo {author} {\bibfnamefont {J.~W.}\ \bibnamefont
  {Freeland}}, \bibinfo {author} {\bibfnamefont {A.~J.}\ \bibnamefont
  {Millis}}, \bibinfo {author} {\bibfnamefont {C.}~\bibnamefont
  {Panagopoulos}}, \ and\ \bibinfo {author} {\bibfnamefont {J.~M.}\
  \bibnamefont {Rondinelli}},\ }\href {\doibase 10.1103/RevModPhys.86.1189}
  {\bibfield  {journal} {\bibinfo  {journal} {Rev. Mod. Phys.}\ }\textbf
  {\bibinfo {volume} {86}},\ \bibinfo {pages} {1189} (\bibinfo {year}
  {2014})}\BibitemShut {NoStop}%
\bibitem [{\citenamefont {Bhattacharya}\ and\ \citenamefont
  {May}(2014)}]{Bhattacharya:2014p65}%
  \BibitemOpen
  \bibfield  {author} {\bibinfo {author} {\bibfnamefont {A.}~\bibnamefont
  {Bhattacharya}}\ and\ \bibinfo {author} {\bibfnamefont {S.~J.}\ \bibnamefont
  {May}},\ }\href {\doibase 10.1146/annurev-matsci-070813-113447} {\bibfield
  {journal} {\bibinfo  {journal} {Annual Review of Materials Research}\
  }\textbf {\bibinfo {volume} {44}},\ \bibinfo {pages} {65} (\bibinfo {year}
  {2014})},\ \Eprint
  {http://arxiv.org/abs/http://dx.doi.org/10.1146/annurev-matsci-070813-113447}
  {http://dx.doi.org/10.1146/annurev-matsci-070813-113447} \BibitemShut
  {NoStop}%
\bibitem [{\citenamefont {Liu}\ \emph {et~al.}(2016)\citenamefont {Liu},
  \citenamefont {Middey}, \citenamefont {Cao}, \citenamefont {Kareev},\ and\
  \citenamefont {Chakhalian}}]{Liu:2016p133}%
  \BibitemOpen
  \bibfield  {author} {\bibinfo {author} {\bibfnamefont {X.}~\bibnamefont
  {Liu}}, \bibinfo {author} {\bibfnamefont {S.}~\bibnamefont {Middey}},
  \bibinfo {author} {\bibfnamefont {Y.}~\bibnamefont {Cao}}, \bibinfo {author}
  {\bibfnamefont {M.}~\bibnamefont {Kareev}}, \ and\ \bibinfo {author}
  {\bibfnamefont {J.}~\bibnamefont {Chakhalian}},\ }\href@noop {} {\bibfield
  {journal} {\bibinfo  {journal} {MRS communications}\ }\textbf {\bibinfo
  {volume} {6}},\ \bibinfo {pages} {133} (\bibinfo {year} {2016})}\BibitemShut
  {NoStop}%
\bibitem [{\citenamefont {Ramesh}\ and\ \citenamefont
  {Schlom}(2019)}]{Ramesh:2019p257}%
  \BibitemOpen
  \bibfield  {author} {\bibinfo {author} {\bibfnamefont {R.}~\bibnamefont
  {Ramesh}}\ and\ \bibinfo {author} {\bibfnamefont {D.~G.}\ \bibnamefont
  {Schlom}},\ }\href {\doibase 10.1038/s41578-019-0095-2} {\bibfield  {journal}
  {\bibinfo  {journal} {Nature Reviews Materials}\ }\textbf {\bibinfo {volume}
  {4}},\ \bibinfo {pages} {257} (\bibinfo {year} {2019})}\BibitemShut {NoStop}%
\bibitem [{\citenamefont {Zhang}\ \emph {et~al.}(2017)\citenamefont {Zhang},
  \citenamefont {Zheng}, \citenamefont {Ren},\ and\ \citenamefont
  {Mitchell}}]{Zhang:2017p2730}%
  \BibitemOpen
  \bibfield  {author} {\bibinfo {author} {\bibfnamefont {J.}~\bibnamefont
  {Zhang}}, \bibinfo {author} {\bibfnamefont {H.}~\bibnamefont {Zheng}},
  \bibinfo {author} {\bibfnamefont {Y.}~\bibnamefont {Ren}}, \ and\ \bibinfo
  {author} {\bibfnamefont {J.}~\bibnamefont {Mitchell}},\ }\href@noop {}
  {\bibfield  {journal} {\bibinfo  {journal} {Crystal Growth \& Design}\
  }\textbf {\bibinfo {volume} {17}},\ \bibinfo {pages} {2730} (\bibinfo {year}
  {2017})}\BibitemShut {NoStop}%
\bibitem [{\citenamefont {Guo}\ \emph {et~al.}(2018)\citenamefont {Guo},
  \citenamefont {Li}, \citenamefont {Zhao}, \citenamefont {Hu}, \citenamefont
  {Chang}, \citenamefont {Kuo}, \citenamefont {Schmidt}, \citenamefont
  {Piovano}, \citenamefont {Pi}, \citenamefont {Sobolev}, \citenamefont
  {Khomskii}, \citenamefont {Tjeng},\ and\ \citenamefont
  {Komarek}}]{Guo:2018p43}%
  \BibitemOpen
  \bibfield  {author} {\bibinfo {author} {\bibfnamefont {H.}~\bibnamefont
  {Guo}}, \bibinfo {author} {\bibfnamefont {Z.}~\bibnamefont {Li}}, \bibinfo
  {author} {\bibfnamefont {L.}~\bibnamefont {Zhao}}, \bibinfo {author}
  {\bibfnamefont {Z.}~\bibnamefont {Hu}}, \bibinfo {author} {\bibfnamefont
  {C.}~\bibnamefont {Chang}}, \bibinfo {author} {\bibfnamefont {C.-Y.}\
  \bibnamefont {Kuo}}, \bibinfo {author} {\bibfnamefont {W.}~\bibnamefont
  {Schmidt}}, \bibinfo {author} {\bibfnamefont {A.}~\bibnamefont {Piovano}},
  \bibinfo {author} {\bibfnamefont {T.}~\bibnamefont {Pi}}, \bibinfo {author}
  {\bibfnamefont {O.}~\bibnamefont {Sobolev}}, \bibinfo {author} {\bibfnamefont
  {D.}~\bibnamefont {Khomskii}}, \bibinfo {author} {\bibfnamefont
  {L.}~\bibnamefont {Tjeng}}, \ and\ \bibinfo {author} {\bibfnamefont
  {A.}~\bibnamefont {Komarek}},\ }\href {\doibase 10.1038/s41467-017-02524-x}
  {\bibfield  {journal} {\bibinfo  {journal} {Nature communications}\ }\textbf
  {\bibinfo {volume} {9}},\ \bibinfo {pages} {43} (\bibinfo {year}
  {2018})}\BibitemShut {NoStop}%
\bibitem [{\citenamefont {Klein}\ \emph {et~al.}(2021)\citenamefont {Klein},
  \citenamefont {Koz{\l}owski}, \citenamefont {Linden}, \citenamefont
  {Lacorre}, \citenamefont {Medarde},\ and\ \citenamefont
  {Gawryluk}}]{Klein:2021}%
  \BibitemOpen
  \bibfield  {author} {\bibinfo {author} {\bibfnamefont {Y.~M.}\ \bibnamefont
  {Klein}}, \bibinfo {author} {\bibfnamefont {M.}~\bibnamefont {Koz{\l}owski}},
  \bibinfo {author} {\bibfnamefont {A.}~\bibnamefont {Linden}}, \bibinfo
  {author} {\bibfnamefont {P.}~\bibnamefont {Lacorre}}, \bibinfo {author}
  {\bibfnamefont {M.}~\bibnamefont {Medarde}}, \ and\ \bibinfo {author}
  {\bibfnamefont {D.~J.}\ \bibnamefont {Gawryluk}},\ }\href@noop {} {\bibfield
  {journal} {\bibinfo  {journal} {Crystal Growth \& Design}\ } (\bibinfo {year}
  {2021})}\BibitemShut {NoStop}%
\bibitem [{\citenamefont {Prasad}\ \emph {et~al.}(1993)\citenamefont {Prasad},
  \citenamefont {Varma}, \citenamefont {Raju}, \citenamefont {Satyalakshmi},
  \citenamefont {Mallya},\ and\ \citenamefont {Hegde}}]{Prasad:1993p1898}%
  \BibitemOpen
  \bibfield  {author} {\bibinfo {author} {\bibfnamefont {K.~V.~R.}\
  \bibnamefont {Prasad}}, \bibinfo {author} {\bibfnamefont {K.~B.~R.}\
  \bibnamefont {Varma}}, \bibinfo {author} {\bibfnamefont {A.~R.}\ \bibnamefont
  {Raju}}, \bibinfo {author} {\bibfnamefont {K.~M.}\ \bibnamefont
  {Satyalakshmi}}, \bibinfo {author} {\bibfnamefont {R.~M.}\ \bibnamefont
  {Mallya}}, \ and\ \bibinfo {author} {\bibfnamefont {M.~S.}\ \bibnamefont
  {Hegde}},\ }\href@noop {} {\bibfield  {journal} {\bibinfo  {journal} {Applied
  Physics Letters}\ }\textbf {\bibinfo {volume} {63}},\ \bibinfo {pages} {1898}
  (\bibinfo {year} {1993})}\BibitemShut {NoStop}%
\bibitem [{\citenamefont {Yang}\ \emph {et~al.}(1995)\citenamefont {Yang},
  \citenamefont {Chen}, \citenamefont {Hong}, \citenamefont {Wu}, \citenamefont
  {Wu},\ and\ \citenamefont {Wu}}]{Yang:1995p2643}%
  \BibitemOpen
  \bibfield  {author} {\bibinfo {author} {\bibfnamefont {C.}~\bibnamefont
  {Yang}}, \bibinfo {author} {\bibfnamefont {M.}~\bibnamefont {Chen}}, \bibinfo
  {author} {\bibfnamefont {T.}~\bibnamefont {Hong}}, \bibinfo {author}
  {\bibfnamefont {C.}~\bibnamefont {Wu}}, \bibinfo {author} {\bibfnamefont
  {J.}~\bibnamefont {Wu}}, \ and\ \bibinfo {author} {\bibfnamefont
  {T.}~\bibnamefont {Wu}},\ }\href@noop {} {\bibfield  {journal} {\bibinfo
  {journal} {Applied Physics Letters}\ }\textbf {\bibinfo {volume} {66}},\
  \bibinfo {pages} {2643} (\bibinfo {year} {1995})}\BibitemShut {NoStop}%
\bibitem [{\citenamefont {DeNatale}\ and\ \citenamefont
  {Kobrin}(1996)}]{DeNatale:1995p2992}%
  \BibitemOpen
  \bibfield  {author} {\bibinfo {author} {\bibfnamefont {J.~F.}\ \bibnamefont
  {DeNatale}}\ and\ \bibinfo {author} {\bibfnamefont {P.~H.}\ \bibnamefont
  {Kobrin}},\ }\href@noop {} {\bibfield  {journal} {\bibinfo  {journal} {J.
  Mater. Res.}\ }\textbf {\bibinfo {volume} {12}},\ \bibinfo {pages} {2992}
  (\bibinfo {year} {1996})}\BibitemShut {NoStop}%
\bibitem [{\citenamefont {Kaul}, \citenamefont {Gorbenko},\ and\ \citenamefont
  {Kamenev}(2004)}]{Kaul:2004p861}%
  \BibitemOpen
  \bibfield  {author} {\bibinfo {author} {\bibfnamefont {A.~R.}\ \bibnamefont
  {Kaul}}, \bibinfo {author} {\bibfnamefont {O.~Y.}\ \bibnamefont {Gorbenko}},
  \ and\ \bibinfo {author} {\bibfnamefont {A.~A.}\ \bibnamefont {Kamenev}},\
  }\href@noop {} {\bibfield  {journal} {\bibinfo  {journal} {Russ. Chem. Rev.}\
  }\textbf {\bibinfo {volume} {73}},\ \bibinfo {pages} {861} (\bibinfo {year}
  {2004})}\BibitemShut {NoStop}%
\bibitem [{\citenamefont {Gorbenko}\ \emph {et~al.}(2002)\citenamefont
  {Gorbenko}, \citenamefont {Samoilenkov}, \citenamefont {Graboy},\ and\
  \citenamefont {Kaul}}]{Gorbenko:2002p4026}%
  \BibitemOpen
  \bibfield  {author} {\bibinfo {author} {\bibfnamefont {O.~Y.}\ \bibnamefont
  {Gorbenko}}, \bibinfo {author} {\bibfnamefont {S.}~\bibnamefont
  {Samoilenkov}}, \bibinfo {author} {\bibfnamefont {I.}~\bibnamefont {Graboy}},
  \ and\ \bibinfo {author} {\bibfnamefont {A.}~\bibnamefont {Kaul}},\
  }\href@noop {} {\bibfield  {journal} {\bibinfo  {journal} {Chemistry of
  materials}\ }\textbf {\bibinfo {volume} {14}},\ \bibinfo {pages} {4026}
  (\bibinfo {year} {2002})}\BibitemShut {NoStop}%
\bibitem [{\citenamefont {Boris}\ \emph {et~al.}(2011)\citenamefont {Boris},
  \citenamefont {Matiks}, \citenamefont {Benckiser}, \citenamefont {Frano},
  \citenamefont {Popovich}, \citenamefont {Hinkov}, \citenamefont {Wochner},
  \citenamefont {Castro-Colin}, \citenamefont {Detemple}, \citenamefont
  {Malik}, \citenamefont {Bernhard}, \citenamefont {Prokscha}, \citenamefont
  {Suter}, \citenamefont {Salman}, \citenamefont {Morenzoni}, \citenamefont
  {Cristiani}, \citenamefont {Habermeier},\ and\ \citenamefont
  {Keimer}}]{Boris:2011p937}%
  \BibitemOpen
  \bibfield  {author} {\bibinfo {author} {\bibfnamefont {A.~V.}\ \bibnamefont
  {Boris}}, \bibinfo {author} {\bibfnamefont {Y.}~\bibnamefont {Matiks}},
  \bibinfo {author} {\bibfnamefont {E.}~\bibnamefont {Benckiser}}, \bibinfo
  {author} {\bibfnamefont {A.}~\bibnamefont {Frano}}, \bibinfo {author}
  {\bibfnamefont {P.}~\bibnamefont {Popovich}}, \bibinfo {author}
  {\bibfnamefont {V.}~\bibnamefont {Hinkov}}, \bibinfo {author} {\bibfnamefont
  {P.}~\bibnamefont {Wochner}}, \bibinfo {author} {\bibfnamefont
  {M.}~\bibnamefont {Castro-Colin}}, \bibinfo {author} {\bibfnamefont
  {E.}~\bibnamefont {Detemple}}, \bibinfo {author} {\bibfnamefont {V.~K.}\
  \bibnamefont {Malik}}, \bibinfo {author} {\bibfnamefont {C.}~\bibnamefont
  {Bernhard}}, \bibinfo {author} {\bibfnamefont {T.}~\bibnamefont {Prokscha}},
  \bibinfo {author} {\bibfnamefont {A.}~\bibnamefont {Suter}}, \bibinfo
  {author} {\bibfnamefont {Z.}~\bibnamefont {Salman}}, \bibinfo {author}
  {\bibfnamefont {E.}~\bibnamefont {Morenzoni}}, \bibinfo {author}
  {\bibfnamefont {G.}~\bibnamefont {Cristiani}}, \bibinfo {author}
  {\bibfnamefont {H.-U.}\ \bibnamefont {Habermeier}}, \ and\ \bibinfo {author}
  {\bibfnamefont {B.}~\bibnamefont {Keimer}},\ }\href {\doibase
  10.1126/science.1202647} {\bibfield  {journal} {\bibinfo  {journal}
  {Science}\ }\textbf {\bibinfo {volume} {332}},\ \bibinfo {pages} {937}
  (\bibinfo {year} {2011})},\ \Eprint
  {http://arxiv.org/abs/http://www.sciencemag.org/content/332/6032/937.full.pdf}
  {http://www.sciencemag.org/content/332/6032/937.full.pdf} \BibitemShut
  {NoStop}%
\bibitem [{\citenamefont {Chakhalian}\ \emph {et~al.}(2011)\citenamefont
  {Chakhalian}, \citenamefont {Rondinelli}, \citenamefont {Liu}, \citenamefont
  {Gray}, \citenamefont {Kareev}, \citenamefont {Moon}, \citenamefont {Prasai},
  \citenamefont {Cohn}, \citenamefont {Varela}, \citenamefont {Tung},
  \citenamefont {Bedzyk}, \citenamefont {Altendorf}, \citenamefont {Strigari},
  \citenamefont {Dabrowski}, \citenamefont {Tjeng}, \citenamefont {Ryan},\ and\
  \citenamefont {Freeland}}]{Chakhalian:2011p116805}%
  \BibitemOpen
  \bibfield  {author} {\bibinfo {author} {\bibfnamefont {J.}~\bibnamefont
  {Chakhalian}}, \bibinfo {author} {\bibfnamefont {J.~M.}\ \bibnamefont
  {Rondinelli}}, \bibinfo {author} {\bibfnamefont {J.}~\bibnamefont {Liu}},
  \bibinfo {author} {\bibfnamefont {B.~A.}\ \bibnamefont {Gray}}, \bibinfo
  {author} {\bibfnamefont {M.}~\bibnamefont {Kareev}}, \bibinfo {author}
  {\bibfnamefont {E.~J.}\ \bibnamefont {Moon}}, \bibinfo {author}
  {\bibfnamefont {N.}~\bibnamefont {Prasai}}, \bibinfo {author} {\bibfnamefont
  {J.~L.}\ \bibnamefont {Cohn}}, \bibinfo {author} {\bibfnamefont
  {M.}~\bibnamefont {Varela}}, \bibinfo {author} {\bibfnamefont {I.~C.}\
  \bibnamefont {Tung}}, \bibinfo {author} {\bibfnamefont {M.~J.}\ \bibnamefont
  {Bedzyk}}, \bibinfo {author} {\bibfnamefont {S.~G.}\ \bibnamefont
  {Altendorf}}, \bibinfo {author} {\bibfnamefont {F.}~\bibnamefont {Strigari}},
  \bibinfo {author} {\bibfnamefont {B.}~\bibnamefont {Dabrowski}}, \bibinfo
  {author} {\bibfnamefont {L.~H.}\ \bibnamefont {Tjeng}}, \bibinfo {author}
  {\bibfnamefont {P.~J.}\ \bibnamefont {Ryan}}, \ and\ \bibinfo {author}
  {\bibfnamefont {J.~W.}\ \bibnamefont {Freeland}},\ }\href {\doibase
  10.1103/PhysRevLett.107.116805} {\bibfield  {journal} {\bibinfo  {journal}
  {Phys. Rev. Lett.}\ }\textbf {\bibinfo {volume} {107}},\ \bibinfo {pages}
  {116805} (\bibinfo {year} {2011})}\BibitemShut {NoStop}%
\bibitem [{\citenamefont {Meyers}\ \emph {et~al.}(2013)\citenamefont {Meyers},
  \citenamefont {Middey}, \citenamefont {Kareev}, \citenamefont {van
  Veenendaal}, \citenamefont {Moon}, \citenamefont {Gray}, \citenamefont {Liu},
  \citenamefont {Freeland},\ and\ \citenamefont
  {Chakhalian}}]{Meyers:2013p075116}%
  \BibitemOpen
  \bibfield  {author} {\bibinfo {author} {\bibfnamefont {D.}~\bibnamefont
  {Meyers}}, \bibinfo {author} {\bibfnamefont {S.}~\bibnamefont {Middey}},
  \bibinfo {author} {\bibfnamefont {M.}~\bibnamefont {Kareev}}, \bibinfo
  {author} {\bibfnamefont {M.}~\bibnamefont {van Veenendaal}}, \bibinfo
  {author} {\bibfnamefont {E.~J.}\ \bibnamefont {Moon}}, \bibinfo {author}
  {\bibfnamefont {B.~A.}\ \bibnamefont {Gray}}, \bibinfo {author}
  {\bibfnamefont {J.}~\bibnamefont {Liu}}, \bibinfo {author} {\bibfnamefont
  {J.~W.}\ \bibnamefont {Freeland}}, \ and\ \bibinfo {author} {\bibfnamefont
  {J.}~\bibnamefont {Chakhalian}},\ }\href {\doibase
  10.1103/PhysRevB.88.075116} {\bibfield  {journal} {\bibinfo  {journal} {Phys.
  Rev. B}\ }\textbf {\bibinfo {volume} {88}},\ \bibinfo {pages} {075116}
  (\bibinfo {year} {2013})}\BibitemShut {NoStop}%
\bibitem [{\citenamefont {Bruno}\ \emph {et~al.}(2013)\citenamefont {Bruno},
  \citenamefont {Rushchanskii}, \citenamefont {Valencia}, \citenamefont
  {Dumont}, \citenamefont {Carr\'et\'ero}, \citenamefont {Jacquet},
  \citenamefont {Abrudan}, \citenamefont {Bl\"ugel}, \citenamefont {Le\ifmmode
  \check{z}\else \v{z}\fi{}ai\ifmmode~\acute{c}\else \'{c}\fi{}}, \citenamefont
  {Bibes},\ and\ \citenamefont {Barth\'el\'emy}}]{Bruno:2013p195108}%
  \BibitemOpen
  \bibfield  {author} {\bibinfo {author} {\bibfnamefont {F.~Y.}\ \bibnamefont
  {Bruno}}, \bibinfo {author} {\bibfnamefont {K.~Z.}\ \bibnamefont
  {Rushchanskii}}, \bibinfo {author} {\bibfnamefont {S.}~\bibnamefont
  {Valencia}}, \bibinfo {author} {\bibfnamefont {Y.}~\bibnamefont {Dumont}},
  \bibinfo {author} {\bibfnamefont {C.}~\bibnamefont {Carr\'et\'ero}}, \bibinfo
  {author} {\bibfnamefont {E.}~\bibnamefont {Jacquet}}, \bibinfo {author}
  {\bibfnamefont {R.}~\bibnamefont {Abrudan}}, \bibinfo {author} {\bibfnamefont
  {S.}~\bibnamefont {Bl\"ugel}}, \bibinfo {author} {\bibfnamefont
  {M.}~\bibnamefont {Le\ifmmode \check{z}\else
  \v{z}\fi{}ai\ifmmode~\acute{c}\else \'{c}\fi{}}}, \bibinfo {author}
  {\bibfnamefont {M.}~\bibnamefont {Bibes}}, \ and\ \bibinfo {author}
  {\bibfnamefont {A.}~\bibnamefont {Barth\'el\'emy}},\ }\href {\doibase
  10.1103/PhysRevB.88.195108} {\bibfield  {journal} {\bibinfo  {journal} {Phys.
  Rev. B}\ }\textbf {\bibinfo {volume} {88}},\ \bibinfo {pages} {195108}
  (\bibinfo {year} {2013})}\BibitemShut {NoStop}%
\bibitem [{\citenamefont {Hepting}\ \emph {et~al.}(2014)\citenamefont
  {Hepting}, \citenamefont {Minola}, \citenamefont {Frano}, \citenamefont
  {Cristiani}, \citenamefont {Logvenov}, \citenamefont {Schierle},
  \citenamefont {Wu}, \citenamefont {Bluschke}, \citenamefont {Weschke},
  \citenamefont {Habermeier}, \citenamefont {Benckiser}, \citenamefont
  {Le~Tacon},\ and\ \citenamefont {Keimer}}]{Hepting:2014p227206}%
  \BibitemOpen
  \bibfield  {author} {\bibinfo {author} {\bibfnamefont {M.}~\bibnamefont
  {Hepting}}, \bibinfo {author} {\bibfnamefont {M.}~\bibnamefont {Minola}},
  \bibinfo {author} {\bibfnamefont {A.}~\bibnamefont {Frano}}, \bibinfo
  {author} {\bibfnamefont {G.}~\bibnamefont {Cristiani}}, \bibinfo {author}
  {\bibfnamefont {G.}~\bibnamefont {Logvenov}}, \bibinfo {author}
  {\bibfnamefont {E.}~\bibnamefont {Schierle}}, \bibinfo {author}
  {\bibfnamefont {M.}~\bibnamefont {Wu}}, \bibinfo {author} {\bibfnamefont
  {M.}~\bibnamefont {Bluschke}}, \bibinfo {author} {\bibfnamefont
  {E.}~\bibnamefont {Weschke}}, \bibinfo {author} {\bibfnamefont {H.-U.}\
  \bibnamefont {Habermeier}}, \bibinfo {author} {\bibfnamefont
  {E.}~\bibnamefont {Benckiser}}, \bibinfo {author} {\bibfnamefont
  {M.}~\bibnamefont {Le~Tacon}}, \ and\ \bibinfo {author} {\bibfnamefont
  {B.}~\bibnamefont {Keimer}},\ }\href {\doibase
  10.1103/PhysRevLett.113.227206} {\bibfield  {journal} {\bibinfo  {journal}
  {Phys. Rev. Lett.}\ }\textbf {\bibinfo {volume} {113}},\ \bibinfo {pages}
  {227206} (\bibinfo {year} {2014})}\BibitemShut {NoStop}%
\bibitem [{\citenamefont {Patel}\ \emph
  {et~al.}(2020{\natexlab{a}})\citenamefont {Patel}, \citenamefont {Ojha},
  \citenamefont {Kumar}, \citenamefont {Saha}, \citenamefont {Mandal},
  \citenamefont {Freeland},\ and\ \citenamefont {Middey}}]{Ranjan:2020p071601}%
  \BibitemOpen
  \bibfield  {author} {\bibinfo {author} {\bibfnamefont {R.~K.}\ \bibnamefont
  {Patel}}, \bibinfo {author} {\bibfnamefont {S.~K.}\ \bibnamefont {Ojha}},
  \bibinfo {author} {\bibfnamefont {S.}~\bibnamefont {Kumar}}, \bibinfo
  {author} {\bibfnamefont {A.}~\bibnamefont {Saha}}, \bibinfo {author}
  {\bibfnamefont {P.}~\bibnamefont {Mandal}}, \bibinfo {author} {\bibfnamefont
  {J.~W.}\ \bibnamefont {Freeland}}, \ and\ \bibinfo {author} {\bibfnamefont
  {S.}~\bibnamefont {Middey}},\ }\href {\doibase 10.1063/1.5133710} {\bibfield
  {journal} {\bibinfo  {journal} {Applied Physics Letters}\ }\textbf {\bibinfo
  {volume} {116}},\ \bibinfo {pages} {071601} (\bibinfo {year}
  {2020}{\natexlab{a}})},\ \Eprint
  {http://arxiv.org/abs/https://doi.org/10.1063/1.5133710}
  {https://doi.org/10.1063/1.5133710} \BibitemShut {NoStop}%
\bibitem [{\citenamefont {Son}\ \emph {et~al.}(2010)\citenamefont {Son},
  \citenamefont {Moetakef}, \citenamefont {LeBeau}, \citenamefont {Ouellette},
  \citenamefont {Balents}, \citenamefont {Allen},\ and\ \citenamefont
  {Stemmer}}]{Son:2010p062114}%
  \BibitemOpen
  \bibfield  {author} {\bibinfo {author} {\bibfnamefont {J.}~\bibnamefont
  {Son}}, \bibinfo {author} {\bibfnamefont {P.}~\bibnamefont {Moetakef}},
  \bibinfo {author} {\bibfnamefont {J.~M.}\ \bibnamefont {LeBeau}}, \bibinfo
  {author} {\bibfnamefont {D.}~\bibnamefont {Ouellette}}, \bibinfo {author}
  {\bibfnamefont {L.}~\bibnamefont {Balents}}, \bibinfo {author} {\bibfnamefont
  {S.~J.}\ \bibnamefont {Allen}}, \ and\ \bibinfo {author} {\bibfnamefont
  {S.}~\bibnamefont {Stemmer}},\ }\href {\doibase
  http://dx.doi.org/10.1063/1.3309713} {\bibfield  {journal} {\bibinfo
  {journal} {Applied Physics Letters}\ }\textbf {\bibinfo {volume} {96}},\
  \bibinfo {pages} {062114} (\bibinfo {year} {2010})}\BibitemShut {NoStop}%
\bibitem [{\citenamefont {Ha}\ \emph {et~al.}(2012)\citenamefont {Ha},
  \citenamefont {Otaki}, \citenamefont {Jaramillo}, \citenamefont {Podpirka},\
  and\ \citenamefont {Ramanathan}}]{Ha:2012p233}%
  \BibitemOpen
  \bibfield  {author} {\bibinfo {author} {\bibfnamefont {S.~D.}\ \bibnamefont
  {Ha}}, \bibinfo {author} {\bibfnamefont {M.}~\bibnamefont {Otaki}}, \bibinfo
  {author} {\bibfnamefont {R.}~\bibnamefont {Jaramillo}}, \bibinfo {author}
  {\bibfnamefont {A.}~\bibnamefont {Podpirka}}, \ and\ \bibinfo {author}
  {\bibfnamefont {S.}~\bibnamefont {Ramanathan}},\ }\href {\doibase
  http://dx.doi.org/10.1016/j.jssc.2012.02.047} {\bibfield  {journal} {\bibinfo
   {journal} {Journal of Solid State Chemistry}\ }\textbf {\bibinfo {volume}
  {190}},\ \bibinfo {pages} {233 } (\bibinfo {year} {2012})}\BibitemShut
  {NoStop}%
\bibitem [{\citenamefont {Hauser}\ \emph {et~al.}(2015)\citenamefont {Hauser},
  \citenamefont {Mikheev}, \citenamefont {Moreno}, \citenamefont {Hwang},
  \citenamefont {Zhang},\ and\ \citenamefont {Stemmer}}]{Hauser:2015p092104}%
  \BibitemOpen
  \bibfield  {author} {\bibinfo {author} {\bibfnamefont {A.~J.}\ \bibnamefont
  {Hauser}}, \bibinfo {author} {\bibfnamefont {E.}~\bibnamefont {Mikheev}},
  \bibinfo {author} {\bibfnamefont {N.~E.}\ \bibnamefont {Moreno}}, \bibinfo
  {author} {\bibfnamefont {J.}~\bibnamefont {Hwang}}, \bibinfo {author}
  {\bibfnamefont {J.~Y.}\ \bibnamefont {Zhang}}, \ and\ \bibinfo {author}
  {\bibfnamefont {S.}~\bibnamefont {Stemmer}},\ }\href {\doibase
  http://dx.doi.org/10.1063/1.4914002} {\bibfield  {journal} {\bibinfo
  {journal} {Applied Physics Letters}\ }\textbf {\bibinfo {volume} {106}},\
  \bibinfo {eid} {092104} (\bibinfo {year} {2015})}\BibitemShut {NoStop}%
\bibitem [{\citenamefont {Catalano}\ \emph {et~al.}(2014)\citenamefont
  {Catalano}, \citenamefont {Gibert}, \citenamefont {Bisogni}, \citenamefont
  {Peil}, \citenamefont {He}, \citenamefont {Sutarto}, \citenamefont {Viret},
  \citenamefont {Zubko}, \citenamefont {Scherwitzl}, \citenamefont {Georges},
  \citenamefont {Sawatzky}, \citenamefont {Schmitt},\ and\ \citenamefont
  {Triscone}}]{Catalano:2014p116110}%
  \BibitemOpen
  \bibfield  {author} {\bibinfo {author} {\bibfnamefont {S.}~\bibnamefont
  {Catalano}}, \bibinfo {author} {\bibfnamefont {M.}~\bibnamefont {Gibert}},
  \bibinfo {author} {\bibfnamefont {V.}~\bibnamefont {Bisogni}}, \bibinfo
  {author} {\bibfnamefont {O.~E.}\ \bibnamefont {Peil}}, \bibinfo {author}
  {\bibfnamefont {F.}~\bibnamefont {He}}, \bibinfo {author} {\bibfnamefont
  {R.}~\bibnamefont {Sutarto}}, \bibinfo {author} {\bibfnamefont
  {M.}~\bibnamefont {Viret}}, \bibinfo {author} {\bibfnamefont
  {P.}~\bibnamefont {Zubko}}, \bibinfo {author} {\bibfnamefont
  {R.}~\bibnamefont {Scherwitzl}}, \bibinfo {author} {\bibfnamefont
  {A.}~\bibnamefont {Georges}}, \bibinfo {author} {\bibfnamefont {G.~A.}\
  \bibnamefont {Sawatzky}}, \bibinfo {author} {\bibfnamefont {T.}~\bibnamefont
  {Schmitt}}, \ and\ \bibinfo {author} {\bibfnamefont {J.-M.}\ \bibnamefont
  {Triscone}},\ }\href {\doibase http://dx.doi.org/10.1063/1.4902138}
  {\bibfield  {journal} {\bibinfo  {journal} {APL Materials}\ }\textbf
  {\bibinfo {volume} {2}},\ \bibinfo {eid} {116110} (\bibinfo {year}
  {2014})}\BibitemShut {NoStop}%
\bibitem [{\citenamefont {Nikolaev}\ \emph {et~al.}(1999)\citenamefont
  {Nikolaev}, \citenamefont {Bhattacharya}, \citenamefont {Kraus},
  \citenamefont {Vas'ko}, \citenamefont {Cooley},\ and\ \citenamefont
  {Goldman}}]{Nikolaev:1999p118}%
  \BibitemOpen
  \bibfield  {author} {\bibinfo {author} {\bibfnamefont {K.~R.}\ \bibnamefont
  {Nikolaev}}, \bibinfo {author} {\bibfnamefont {A.}~\bibnamefont
  {Bhattacharya}}, \bibinfo {author} {\bibfnamefont {P.~A.}\ \bibnamefont
  {Kraus}}, \bibinfo {author} {\bibfnamefont {V.~A.}\ \bibnamefont {Vas'ko}},
  \bibinfo {author} {\bibfnamefont {W.~K.}\ \bibnamefont {Cooley}}, \ and\
  \bibinfo {author} {\bibfnamefont {A.~M.}\ \bibnamefont {Goldman}},\
  }\href@noop {} {\bibfield  {journal} {\bibinfo  {journal} {Applied Physics
  Letters}\ }\textbf {\bibinfo {volume} {75}},\ \bibinfo {pages} {118}
  (\bibinfo {year} {1999})}\BibitemShut {NoStop}%
\bibitem [{\citenamefont {May}\ \emph {et~al.}(2010)\citenamefont {May},
  \citenamefont {Kim}, \citenamefont {Rondinelli}, \citenamefont {Karapetrova},
  \citenamefont {Spaldin}, \citenamefont {Bhattacharya},\ and\ \citenamefont
  {Ryan}}]{May:2010p014110}%
  \BibitemOpen
  \bibfield  {author} {\bibinfo {author} {\bibfnamefont {S.~J.}\ \bibnamefont
  {May}}, \bibinfo {author} {\bibfnamefont {J.-W.}\ \bibnamefont {Kim}},
  \bibinfo {author} {\bibfnamefont {J.~M.}\ \bibnamefont {Rondinelli}},
  \bibinfo {author} {\bibfnamefont {E.}~\bibnamefont {Karapetrova}}, \bibinfo
  {author} {\bibfnamefont {N.~A.}\ \bibnamefont {Spaldin}}, \bibinfo {author}
  {\bibfnamefont {A.}~\bibnamefont {Bhattacharya}}, \ and\ \bibinfo {author}
  {\bibfnamefont {P.~J.}\ \bibnamefont {Ryan}},\ }\href {\doibase
  10.1103/PhysRevB.82.014110} {\bibfield  {journal} {\bibinfo  {journal} {Phys.
  Rev. B}\ }\textbf {\bibinfo {volume} {82}},\ \bibinfo {pages} {014110}
  (\bibinfo {year} {2010})}\BibitemShut {NoStop}%
\bibitem [{\citenamefont {Feigl}\ \emph {et~al.}(2013)\citenamefont {Feigl},
  \citenamefont {Schultz}, \citenamefont {Ohya}, \citenamefont {Ouellette},
  \citenamefont {Kozhanov},\ and\ \citenamefont {Palmstram}}]{Feigl:2013p51}%
  \BibitemOpen
  \bibfield  {author} {\bibinfo {author} {\bibfnamefont {L.}~\bibnamefont
  {Feigl}}, \bibinfo {author} {\bibfnamefont {B.}~\bibnamefont {Schultz}},
  \bibinfo {author} {\bibfnamefont {S.}~\bibnamefont {Ohya}}, \bibinfo {author}
  {\bibfnamefont {D.}~\bibnamefont {Ouellette}}, \bibinfo {author}
  {\bibfnamefont {A.}~\bibnamefont {Kozhanov}}, \ and\ \bibinfo {author}
  {\bibfnamefont {C.}~\bibnamefont {Palmstram}},\ }\href {\doibase
  http://dx.doi.org/10.1016/j.jcrysgro.2012.12.018} {\bibfield  {journal}
  {\bibinfo  {journal} {Journal of Crystal Growth}\ }\textbf {\bibinfo {volume}
  {366}},\ \bibinfo {pages} {51 } (\bibinfo {year} {2013})}\BibitemShut
  {NoStop}%
\bibitem [{\citenamefont {Chaloupka}\ and\ \citenamefont
  {Khaliullin}(2008)}]{Chaloupka:2008p016404}%
  \BibitemOpen
  \bibfield  {author} {\bibinfo {author} {\bibfnamefont {J.~c.~v.}\
  \bibnamefont {Chaloupka}}\ and\ \bibinfo {author} {\bibfnamefont
  {G.}~\bibnamefont {Khaliullin}},\ }\href {\doibase
  10.1103/PhysRevLett.100.016404} {\bibfield  {journal} {\bibinfo  {journal}
  {Phys. Rev. Lett.}\ }\textbf {\bibinfo {volume} {100}},\ \bibinfo {pages}
  {016404} (\bibinfo {year} {2008})}\BibitemShut {NoStop}%
\bibitem [{\citenamefont {Hansmann}\ \emph {et~al.}(2009)\citenamefont
  {Hansmann}, \citenamefont {Yang}, \citenamefont {Toschi}, \citenamefont
  {Khaliullin}, \citenamefont {Andersen},\ and\ \citenamefont
  {Held}}]{Hansmann:2009p016401}%
  \BibitemOpen
  \bibfield  {author} {\bibinfo {author} {\bibfnamefont {P.}~\bibnamefont
  {Hansmann}}, \bibinfo {author} {\bibfnamefont {X.}~\bibnamefont {Yang}},
  \bibinfo {author} {\bibfnamefont {A.}~\bibnamefont {Toschi}}, \bibinfo
  {author} {\bibfnamefont {G.}~\bibnamefont {Khaliullin}}, \bibinfo {author}
  {\bibfnamefont {O.~K.}\ \bibnamefont {Andersen}}, \ and\ \bibinfo {author}
  {\bibfnamefont {K.}~\bibnamefont {Held}},\ }\href {\doibase
  10.1103/PhysRevLett.103.016401} {\bibfield  {journal} {\bibinfo  {journal}
  {Phys. Rev. Lett.}\ }\textbf {\bibinfo {volume} {103}},\ \bibinfo {pages}
  {016401} (\bibinfo {year} {2009})}\BibitemShut {NoStop}%
\bibitem [{\citenamefont {Middey}\ \emph {et~al.}(2016)\citenamefont {Middey},
  \citenamefont {Chakhalian}, \citenamefont {Mahadevan}, \citenamefont
  {Freeland}, \citenamefont {Millis},\ and\ \citenamefont
  {Sarma}}]{Middey:2016p305}%
  \BibitemOpen
  \bibfield  {author} {\bibinfo {author} {\bibfnamefont {S.}~\bibnamefont
  {Middey}}, \bibinfo {author} {\bibfnamefont {J.}~\bibnamefont {Chakhalian}},
  \bibinfo {author} {\bibfnamefont {P.}~\bibnamefont {Mahadevan}}, \bibinfo
  {author} {\bibfnamefont {J.~W.}\ \bibnamefont {Freeland}}, \bibinfo {author}
  {\bibfnamefont {A.~J.}\ \bibnamefont {Millis}}, \ and\ \bibinfo {author}
  {\bibfnamefont {D.~D.}\ \bibnamefont {Sarma}},\ }\href {\doibase
  10.1146/annurev-matsci-070115-032057} {\bibfield  {journal} {\bibinfo
  {journal} {Annual Review of Materials Research}\ }\textbf {\bibinfo {volume}
  {46}},\ \bibinfo {pages} {305} (\bibinfo {year} {2016})}\BibitemShut
  {NoStop}%
\bibitem [{\citenamefont {Catalano}\ \emph {et~al.}(2018)\citenamefont
  {Catalano}, \citenamefont {Gibert}, \citenamefont {Fowlie}, \citenamefont
  {{\'I}{\~n}iguez}, \citenamefont {Triscone},\ and\ \citenamefont
  {Kreisel}}]{Catalano:2018p046501}%
  \BibitemOpen
  \bibfield  {author} {\bibinfo {author} {\bibfnamefont {S.}~\bibnamefont
  {Catalano}}, \bibinfo {author} {\bibfnamefont {M.}~\bibnamefont {Gibert}},
  \bibinfo {author} {\bibfnamefont {J.}~\bibnamefont {Fowlie}}, \bibinfo
  {author} {\bibfnamefont {J.}~\bibnamefont {{\'I}{\~n}iguez}}, \bibinfo
  {author} {\bibfnamefont {J.-M.}\ \bibnamefont {Triscone}}, \ and\ \bibinfo
  {author} {\bibfnamefont {J.}~\bibnamefont {Kreisel}},\ }\href {\doibase
  10.1088/1361-6633/aaa37a} {\bibfield  {journal} {\bibinfo  {journal} {Reports
  on Progress in Physics}\ }\textbf {\bibinfo {volume} {81}},\ \bibinfo {pages}
  {046501} (\bibinfo {year} {2018})}\BibitemShut {NoStop}%
\bibitem [{\citenamefont {Kim}\ \emph {et~al.}(2020)\citenamefont {Kim},
  \citenamefont {Choi}, \citenamefont {Middey}, \citenamefont {Meyers},
  \citenamefont {Chakhalian}, \citenamefont {Shafer}, \citenamefont {Park},\
  and\ \citenamefont {Ryan}}]{Kim:2020p127601}%
  \BibitemOpen
  \bibfield  {author} {\bibinfo {author} {\bibfnamefont {J.-W.}\ \bibnamefont
  {Kim}}, \bibinfo {author} {\bibfnamefont {Y.}~\bibnamefont {Choi}}, \bibinfo
  {author} {\bibfnamefont {S.}~\bibnamefont {Middey}}, \bibinfo {author}
  {\bibfnamefont {D.}~\bibnamefont {Meyers}}, \bibinfo {author} {\bibfnamefont
  {J.}~\bibnamefont {Chakhalian}}, \bibinfo {author} {\bibfnamefont
  {P.}~\bibnamefont {Shafer}}, \bibinfo {author} {\bibfnamefont
  {H.}~\bibnamefont {Park}}, \ and\ \bibinfo {author} {\bibfnamefont {P.~J.}\
  \bibnamefont {Ryan}},\ }\href {\doibase 10.1103/PhysRevLett.124.127601}
  {\bibfield  {journal} {\bibinfo  {journal} {Phys. Rev. Lett.}\ }\textbf
  {\bibinfo {volume} {124}},\ \bibinfo {pages} {127601} (\bibinfo {year}
  {2020})}\BibitemShut {NoStop}%
\bibitem [{\citenamefont {Patel}\ \emph
  {et~al.}(2020{\natexlab{b}})\citenamefont {Patel}, \citenamefont {Meyers},
  \citenamefont {Liu}, \citenamefont {Mandal}, \citenamefont {Kareev},
  \citenamefont {Shafer}, \citenamefont {Kim}, \citenamefont {Ryan},
  \citenamefont {Middey},\ and\ \citenamefont
  {Chakhalian}}]{Patel:2020p041113}%
  \BibitemOpen
  \bibfield  {author} {\bibinfo {author} {\bibfnamefont {R.~K.}\ \bibnamefont
  {Patel}}, \bibinfo {author} {\bibfnamefont {D.}~\bibnamefont {Meyers}},
  \bibinfo {author} {\bibfnamefont {X.}~\bibnamefont {Liu}}, \bibinfo {author}
  {\bibfnamefont {P.}~\bibnamefont {Mandal}}, \bibinfo {author} {\bibfnamefont
  {M.}~\bibnamefont {Kareev}}, \bibinfo {author} {\bibfnamefont
  {P.}~\bibnamefont {Shafer}}, \bibinfo {author} {\bibfnamefont {J.-W.}\
  \bibnamefont {Kim}}, \bibinfo {author} {\bibfnamefont {P.}~\bibnamefont
  {Ryan}}, \bibinfo {author} {\bibfnamefont {S.}~\bibnamefont {Middey}}, \ and\
  \bibinfo {author} {\bibfnamefont {J.}~\bibnamefont {Chakhalian}},\
  }\href@noop {} {\bibfield  {journal} {\bibinfo  {journal} {APL Materials}\
  }\textbf {\bibinfo {volume} {8}},\ \bibinfo {pages} {041113} (\bibinfo {year}
  {2020}{\natexlab{b}})}\BibitemShut {NoStop}%
\bibitem [{\citenamefont {Glazer}(1972)}]{Glazer:1972p09401}%
  \BibitemOpen
  \bibfield  {author} {\bibinfo {author} {\bibfnamefont {A.~M.}\ \bibnamefont
  {Glazer}},\ }\href {\doibase 10.1107/S0567740872007976} {\bibfield  {journal}
  {\bibinfo  {journal} {Acta Crystallographica Section B}\ }\textbf {\bibinfo
  {volume} {28}},\ \bibinfo {pages} {3384} (\bibinfo {year}
  {1972})}\BibitemShut {NoStop}%
\bibitem [{\citenamefont {Tung}\ \emph {et~al.}(2013)\citenamefont {Tung},
  \citenamefont {Balachandran}, \citenamefont {Liu}, \citenamefont {Gray},
  \citenamefont {Karapetrova}, \citenamefont {Lee}, \citenamefont {Chakhalian},
  \citenamefont {Bedzyk}, \citenamefont {Rondinelli},\ and\ \citenamefont
  {Freeland}}]{Tung:2013p205112}%
  \BibitemOpen
  \bibfield  {author} {\bibinfo {author} {\bibfnamefont {I.~C.}\ \bibnamefont
  {Tung}}, \bibinfo {author} {\bibfnamefont {P.~V.}\ \bibnamefont
  {Balachandran}}, \bibinfo {author} {\bibfnamefont {J.}~\bibnamefont {Liu}},
  \bibinfo {author} {\bibfnamefont {B.~A.}\ \bibnamefont {Gray}}, \bibinfo
  {author} {\bibfnamefont {E.~A.}\ \bibnamefont {Karapetrova}}, \bibinfo
  {author} {\bibfnamefont {J.~H.}\ \bibnamefont {Lee}}, \bibinfo {author}
  {\bibfnamefont {J.}~\bibnamefont {Chakhalian}}, \bibinfo {author}
  {\bibfnamefont {M.~J.}\ \bibnamefont {Bedzyk}}, \bibinfo {author}
  {\bibfnamefont {J.~M.}\ \bibnamefont {Rondinelli}}, \ and\ \bibinfo {author}
  {\bibfnamefont {J.~W.}\ \bibnamefont {Freeland}},\ }\href {\doibase
  10.1103/PhysRevB.88.205112} {\bibfield  {journal} {\bibinfo  {journal} {Phys.
  Rev. B}\ }\textbf {\bibinfo {volume} {88}},\ \bibinfo {pages} {205112}
  (\bibinfo {year} {2013})}\BibitemShut {NoStop}%
\bibitem [{\citenamefont {Middey}\ \emph
  {et~al.}(2018{\natexlab{a}})\citenamefont {Middey}, \citenamefont {Meyers},
  \citenamefont {Kareev}, \citenamefont {Cao}, \citenamefont {Liu},
  \citenamefont {Shafer}, \citenamefont {Freeland}, \citenamefont {Kim},
  \citenamefont {Ryan},\ and\ \citenamefont {Chakhalian}}]{Middey:2018p156801}%
  \BibitemOpen
  \bibfield  {author} {\bibinfo {author} {\bibfnamefont {S.}~\bibnamefont
  {Middey}}, \bibinfo {author} {\bibfnamefont {D.}~\bibnamefont {Meyers}},
  \bibinfo {author} {\bibfnamefont {M.}~\bibnamefont {Kareev}}, \bibinfo
  {author} {\bibfnamefont {Y.}~\bibnamefont {Cao}}, \bibinfo {author}
  {\bibfnamefont {X.}~\bibnamefont {Liu}}, \bibinfo {author} {\bibfnamefont
  {P.}~\bibnamefont {Shafer}}, \bibinfo {author} {\bibfnamefont {J.~W.}\
  \bibnamefont {Freeland}}, \bibinfo {author} {\bibfnamefont {J.-W.}\
  \bibnamefont {Kim}}, \bibinfo {author} {\bibfnamefont {P.~J.}\ \bibnamefont
  {Ryan}}, \ and\ \bibinfo {author} {\bibfnamefont {J.}~\bibnamefont
  {Chakhalian}},\ }\href {\doibase 10.1103/PhysRevLett.120.156801} {\bibfield
  {journal} {\bibinfo  {journal} {Phys. Rev. Lett.}\ }\textbf {\bibinfo
  {volume} {120}},\ \bibinfo {pages} {156801} (\bibinfo {year}
  {2018}{\natexlab{a}})}\BibitemShut {NoStop}%
\bibitem [{\citenamefont {Middey}\ \emph
  {et~al.}(2018{\natexlab{b}})\citenamefont {Middey}, \citenamefont {Meyers},
  \citenamefont {Ojha}, \citenamefont {Kareev}, \citenamefont {Liu},
  \citenamefont {Cao}, \citenamefont {Freeland},\ and\ \citenamefont
  {Chakhalian}}]{Middey:2018p045115}%
  \BibitemOpen
  \bibfield  {author} {\bibinfo {author} {\bibfnamefont {S.}~\bibnamefont
  {Middey}}, \bibinfo {author} {\bibfnamefont {D.}~\bibnamefont {Meyers}},
  \bibinfo {author} {\bibfnamefont {S.~K.}\ \bibnamefont {Ojha}}, \bibinfo
  {author} {\bibfnamefont {M.}~\bibnamefont {Kareev}}, \bibinfo {author}
  {\bibfnamefont {X.}~\bibnamefont {Liu}}, \bibinfo {author} {\bibfnamefont
  {Y.}~\bibnamefont {Cao}}, \bibinfo {author} {\bibfnamefont {J.~W.}\
  \bibnamefont {Freeland}}, \ and\ \bibinfo {author} {\bibfnamefont
  {J.}~\bibnamefont {Chakhalian}},\ }\href {\doibase
  10.1103/PhysRevB.98.045115} {\bibfield  {journal} {\bibinfo  {journal} {Phys.
  Rev. B}\ }\textbf {\bibinfo {volume} {98}},\ \bibinfo {pages} {045115}
  (\bibinfo {year} {2018}{\natexlab{b}})}\BibitemShut {NoStop}%
\bibitem [{\citenamefont {Middey}\ \emph
  {et~al.}(2018{\natexlab{c}})\citenamefont {Middey}, \citenamefont {Meyers},
  \citenamefont {Kumar~Patel}, \citenamefont {Liu}, \citenamefont {Kareev},
  \citenamefont {Shafer}, \citenamefont {Kim}, \citenamefont {Ryan},\ and\
  \citenamefont {Chakhalian}}]{Middey:2018p081602}%
  \BibitemOpen
  \bibfield  {author} {\bibinfo {author} {\bibfnamefont {S.}~\bibnamefont
  {Middey}}, \bibinfo {author} {\bibfnamefont {D.}~\bibnamefont {Meyers}},
  \bibinfo {author} {\bibfnamefont {R.}~\bibnamefont {Kumar~Patel}}, \bibinfo
  {author} {\bibfnamefont {X.}~\bibnamefont {Liu}}, \bibinfo {author}
  {\bibfnamefont {M.}~\bibnamefont {Kareev}}, \bibinfo {author} {\bibfnamefont
  {P.}~\bibnamefont {Shafer}}, \bibinfo {author} {\bibfnamefont {J.-W.}\
  \bibnamefont {Kim}}, \bibinfo {author} {\bibfnamefont {P.~J.}\ \bibnamefont
  {Ryan}}, \ and\ \bibinfo {author} {\bibfnamefont {J.}~\bibnamefont
  {Chakhalian}},\ }\href {\doibase 10.1063/1.5045756} {\bibfield  {journal}
  {\bibinfo  {journal} {Applied Physics Letters}\ }\textbf {\bibinfo {volume}
  {113}},\ \bibinfo {pages} {081602} (\bibinfo {year}
  {2018}{\natexlab{c}})}\BibitemShut {NoStop}%
\bibitem [{\citenamefont {Green}, \citenamefont {Haverkort},\ and\
  \citenamefont {Sawatzky}(2016)}]{Green:2016p195127}%
  \BibitemOpen
  \bibfield  {author} {\bibinfo {author} {\bibfnamefont {R.~J.}\ \bibnamefont
  {Green}}, \bibinfo {author} {\bibfnamefont {M.~W.}\ \bibnamefont
  {Haverkort}}, \ and\ \bibinfo {author} {\bibfnamefont {G.~A.}\ \bibnamefont
  {Sawatzky}},\ }\href {\doibase 10.1103/PhysRevB.94.195127} {\bibfield
  {journal} {\bibinfo  {journal} {Phys. Rev. B}\ }\textbf {\bibinfo {volume}
  {94}},\ \bibinfo {pages} {195127} (\bibinfo {year} {2016})}\BibitemShut
  {NoStop}%
\bibitem [{\citenamefont {Lee}\ \emph {et~al.}(2021)\citenamefont {Lee},
  \citenamefont {Kim}, \citenamefont {Jeong}, \citenamefont {Yang},
  \citenamefont {Kim}, \citenamefont {Cho}, \citenamefont {Choi}, \citenamefont
  {Kim}, \citenamefont {Choi}, \citenamefont {Lee} \emph
  {et~al.}}]{Lee:2021p54466}%
  \BibitemOpen
  \bibfield  {author} {\bibinfo {author} {\bibfnamefont {J.}~\bibnamefont
  {Lee}}, \bibinfo {author} {\bibfnamefont {G.-Y.}\ \bibnamefont {Kim}},
  \bibinfo {author} {\bibfnamefont {S.}~\bibnamefont {Jeong}}, \bibinfo
  {author} {\bibfnamefont {M.}~\bibnamefont {Yang}}, \bibinfo {author}
  {\bibfnamefont {J.-W.}\ \bibnamefont {Kim}}, \bibinfo {author} {\bibfnamefont
  {B.-G.}\ \bibnamefont {Cho}}, \bibinfo {author} {\bibfnamefont
  {Y.}~\bibnamefont {Choi}}, \bibinfo {author} {\bibfnamefont {S.}~\bibnamefont
  {Kim}}, \bibinfo {author} {\bibfnamefont {J.~S.}\ \bibnamefont {Choi}},
  \bibinfo {author} {\bibfnamefont {T.~K.}\ \bibnamefont {Lee}},  \emph
  {et~al.},\ }\href@noop {} {\bibfield  {journal} {\bibinfo  {journal} {ACS
  Applied Materials \& Interfaces}\ }\textbf {\bibinfo {volume} {13}},\
  \bibinfo {pages} {54466-54475} (\bibinfo {year} {2021})}\BibitemShut
  {NoStop}%
\bibitem [{\citenamefont {Dom{\'\i}nguez}\ \emph {et~al.}(2020)\citenamefont
  {Dom{\'\i}nguez}, \citenamefont {Georgescu}, \citenamefont {Mundet},
  \citenamefont {Zhang}, \citenamefont {Fowlie}, \citenamefont {Mercy},
  \citenamefont {Waelchli}, \citenamefont {Catalano}, \citenamefont
  {Alexander}, \citenamefont {Ghosez} \emph {et~al.}}]{Dominguez:2020p1182}%
  \BibitemOpen
  \bibfield  {author} {\bibinfo {author} {\bibfnamefont {C.}~\bibnamefont
  {Dom{\'\i}nguez}}, \bibinfo {author} {\bibfnamefont {A.~B.}\ \bibnamefont
  {Georgescu}}, \bibinfo {author} {\bibfnamefont {B.}~\bibnamefont {Mundet}},
  \bibinfo {author} {\bibfnamefont {Y.}~\bibnamefont {Zhang}}, \bibinfo
  {author} {\bibfnamefont {J.}~\bibnamefont {Fowlie}}, \bibinfo {author}
  {\bibfnamefont {A.}~\bibnamefont {Mercy}}, \bibinfo {author} {\bibfnamefont
  {A.}~\bibnamefont {Waelchli}}, \bibinfo {author} {\bibfnamefont
  {S.}~\bibnamefont {Catalano}}, \bibinfo {author} {\bibfnamefont {D.~T.}\
  \bibnamefont {Alexander}}, \bibinfo {author} {\bibfnamefont {P.}~\bibnamefont
  {Ghosez}},  \emph {et~al.},\ }\href@noop {} {\bibfield  {journal} {\bibinfo
  {journal} {Nature Materials}\ }\textbf {\bibinfo {volume} {19}},\ \bibinfo
  {pages} {1182} (\bibinfo {year} {2020})}\BibitemShut {NoStop}%
\bibitem [{\citenamefont {Cao}\ \emph {et~al.}(2016)\citenamefont {Cao},
  \citenamefont {Liu}, \citenamefont {Kareev}, \citenamefont {Choudhury},
  \citenamefont {Middey}, \citenamefont {Meyers}, \citenamefont {Kim},
  \citenamefont {Ryan}, \citenamefont {Freeland},\ and\ \citenamefont
  {Chakhalian}}]{Cao:2016p10418}%
  \BibitemOpen
  \bibfield  {author} {\bibinfo {author} {\bibfnamefont {Y.}~\bibnamefont
  {Cao}}, \bibinfo {author} {\bibfnamefont {X.}~\bibnamefont {Liu}}, \bibinfo
  {author} {\bibfnamefont {M.}~\bibnamefont {Kareev}}, \bibinfo {author}
  {\bibfnamefont {D.}~\bibnamefont {Choudhury}}, \bibinfo {author}
  {\bibfnamefont {S.}~\bibnamefont {Middey}}, \bibinfo {author} {\bibfnamefont
  {D.}~\bibnamefont {Meyers}}, \bibinfo {author} {\bibfnamefont
  {J.}~\bibnamefont {Kim}}, \bibinfo {author} {\bibfnamefont {P.}~\bibnamefont
  {Ryan}}, \bibinfo {author} {\bibfnamefont {J.}~\bibnamefont {Freeland}}, \
  and\ \bibinfo {author} {\bibfnamefont {J.}~\bibnamefont {Chakhalian}},\
  }\href@noop {} {\bibfield  {journal} {\bibinfo  {journal} {Nature
  Communications}\ }\textbf {\bibinfo {volume} {7}},\ \bibinfo {pages} {10418}
  (\bibinfo {year} {2016})}\BibitemShut {NoStop}%
\bibitem [{\citenamefont {Chen}, \citenamefont {Millis},\ and\ \citenamefont
  {Marianetti}(2013)}]{Chen:2013p116403}%
  \BibitemOpen
  \bibfield  {author} {\bibinfo {author} {\bibfnamefont {H.}~\bibnamefont
  {Chen}}, \bibinfo {author} {\bibfnamefont {A.~J.}\ \bibnamefont {Millis}}, \
  and\ \bibinfo {author} {\bibfnamefont {C.~A.}\ \bibnamefont {Marianetti}},\
  }\href {\doibase 10.1103/PhysRevLett.111.116403} {\bibfield  {journal}
  {\bibinfo  {journal} {Phys. Rev. Lett.}\ }\textbf {\bibinfo {volume} {111}},\
  \bibinfo {pages} {116403} (\bibinfo {year} {2013})}\BibitemShut {NoStop}%
\bibitem [{\citenamefont {Grisolia}\ \emph {et~al.}(2016)\citenamefont
  {Grisolia}, \citenamefont {Varignon}, \citenamefont {Sanchez-Santolino},
  \citenamefont {Arora}, \citenamefont {Valencia}, \citenamefont {Varela},
  \citenamefont {Abrudan}, \citenamefont {Weschke}, \citenamefont {Schierle},
  \citenamefont {Rault} \emph {et~al.}}]{Grisolia:2016p484}%
  \BibitemOpen
  \bibfield  {author} {\bibinfo {author} {\bibfnamefont {M.}~\bibnamefont
  {Grisolia}}, \bibinfo {author} {\bibfnamefont {J.}~\bibnamefont {Varignon}},
  \bibinfo {author} {\bibfnamefont {G.}~\bibnamefont {Sanchez-Santolino}},
  \bibinfo {author} {\bibfnamefont {A.}~\bibnamefont {Arora}}, \bibinfo
  {author} {\bibfnamefont {S.}~\bibnamefont {Valencia}}, \bibinfo {author}
  {\bibfnamefont {M.}~\bibnamefont {Varela}}, \bibinfo {author} {\bibfnamefont
  {R.}~\bibnamefont {Abrudan}}, \bibinfo {author} {\bibfnamefont
  {E.}~\bibnamefont {Weschke}}, \bibinfo {author} {\bibfnamefont
  {E.}~\bibnamefont {Schierle}}, \bibinfo {author} {\bibfnamefont
  {J.}~\bibnamefont {Rault}},  \emph {et~al.},\ }\href@noop {} {\bibfield
  {journal} {\bibinfo  {journal} {Nature physics}\ }\textbf {\bibinfo {volume}
  {12}},\ \bibinfo {pages} {484} (\bibinfo {year} {2016})}\BibitemShut
  {NoStop}%
\bibitem [{\citenamefont {Chen}\ \emph {et~al.}(2013)\citenamefont {Chen},
  \citenamefont {Kumah}, \citenamefont {Disa}, \citenamefont {Walker},
  \citenamefont {Ahn},\ and\ \citenamefont {Ismail-Beigi}}]{Chen:2013p186402}%
  \BibitemOpen
  \bibfield  {author} {\bibinfo {author} {\bibfnamefont {H.}~\bibnamefont
  {Chen}}, \bibinfo {author} {\bibfnamefont {D.~P.}\ \bibnamefont {Kumah}},
  \bibinfo {author} {\bibfnamefont {A.~S.}\ \bibnamefont {Disa}}, \bibinfo
  {author} {\bibfnamefont {F.~J.}\ \bibnamefont {Walker}}, \bibinfo {author}
  {\bibfnamefont {C.~H.}\ \bibnamefont {Ahn}}, \ and\ \bibinfo {author}
  {\bibfnamefont {S.}~\bibnamefont {Ismail-Beigi}},\ }\href {\doibase
  10.1103/PhysRevLett.110.186402} {\bibfield  {journal} {\bibinfo  {journal}
  {Phys. Rev. Lett.}\ }\textbf {\bibinfo {volume} {110}},\ \bibinfo {pages}
  {186402} (\bibinfo {year} {2013})}\BibitemShut {NoStop}%
\bibitem [{\citenamefont {Disa}\ \emph {et~al.}(2015)\citenamefont {Disa},
  \citenamefont {Kumah}, \citenamefont {Malashevich}, \citenamefont {Chen},
  \citenamefont {Arena}, \citenamefont {Specht}, \citenamefont {Ismail-Beigi},
  \citenamefont {Walker},\ and\ \citenamefont {Ahn}}]{Disa:2015p026801}%
  \BibitemOpen
  \bibfield  {author} {\bibinfo {author} {\bibfnamefont {A.~S.}\ \bibnamefont
  {Disa}}, \bibinfo {author} {\bibfnamefont {D.~P.}\ \bibnamefont {Kumah}},
  \bibinfo {author} {\bibfnamefont {A.}~\bibnamefont {Malashevich}}, \bibinfo
  {author} {\bibfnamefont {H.}~\bibnamefont {Chen}}, \bibinfo {author}
  {\bibfnamefont {D.~A.}\ \bibnamefont {Arena}}, \bibinfo {author}
  {\bibfnamefont {E.~D.}\ \bibnamefont {Specht}}, \bibinfo {author}
  {\bibfnamefont {S.}~\bibnamefont {Ismail-Beigi}}, \bibinfo {author}
  {\bibfnamefont {F.~J.}\ \bibnamefont {Walker}}, \ and\ \bibinfo {author}
  {\bibfnamefont {C.~H.}\ \bibnamefont {Ahn}},\ }\href {\doibase
  10.1103/PhysRevLett.114.026801} {\bibfield  {journal} {\bibinfo  {journal}
  {Phys. Rev. Lett.}\ }\textbf {\bibinfo {volume} {114}},\ \bibinfo {pages}
  {026801} (\bibinfo {year} {2015})}\BibitemShut {NoStop}%
\bibitem [{\citenamefont {Kaiser}\ \emph {et~al.}(2011)\citenamefont {Kaiser},
  \citenamefont {Gray}, \citenamefont {Conti}, \citenamefont {Son},
  \citenamefont {Greer}, \citenamefont {Perona}, \citenamefont {Rattanachata},
  \citenamefont {Saw}, \citenamefont {Bostwick}, \citenamefont {Yang},
  \citenamefont {Yang}, \citenamefont {Gullikson}, \citenamefont {Kortright},
  \citenamefont {Stemmer},\ and\ \citenamefont {Fadley}}]{Kaiser:2011p116402}%
  \BibitemOpen
  \bibfield  {author} {\bibinfo {author} {\bibfnamefont {A.~M.}\ \bibnamefont
  {Kaiser}}, \bibinfo {author} {\bibfnamefont {A.~X.}\ \bibnamefont {Gray}},
  \bibinfo {author} {\bibfnamefont {G.}~\bibnamefont {Conti}}, \bibinfo
  {author} {\bibfnamefont {J.}~\bibnamefont {Son}}, \bibinfo {author}
  {\bibfnamefont {A.}~\bibnamefont {Greer}}, \bibinfo {author} {\bibfnamefont
  {A.}~\bibnamefont {Perona}}, \bibinfo {author} {\bibfnamefont
  {A.}~\bibnamefont {Rattanachata}}, \bibinfo {author} {\bibfnamefont {A.~Y.}\
  \bibnamefont {Saw}}, \bibinfo {author} {\bibfnamefont {A.}~\bibnamefont
  {Bostwick}}, \bibinfo {author} {\bibfnamefont {S.}~\bibnamefont {Yang}},
  \bibinfo {author} {\bibfnamefont {S.-H.}\ \bibnamefont {Yang}}, \bibinfo
  {author} {\bibfnamefont {E.~M.}\ \bibnamefont {Gullikson}}, \bibinfo {author}
  {\bibfnamefont {J.~B.}\ \bibnamefont {Kortright}}, \bibinfo {author}
  {\bibfnamefont {S.}~\bibnamefont {Stemmer}}, \ and\ \bibinfo {author}
  {\bibfnamefont {C.~S.}\ \bibnamefont {Fadley}},\ }\href {\doibase
  10.1103/PhysRevLett.107.116402} {\bibfield  {journal} {\bibinfo  {journal}
  {Phys. Rev. Lett.}\ }\textbf {\bibinfo {volume} {107}},\ \bibinfo {pages}
  {116402} (\bibinfo {year} {2011})}\BibitemShut {NoStop}%
\bibitem [{\citenamefont {Chen}\ \emph {et~al.}(2021)\citenamefont {Chen},
  \citenamefont {Gauquelin}, \citenamefont {Green}, \citenamefont {Lee},
  \citenamefont {Piamonteze}, \citenamefont {Spreitzer}, \citenamefont
  {Jannis}, \citenamefont {Verbeeck}, \citenamefont {Bibes}, \citenamefont
  {Huijben} \emph {et~al.}}]{Chen:2021p1295}%
  \BibitemOpen
  \bibfield  {author} {\bibinfo {author} {\bibfnamefont {B.}~\bibnamefont
  {Chen}}, \bibinfo {author} {\bibfnamefont {N.}~\bibnamefont {Gauquelin}},
  \bibinfo {author} {\bibfnamefont {R.~J.}\ \bibnamefont {Green}}, \bibinfo
  {author} {\bibfnamefont {J.~H.}\ \bibnamefont {Lee}}, \bibinfo {author}
  {\bibfnamefont {C.}~\bibnamefont {Piamonteze}}, \bibinfo {author}
  {\bibfnamefont {M.}~\bibnamefont {Spreitzer}}, \bibinfo {author}
  {\bibfnamefont {D.}~\bibnamefont {Jannis}}, \bibinfo {author} {\bibfnamefont
  {J.}~\bibnamefont {Verbeeck}}, \bibinfo {author} {\bibfnamefont
  {M.}~\bibnamefont {Bibes}}, \bibinfo {author} {\bibfnamefont
  {M.}~\bibnamefont {Huijben}},  \emph {et~al.},\ }\href@noop {} {\bibfield
  {journal} {\bibinfo  {journal} {Nano letters}\ }\textbf {\bibinfo {volume}
  {21}},\ \bibinfo {pages} {1295} (\bibinfo {year} {2021})}\BibitemShut
  {NoStop}%
\bibitem [{\citenamefont {Liao}\ \emph {et~al.}(2018)\citenamefont {Liao},
  \citenamefont {Gauquelin}, \citenamefont {Green}, \citenamefont
  {M{\"u}ller-Caspary}, \citenamefont {Lobato}, \citenamefont {Li},
  \citenamefont {Van~Aert}, \citenamefont {Verbeeck}, \citenamefont {Huijben},
  \citenamefont {Grisolia} \emph {et~al.}}]{Liao:2018p9515}%
  \BibitemOpen
  \bibfield  {author} {\bibinfo {author} {\bibfnamefont {Z.}~\bibnamefont
  {Liao}}, \bibinfo {author} {\bibfnamefont {N.}~\bibnamefont {Gauquelin}},
  \bibinfo {author} {\bibfnamefont {R.~J.}\ \bibnamefont {Green}}, \bibinfo
  {author} {\bibfnamefont {K.}~\bibnamefont {M{\"u}ller-Caspary}}, \bibinfo
  {author} {\bibfnamefont {I.}~\bibnamefont {Lobato}}, \bibinfo {author}
  {\bibfnamefont {L.}~\bibnamefont {Li}}, \bibinfo {author} {\bibfnamefont
  {S.}~\bibnamefont {Van~Aert}}, \bibinfo {author} {\bibfnamefont
  {J.}~\bibnamefont {Verbeeck}}, \bibinfo {author} {\bibfnamefont
  {M.}~\bibnamefont {Huijben}}, \bibinfo {author} {\bibfnamefont {M.~N.}\
  \bibnamefont {Grisolia}},  \emph {et~al.},\ }\href@noop {} {\bibfield
  {journal} {\bibinfo  {journal} {Proceedings of the National Academy of
  Sciences}\ }\textbf {\bibinfo {volume} {115}},\ \bibinfo {pages} {9515}
  (\bibinfo {year} {2018})}\BibitemShut {NoStop}%
\bibitem [{\citenamefont {Zeng}, \citenamefont {Greenblatt},\ and\
  \citenamefont {Croft}(1999)}]{Zeng:1999p8784}%
  \BibitemOpen
  \bibfield  {author} {\bibinfo {author} {\bibfnamefont {Z.}~\bibnamefont
  {Zeng}}, \bibinfo {author} {\bibfnamefont {M.}~\bibnamefont {Greenblatt}}, \
  and\ \bibinfo {author} {\bibfnamefont {M.}~\bibnamefont {Croft}},\ }\href
  {\doibase 10.1103/PhysRevB.59.8784} {\bibfield  {journal} {\bibinfo
  {journal} {Phys. Rev. B}\ }\textbf {\bibinfo {volume} {59}},\ \bibinfo
  {pages} {8784} (\bibinfo {year} {1999})}\BibitemShut {NoStop}%
\bibitem [{\citenamefont {Grutter}\ \emph {et~al.}(2013)\citenamefont
  {Grutter}, \citenamefont {Yang}, \citenamefont {Kirby}, \citenamefont
  {Fitzsimmons}, \citenamefont {Aguiar}, \citenamefont {Browning},
  \citenamefont {Jenkins}, \citenamefont {Arenholz}, \citenamefont {Mehta},
  \citenamefont {Alaan},\ and\ \citenamefont {Suzuki}}]{Grutter:2013p087202}%
  \BibitemOpen
  \bibfield  {author} {\bibinfo {author} {\bibfnamefont {A.~J.}\ \bibnamefont
  {Grutter}}, \bibinfo {author} {\bibfnamefont {H.}~\bibnamefont {Yang}},
  \bibinfo {author} {\bibfnamefont {B.~J.}\ \bibnamefont {Kirby}}, \bibinfo
  {author} {\bibfnamefont {M.~R.}\ \bibnamefont {Fitzsimmons}}, \bibinfo
  {author} {\bibfnamefont {J.~A.}\ \bibnamefont {Aguiar}}, \bibinfo {author}
  {\bibfnamefont {N.~D.}\ \bibnamefont {Browning}}, \bibinfo {author}
  {\bibfnamefont {C.~A.}\ \bibnamefont {Jenkins}}, \bibinfo {author}
  {\bibfnamefont {E.}~\bibnamefont {Arenholz}}, \bibinfo {author}
  {\bibfnamefont {V.~V.}\ \bibnamefont {Mehta}}, \bibinfo {author}
  {\bibfnamefont {U.~S.}\ \bibnamefont {Alaan}}, \ and\ \bibinfo {author}
  {\bibfnamefont {Y.}~\bibnamefont {Suzuki}},\ }\href {\doibase
  10.1103/PhysRevLett.111.087202} {\bibfield  {journal} {\bibinfo  {journal}
  {Phys. Rev. Lett.}\ }\textbf {\bibinfo {volume} {111}},\ \bibinfo {pages}
  {087202} (\bibinfo {year} {2013})}\BibitemShut {NoStop}%
\bibitem [{\citenamefont {Hoffman}\ \emph {et~al.}(2013)\citenamefont
  {Hoffman}, \citenamefont {Tung}, \citenamefont {Nelson-Cheeseman},
  \citenamefont {Liu}, \citenamefont {Freeland},\ and\ \citenamefont
  {Bhattacharya}}]{Hoffman:2013p144411}%
  \BibitemOpen
  \bibfield  {author} {\bibinfo {author} {\bibfnamefont {J.}~\bibnamefont
  {Hoffman}}, \bibinfo {author} {\bibfnamefont {I.~C.}\ \bibnamefont {Tung}},
  \bibinfo {author} {\bibfnamefont {B.~B.}\ \bibnamefont {Nelson-Cheeseman}},
  \bibinfo {author} {\bibfnamefont {M.}~\bibnamefont {Liu}}, \bibinfo {author}
  {\bibfnamefont {J.~W.}\ \bibnamefont {Freeland}}, \ and\ \bibinfo {author}
  {\bibfnamefont {A.}~\bibnamefont {Bhattacharya}},\ }\href {\doibase
  10.1103/PhysRevB.88.144411} {\bibfield  {journal} {\bibinfo  {journal} {Phys.
  Rev. B}\ }\textbf {\bibinfo {volume} {88}},\ \bibinfo {pages} {144411}
  (\bibinfo {year} {2013})}\BibitemShut {NoStop}%
 \bibitem{Gibert:2012p195} M. Gibert, P. Zubko, R. Scherwitzl, J. Iniguez, and J.-M. Triscone,  Nat Mater.  {\bf 11}, 195 (2012).
\bibitem [{\citenamefont {Piamonteze}\ \emph {et~al.}(2015)\citenamefont
  {Piamonteze}, \citenamefont {Gibert}, \citenamefont {Heidler}, \citenamefont
  {Dreiser}, \citenamefont {Rusponi}, \citenamefont {Brune}, \citenamefont
  {Triscone}, \citenamefont {Nolting},\ and\ \citenamefont
  {Staub}}]{Piamonteze:2015p014426}%
  \BibitemOpen
  \bibfield  {author} {\bibinfo {author} {\bibfnamefont {C.}~\bibnamefont
  {Piamonteze}}, \bibinfo {author} {\bibfnamefont {M.}~\bibnamefont {Gibert}},
  \bibinfo {author} {\bibfnamefont {J.}~\bibnamefont {Heidler}}, \bibinfo
  {author} {\bibfnamefont {J.}~\bibnamefont {Dreiser}}, \bibinfo {author}
  {\bibfnamefont {S.}~\bibnamefont {Rusponi}}, \bibinfo {author} {\bibfnamefont
  {H.}~\bibnamefont {Brune}}, \bibinfo {author} {\bibfnamefont {J.-M.}\
  \bibnamefont {Triscone}}, \bibinfo {author} {\bibfnamefont {F.}~\bibnamefont
  {Nolting}}, \ and\ \bibinfo {author} {\bibfnamefont {U.}~\bibnamefont
  {Staub}},\ }\href {\doibase 10.1103/PhysRevB.92.014426} {\bibfield  {journal}
  {\bibinfo  {journal} {Phys. Rev. B}\ }\textbf {\bibinfo {volume} {92}},\
  \bibinfo {pages} {014426} (\bibinfo {year} {2015})}\BibitemShut {NoStop}%
\bibitem [{\citenamefont {Hoffman}\ \emph {et~al.}(2016)\citenamefont
  {Hoffman}, \citenamefont {Kirby}, \citenamefont {Kwon}, \citenamefont
  {Fabbris}, \citenamefont {Meyers}, \citenamefont {Freeland}, \citenamefont
  {Martin}, \citenamefont {Heinonen}, \citenamefont {Steadman}, \citenamefont
  {Zhou}, \citenamefont {Schlep\"utz}, \citenamefont {Dean}, \citenamefont
  {te~Velthuis}, \citenamefont {Zuo},\ and\ \citenamefont
  {Bhattacharya}}]{Hoffman:2016p041038}%
  \BibitemOpen
  \bibfield  {author} {\bibinfo {author} {\bibfnamefont {J.~D.}\ \bibnamefont
  {Hoffman}}, \bibinfo {author} {\bibfnamefont {B.~J.}\ \bibnamefont {Kirby}},
  \bibinfo {author} {\bibfnamefont {J.}~\bibnamefont {Kwon}}, \bibinfo {author}
  {\bibfnamefont {G.}~\bibnamefont {Fabbris}}, \bibinfo {author} {\bibfnamefont
  {D.}~\bibnamefont {Meyers}}, \bibinfo {author} {\bibfnamefont {J.~W.}\
  \bibnamefont {Freeland}}, \bibinfo {author} {\bibfnamefont {I.}~\bibnamefont
  {Martin}}, \bibinfo {author} {\bibfnamefont {O.~G.}\ \bibnamefont
  {Heinonen}}, \bibinfo {author} {\bibfnamefont {P.}~\bibnamefont {Steadman}},
  \bibinfo {author} {\bibfnamefont {H.}~\bibnamefont {Zhou}}, \bibinfo {author}
  {\bibfnamefont {C.~M.}\ \bibnamefont {Schlep\"utz}}, \bibinfo {author}
  {\bibfnamefont {M.~P.~M.}\ \bibnamefont {Dean}}, \bibinfo {author}
  {\bibfnamefont {S.~G.~E.}\ \bibnamefont {te~Velthuis}}, \bibinfo {author}
  {\bibfnamefont {J.-M.}\ \bibnamefont {Zuo}}, \ and\ \bibinfo {author}
  {\bibfnamefont {A.}~\bibnamefont {Bhattacharya}},\ }\href {\doibase
  10.1103/PhysRevX.6.041038} {\bibfield  {journal} {\bibinfo  {journal} {Phys.
  Rev. X}\ }\textbf {\bibinfo {volume} {6}},\ \bibinfo {pages} {041038}
  (\bibinfo {year} {2016})}\BibitemShut {NoStop}%
\bibitem [{\citenamefont {Fabbris}\ \emph {et~al.}(2018)\citenamefont
  {Fabbris}, \citenamefont {Jaouen}, \citenamefont {Meyers}, \citenamefont
  {Feng}, \citenamefont {Hoffman}, \citenamefont {Sutarto}, \citenamefont
  {Chiuzb\ifmmode~\u{a}\else \u{a}\fi{}ian}, \citenamefont {Bhattacharya},\
  and\ \citenamefont {Dean}}]{Fabbris:2018p180401}%
  \BibitemOpen
  \bibfield  {author} {\bibinfo {author} {\bibfnamefont {G.}~\bibnamefont
  {Fabbris}}, \bibinfo {author} {\bibfnamefont {N.}~\bibnamefont {Jaouen}},
  \bibinfo {author} {\bibfnamefont {D.}~\bibnamefont {Meyers}}, \bibinfo
  {author} {\bibfnamefont {J.}~\bibnamefont {Feng}}, \bibinfo {author}
  {\bibfnamefont {J.~D.}\ \bibnamefont {Hoffman}}, \bibinfo {author}
  {\bibfnamefont {R.}~\bibnamefont {Sutarto}}, \bibinfo {author} {\bibfnamefont
  {S.~G.}\ \bibnamefont {Chiuzb\ifmmode~\u{a}\else \u{a}\fi{}ian}}, \bibinfo
  {author} {\bibfnamefont {A.}~\bibnamefont {Bhattacharya}}, \ and\ \bibinfo
  {author} {\bibfnamefont {M.~P.~M.}\ \bibnamefont {Dean}},\ }\href {\doibase
  10.1103/PhysRevB.98.180401} {\bibfield  {journal} {\bibinfo  {journal} {Phys.
  Rev. B}\ }\textbf {\bibinfo {volume} {98}},\ \bibinfo {pages} {180401}
  (\bibinfo {year} {2018})}\BibitemShut {NoStop}%
\bibitem [{\citenamefont {Xu}\ \emph {et~al.}(2018)\citenamefont {Xu},
  \citenamefont {Hu}, \citenamefont {Wu}, \citenamefont {Wang}, \citenamefont
  {Wu},\ and\ \citenamefont {Chen}}]{Xu:2018p30803}%
  \BibitemOpen
  \bibfield  {author} {\bibinfo {author} {\bibfnamefont {Z.}~\bibnamefont
  {Xu}}, \bibinfo {author} {\bibfnamefont {S.}~\bibnamefont {Hu}}, \bibinfo
  {author} {\bibfnamefont {R.}~\bibnamefont {Wu}}, \bibinfo {author}
  {\bibfnamefont {J.-O.}\ \bibnamefont {Wang}}, \bibinfo {author}
  {\bibfnamefont {T.}~\bibnamefont {Wu}}, \ and\ \bibinfo {author}
  {\bibfnamefont {L.}~\bibnamefont {Chen}},\ }\href@noop {} {\bibfield
  {journal} {\bibinfo  {journal} {ACS applied materials \& interfaces}\
  }\textbf {\bibinfo {volume} {10}},\ \bibinfo {pages} {30803} (\bibinfo {year}
  {2018})}\BibitemShut {NoStop}%
\bibitem [{\citenamefont {Chen}\ \emph {et~al.}(2020)\citenamefont {Chen},
  \citenamefont {Luo}, \citenamefont {Chen}, \citenamefont {Abrudan},
  \citenamefont {Koster}, \citenamefont {Mishra},\ and\ \citenamefont
  {Radu}}]{Chen:2020p054408}%
  \BibitemOpen
  \bibfield  {author} {\bibinfo {author} {\bibfnamefont {K.}~\bibnamefont
  {Chen}}, \bibinfo {author} {\bibfnamefont {C.}~\bibnamefont {Luo}}, \bibinfo
  {author} {\bibfnamefont {B.~B.}\ \bibnamefont {Chen}}, \bibinfo {author}
  {\bibfnamefont {R.~M.}\ \bibnamefont {Abrudan}}, \bibinfo {author}
  {\bibfnamefont {G.}~\bibnamefont {Koster}}, \bibinfo {author} {\bibfnamefont
  {S.~K.}\ \bibnamefont {Mishra}}, \ and\ \bibinfo {author} {\bibfnamefont
  {F.}~\bibnamefont {Radu}},\ }\href {\doibase
  10.1103/PhysRevMaterials.4.054408} {\bibfield  {journal} {\bibinfo  {journal}
  {Phys. Rev. Materials}\ }\textbf {\bibinfo {volume} {4}},\ \bibinfo {pages}
  {054408} (\bibinfo {year} {2020})}\BibitemShut {NoStop}%
\bibitem [{\citenamefont {Wrobel}\ \emph {et~al.}(2018)\citenamefont {Wrobel},
  \citenamefont {Geisler}, \citenamefont {Wang}, \citenamefont {Christiani},
  \citenamefont {Logvenov}, \citenamefont {Bluschke}, \citenamefont {Schierle},
  \citenamefont {van Aken}, \citenamefont {Keimer}, \citenamefont {Pentcheva},\
  and\ \citenamefont {Benckiser}}]{Wrobel:2018p035001}%
  \BibitemOpen
  \bibfield  {author} {\bibinfo {author} {\bibfnamefont {F.}~\bibnamefont
  {Wrobel}}, \bibinfo {author} {\bibfnamefont {B.}~\bibnamefont {Geisler}},
  \bibinfo {author} {\bibfnamefont {Y.}~\bibnamefont {Wang}}, \bibinfo {author}
  {\bibfnamefont {G.}~\bibnamefont {Christiani}}, \bibinfo {author}
  {\bibfnamefont {G.}~\bibnamefont {Logvenov}}, \bibinfo {author}
  {\bibfnamefont {M.}~\bibnamefont {Bluschke}}, \bibinfo {author}
  {\bibfnamefont {E.}~\bibnamefont {Schierle}}, \bibinfo {author}
  {\bibfnamefont {P.~A.}\ \bibnamefont {van Aken}}, \bibinfo {author}
  {\bibfnamefont {B.}~\bibnamefont {Keimer}}, \bibinfo {author} {\bibfnamefont
  {R.}~\bibnamefont {Pentcheva}}, \ and\ \bibinfo {author} {\bibfnamefont
  {E.}~\bibnamefont {Benckiser}},\ }\href {\doibase
  10.1103/PhysRevMaterials.2.035001} {\bibfield  {journal} {\bibinfo  {journal}
  {Phys. Rev. Materials}\ }\textbf {\bibinfo {volume} {2}},\ \bibinfo {pages}
  {035001} (\bibinfo {year} {2018})}\BibitemShut {NoStop}%
\bibitem [{\citenamefont {Witczak-Krempa}\ \emph {et~al.}(2014)\citenamefont
  {Witczak-Krempa}, \citenamefont {Chen}, \citenamefont {Kim},\ and\
  \citenamefont {Balents}}]{Witczak:2014p57}%
  \BibitemOpen
  \bibfield  {author} {\bibinfo {author} {\bibfnamefont {W.}~\bibnamefont
  {Witczak-Krempa}}, \bibinfo {author} {\bibfnamefont {G.}~\bibnamefont
  {Chen}}, \bibinfo {author} {\bibfnamefont {Y.~B.}\ \bibnamefont {Kim}}, \
  and\ \bibinfo {author} {\bibfnamefont {L.}~\bibnamefont {Balents}},\
  }\href@noop {} {\bibfield  {journal} {\bibinfo  {journal} {Annu. Rev.
  Condens. Matter Phys.}\ }\textbf {\bibinfo {volume} {5}},\ \bibinfo {pages}
  {57} (\bibinfo {year} {2014})}\BibitemShut {NoStop}%
\bibitem [{\citenamefont {Kim}\ \emph {et~al.}(2008)\citenamefont {Kim},
  \citenamefont {Jin}, \citenamefont {Moon}, \citenamefont {Kim}, \citenamefont
  {Park}, \citenamefont {Leem}, \citenamefont {Yu}, \citenamefont {Noh},
  \citenamefont {Kim}, \citenamefont {Oh}, \citenamefont {Park}, \citenamefont
  {Durairaj}, \citenamefont {Cao},\ and\ \citenamefont
  {Rotenberg}}]{Kim:2008p076402}%
  \BibitemOpen
  \bibfield  {author} {\bibinfo {author} {\bibfnamefont {B.~J.}\ \bibnamefont
  {Kim}}, \bibinfo {author} {\bibfnamefont {H.}~\bibnamefont {Jin}}, \bibinfo
  {author} {\bibfnamefont {S.~J.}\ \bibnamefont {Moon}}, \bibinfo {author}
  {\bibfnamefont {J.-Y.}\ \bibnamefont {Kim}}, \bibinfo {author} {\bibfnamefont
  {B.-G.}\ \bibnamefont {Park}}, \bibinfo {author} {\bibfnamefont {C.~S.}\
  \bibnamefont {Leem}}, \bibinfo {author} {\bibfnamefont {J.}~\bibnamefont
  {Yu}}, \bibinfo {author} {\bibfnamefont {T.~W.}\ \bibnamefont {Noh}},
  \bibinfo {author} {\bibfnamefont {C.}~\bibnamefont {Kim}}, \bibinfo {author}
  {\bibfnamefont {S.-J.}\ \bibnamefont {Oh}}, \bibinfo {author} {\bibfnamefont
  {J.-H.}\ \bibnamefont {Park}}, \bibinfo {author} {\bibfnamefont
  {V.}~\bibnamefont {Durairaj}}, \bibinfo {author} {\bibfnamefont
  {G.}~\bibnamefont {Cao}}, \ and\ \bibinfo {author} {\bibfnamefont
  {E.}~\bibnamefont {Rotenberg}},\ }\href {\doibase
  10.1103/PhysRevLett.101.076402} {\bibfield  {journal} {\bibinfo  {journal}
  {Phys. Rev. Lett.}\ }\textbf {\bibinfo {volume} {101}},\ \bibinfo {pages}
  {076402} (\bibinfo {year} {2008})}\BibitemShut {NoStop}%
\bibitem [{\citenamefont {Moon}\ \emph {et~al.}(2008)\citenamefont {Moon},
  \citenamefont {Jin}, \citenamefont {Kim}, \citenamefont {Choi}, \citenamefont
  {Lee}, \citenamefont {Yu}, \citenamefont {Cao}, \citenamefont {Sumi},
  \citenamefont {Funakubo}, \citenamefont {Bernhard},\ and\ \citenamefont
  {Noh}}]{Moon:2008p226402}%
  \BibitemOpen
  \bibfield  {author} {\bibinfo {author} {\bibfnamefont {S.~J.}\ \bibnamefont
  {Moon}}, \bibinfo {author} {\bibfnamefont {H.}~\bibnamefont {Jin}}, \bibinfo
  {author} {\bibfnamefont {K.~W.}\ \bibnamefont {Kim}}, \bibinfo {author}
  {\bibfnamefont {W.~S.}\ \bibnamefont {Choi}}, \bibinfo {author}
  {\bibfnamefont {Y.~S.}\ \bibnamefont {Lee}}, \bibinfo {author} {\bibfnamefont
  {J.}~\bibnamefont {Yu}}, \bibinfo {author} {\bibfnamefont {G.}~\bibnamefont
  {Cao}}, \bibinfo {author} {\bibfnamefont {A.}~\bibnamefont {Sumi}}, \bibinfo
  {author} {\bibfnamefont {H.}~\bibnamefont {Funakubo}}, \bibinfo {author}
  {\bibfnamefont {C.}~\bibnamefont {Bernhard}}, \ and\ \bibinfo {author}
  {\bibfnamefont {T.~W.}\ \bibnamefont {Noh}},\ }\href {\doibase
  10.1103/PhysRevLett.101.226402} {\bibfield  {journal} {\bibinfo  {journal}
  {Phys. Rev. Lett.}\ }\textbf {\bibinfo {volume} {101}},\ \bibinfo {pages}
  {226402} (\bibinfo {year} {2008})}\BibitemShut {NoStop}%
\bibitem [{\citenamefont {Liu}\ \emph {et~al.}(2019)\citenamefont {Liu},
  \citenamefont {Kotiuga}, \citenamefont {Kim}, \citenamefont {N'Diaye},
  \citenamefont {Choi}, \citenamefont {Zhang}, \citenamefont {Cao},
  \citenamefont {Kareev}, \citenamefont {Wen}, \citenamefont {Pal},
  \citenamefont {Freeland}, \citenamefont {Gu}, \citenamefont {Haskel},
  \citenamefont {Shafer}, \citenamefont {Arenholz}, \citenamefont {Haule},
  \citenamefont {Vanderbilt}, \citenamefont {Rabe},\ and\ \citenamefont
  {Chakhalian}}]{Liu:2019p19863}%
  \BibitemOpen
  \bibfield  {author} {\bibinfo {author} {\bibfnamefont {X.}~\bibnamefont
  {Liu}}, \bibinfo {author} {\bibfnamefont {M.}~\bibnamefont {Kotiuga}},
  \bibinfo {author} {\bibfnamefont {H.-S.}\ \bibnamefont {Kim}}, \bibinfo
  {author} {\bibfnamefont {A.~T.}\ \bibnamefont {N'Diaye}}, \bibinfo {author}
  {\bibfnamefont {Y.}~\bibnamefont {Choi}}, \bibinfo {author} {\bibfnamefont
  {Q.}~\bibnamefont {Zhang}}, \bibinfo {author} {\bibfnamefont
  {Y.}~\bibnamefont {Cao}}, \bibinfo {author} {\bibfnamefont {M.}~\bibnamefont
  {Kareev}}, \bibinfo {author} {\bibfnamefont {F.}~\bibnamefont {Wen}},
  \bibinfo {author} {\bibfnamefont {B.}~\bibnamefont {Pal}}, \bibinfo {author}
  {\bibfnamefont {J.~W.}\ \bibnamefont {Freeland}}, \bibinfo {author}
  {\bibfnamefont {L.}~\bibnamefont {Gu}}, \bibinfo {author} {\bibfnamefont
  {D.}~\bibnamefont {Haskel}}, \bibinfo {author} {\bibfnamefont
  {P.}~\bibnamefont {Shafer}}, \bibinfo {author} {\bibfnamefont
  {E.}~\bibnamefont {Arenholz}}, \bibinfo {author} {\bibfnamefont
  {K.}~\bibnamefont {Haule}}, \bibinfo {author} {\bibfnamefont
  {D.}~\bibnamefont {Vanderbilt}}, \bibinfo {author} {\bibfnamefont {K.~M.}\
  \bibnamefont {Rabe}}, \ and\ \bibinfo {author} {\bibfnamefont
  {J.}~\bibnamefont {Chakhalian}},\ }\href {\doibase 10.1073/pnas.1907043116}
  {\bibfield  {journal} {\bibinfo  {journal} {Proceedings of the National
  Academy of Sciences}\ }\textbf {\bibinfo {volume} {116}},\ \bibinfo {pages}
  {19863} (\bibinfo {year} {2019})}\BibitemShut {NoStop}%
\bibitem [{\citenamefont {Liao}\ \emph {et~al.}(2019)\citenamefont {Liao},
  \citenamefont {Skoropata}, \citenamefont {Freeland}, \citenamefont {Guo},
  \citenamefont {Desautels}, \citenamefont {Gao}, \citenamefont {Sohn},
  \citenamefont {Rastogi}, \citenamefont {Ward}, \citenamefont {Zou} \emph
  {et~al.}}]{Liao:2019p589}%
  \BibitemOpen
  \bibfield  {author} {\bibinfo {author} {\bibfnamefont {Z.}~\bibnamefont
  {Liao}}, \bibinfo {author} {\bibfnamefont {E.}~\bibnamefont {Skoropata}},
  \bibinfo {author} {\bibfnamefont {J.}~\bibnamefont {Freeland}}, \bibinfo
  {author} {\bibfnamefont {E.-J.}\ \bibnamefont {Guo}}, \bibinfo {author}
  {\bibfnamefont {R.}~\bibnamefont {Desautels}}, \bibinfo {author}
  {\bibfnamefont {X.}~\bibnamefont {Gao}}, \bibinfo {author} {\bibfnamefont
  {C.}~\bibnamefont {Sohn}}, \bibinfo {author} {\bibfnamefont {A.}~\bibnamefont
  {Rastogi}}, \bibinfo {author} {\bibfnamefont {T.~Z.}\ \bibnamefont {Ward}},
  \bibinfo {author} {\bibfnamefont {T.}~\bibnamefont {Zou}},  \emph {et~al.},\
  }\href@noop {} {\bibfield  {journal} {\bibinfo  {journal} {Nature
  communications}\ }\textbf {\bibinfo {volume} {10}},\ \bibinfo {pages} {589}
  (\bibinfo {year} {2019})}\BibitemShut {NoStop}%
\bibitem [{\citenamefont {Ortiz}\ \emph {et~al.}(2021)\citenamefont {Ortiz},
  \citenamefont {Menke}, \citenamefont {Misj\'ak}, \citenamefont {Mantadakis},
  \citenamefont {F\"ursich}, \citenamefont {Schierle}, \citenamefont
  {Logvenov}, \citenamefont {Kaiser}, \citenamefont {Keimer}, \citenamefont
  {Hansmann},\ and\ \citenamefont {Benckiser}}]{Ortiz:2021p165137}%
  \BibitemOpen
  \bibfield  {author} {\bibinfo {author} {\bibfnamefont {R.~A.}\ \bibnamefont
  {Ortiz}}, \bibinfo {author} {\bibfnamefont {H.}~\bibnamefont {Menke}},
  \bibinfo {author} {\bibfnamefont {F.}~\bibnamefont {Misj\'ak}}, \bibinfo
  {author} {\bibfnamefont {D.~T.}\ \bibnamefont {Mantadakis}}, \bibinfo
  {author} {\bibfnamefont {K.}~\bibnamefont {F\"ursich}}, \bibinfo {author}
  {\bibfnamefont {E.}~\bibnamefont {Schierle}}, \bibinfo {author}
  {\bibfnamefont {G.}~\bibnamefont {Logvenov}}, \bibinfo {author}
  {\bibfnamefont {U.}~\bibnamefont {Kaiser}}, \bibinfo {author} {\bibfnamefont
  {B.}~\bibnamefont {Keimer}}, \bibinfo {author} {\bibfnamefont
  {P.}~\bibnamefont {Hansmann}}, \ and\ \bibinfo {author} {\bibfnamefont
  {E.}~\bibnamefont {Benckiser}},\ }\href {\doibase
  10.1103/PhysRevB.104.165137} {\bibfield  {journal} {\bibinfo  {journal}
  {Phys. Rev. B}\ }\textbf {\bibinfo {volume} {104}},\ \bibinfo {pages}
  {165137} (\bibinfo {year} {2021})}\BibitemShut {NoStop}%
\bibitem [{\citenamefont {Cheong}\ \emph {et~al.}(1994)\citenamefont {Cheong},
  \citenamefont {Hwang}, \citenamefont {Batlogg}, \citenamefont {Cooper},\ and\
  \citenamefont {Canfield}}]{Cheong:1994p1087}%
  \BibitemOpen
  \bibfield  {author} {\bibinfo {author} {\bibfnamefont {S.-W.}\ \bibnamefont
  {Cheong}}, \bibinfo {author} {\bibfnamefont {H.}~\bibnamefont {Hwang}},
  \bibinfo {author} {\bibfnamefont {B.}~\bibnamefont {Batlogg}}, \bibinfo
  {author} {\bibfnamefont {A.}~\bibnamefont {Cooper}}, \ and\ \bibinfo {author}
  {\bibfnamefont {P.}~\bibnamefont {Canfield}},\ }\href@noop {} {\bibfield
  {journal} {\bibinfo  {journal} {Physica B: Condensed Matter}\ }\textbf
  {\bibinfo {volume} {194}},\ \bibinfo {pages} {1087} (\bibinfo {year}
  {1994})}\BibitemShut {NoStop}%
\bibitem [{\citenamefont {Garc{\'\i}a-Mu{\~n}oz}\ \emph
  {et~al.}(1995)\citenamefont {Garc{\'\i}a-Mu{\~n}oz}, \citenamefont {Suaaidi},
  \citenamefont {Mart{\'\i}nez-Lope},\ and\ \citenamefont
  {Alonso}}]{Garcia:1995p13563}%
  \BibitemOpen
  \bibfield  {author} {\bibinfo {author} {\bibfnamefont {J.}~\bibnamefont
  {Garc{\'\i}a-Mu{\~n}oz}}, \bibinfo {author} {\bibfnamefont {M.}~\bibnamefont
  {Suaaidi}}, \bibinfo {author} {\bibfnamefont {M.}~\bibnamefont
  {Mart{\'\i}nez-Lope}}, \ and\ \bibinfo {author} {\bibfnamefont
  {J.}~\bibnamefont {Alonso}},\ }\href@noop {} {\bibfield  {journal} {\bibinfo
  {journal} {Physical Review B}\ }\textbf {\bibinfo {volume} {52}},\ \bibinfo
  {pages} {13563} (\bibinfo {year} {1995})}\BibitemShut {NoStop}%
\bibitem [{\citenamefont {Xiang}\ \emph {et~al.}(2010)\citenamefont {Xiang},
  \citenamefont {Asanuma}, \citenamefont {Yamada}, \citenamefont {Inoue},
  \citenamefont {Akoh},\ and\ \citenamefont {Sawa}}]{Xiang:2010p032114}%
  \BibitemOpen
  \bibfield  {author} {\bibinfo {author} {\bibfnamefont {P.-H.}\ \bibnamefont
  {Xiang}}, \bibinfo {author} {\bibfnamefont {S.}~\bibnamefont {Asanuma}},
  \bibinfo {author} {\bibfnamefont {H.}~\bibnamefont {Yamada}}, \bibinfo
  {author} {\bibfnamefont {I.}~\bibnamefont {Inoue}}, \bibinfo {author}
  {\bibfnamefont {H.}~\bibnamefont {Akoh}}, \ and\ \bibinfo {author}
  {\bibfnamefont {A.}~\bibnamefont {Sawa}},\ }\href@noop {} {\bibfield
  {journal} {\bibinfo  {journal} {Applied Physics Letters}\ }\textbf {\bibinfo
  {volume} {97}},\ \bibinfo {pages} {032114} (\bibinfo {year}
  {2010})}\BibitemShut {NoStop}%
\bibitem [{\citenamefont {Shi}, \citenamefont {Zhou},\ and\ \citenamefont
  {Ramanathan}(2014)}]{Shi:2014p4860}%
  \BibitemOpen
  \bibfield  {author} {\bibinfo {author} {\bibfnamefont {J.}~\bibnamefont
  {Shi}}, \bibinfo {author} {\bibfnamefont {Y.}~\bibnamefont {Zhou}}, \ and\
  \bibinfo {author} {\bibfnamefont {S.}~\bibnamefont {Ramanathan}},\
  }\href@noop {} {\bibfield  {journal} {\bibinfo  {journal} {Nature
  communications}\ }\textbf {\bibinfo {volume} {5}},\ \bibinfo {pages} {1}
  (\bibinfo {year} {2014})}\BibitemShut {NoStop}%
\bibitem [{\citenamefont {Asanuma}\ \emph {et~al.}(2010)\citenamefont
  {Asanuma}, \citenamefont {Xiang}, \citenamefont {Yamada}, \citenamefont
  {Sato}, \citenamefont {Inoue}, \citenamefont {Akoh}, \citenamefont {Sawa},
  \citenamefont {Ueno}, \citenamefont {Shimotani}, \citenamefont {Yuan},
  \citenamefont {Kawasaki},\ and\ \citenamefont {Iwasa}}]{Asanuma:2010p142110}%
  \BibitemOpen
  \bibfield  {author} {\bibinfo {author} {\bibfnamefont {S.}~\bibnamefont
  {Asanuma}}, \bibinfo {author} {\bibfnamefont {P.-H.}\ \bibnamefont {Xiang}},
  \bibinfo {author} {\bibfnamefont {H.}~\bibnamefont {Yamada}}, \bibinfo
  {author} {\bibfnamefont {H.}~\bibnamefont {Sato}}, \bibinfo {author}
  {\bibfnamefont {I.~H.}\ \bibnamefont {Inoue}}, \bibinfo {author}
  {\bibfnamefont {H.}~\bibnamefont {Akoh}}, \bibinfo {author} {\bibfnamefont
  {A.}~\bibnamefont {Sawa}}, \bibinfo {author} {\bibfnamefont {K.}~\bibnamefont
  {Ueno}}, \bibinfo {author} {\bibfnamefont {H.}~\bibnamefont {Shimotani}},
  \bibinfo {author} {\bibfnamefont {H.}~\bibnamefont {Yuan}}, \bibinfo {author}
  {\bibfnamefont {M.}~\bibnamefont {Kawasaki}}, \ and\ \bibinfo {author}
  {\bibfnamefont {Y.}~\bibnamefont {Iwasa}},\ }\href {\doibase
  10.1063/1.3496458} {\bibfield  {journal} {\bibinfo  {journal} {Applied
  Physics Letters}\ }\textbf {\bibinfo {volume} {97}},\ \bibinfo {pages}
  {142110} (\bibinfo {year} {2010})},\ \Eprint
  {http://arxiv.org/abs/https://doi.org/10.1063/1.3496458}
  {https://doi.org/10.1063/1.3496458} \BibitemShut {NoStop}%
\bibitem [{\citenamefont {Scherwitzl}\ \emph {et~al.}(2010)\citenamefont
  {Scherwitzl}, \citenamefont {Zubko}, \citenamefont {Lezama}, \citenamefont
  {Ono}, \citenamefont {Morpurgo}, \citenamefont {Catalan},\ and\ \citenamefont
  {Triscone}}]{Scherwitzl:2010p5517}%
  \BibitemOpen
  \bibfield  {author} {\bibinfo {author} {\bibfnamefont {R.}~\bibnamefont
  {Scherwitzl}}, \bibinfo {author} {\bibfnamefont {P.}~\bibnamefont {Zubko}},
  \bibinfo {author} {\bibfnamefont {I.~G.}\ \bibnamefont {Lezama}}, \bibinfo
  {author} {\bibfnamefont {S.}~\bibnamefont {Ono}}, \bibinfo {author}
  {\bibfnamefont {A.~F.}\ \bibnamefont {Morpurgo}}, \bibinfo {author}
  {\bibfnamefont {G.}~\bibnamefont {Catalan}}, \ and\ \bibinfo {author}
  {\bibfnamefont {J.-M.}\ \bibnamefont {Triscone}},\ }\href@noop {} {\bibfield
  {journal} {\bibinfo  {journal} {Advanced Materials}\ }\textbf {\bibinfo
  {volume} {22}},\ \bibinfo {pages} {5517} (\bibinfo {year}
  {2010})}\BibitemShut {NoStop}%
\bibitem [{\citenamefont {Bubel}\ \emph {et~al.}(2015)\citenamefont {Bubel},
  \citenamefont {Hauser}, \citenamefont {Glaudell}, \citenamefont {Mates},
  \citenamefont {Stemmer},\ and\ \citenamefont
  {Chabinyc}}]{Simon:20115p122102}%
  \BibitemOpen
  \bibfield  {author} {\bibinfo {author} {\bibfnamefont {S.}~\bibnamefont
  {Bubel}}, \bibinfo {author} {\bibfnamefont {A.~J.}\ \bibnamefont {Hauser}},
  \bibinfo {author} {\bibfnamefont {A.~M.}\ \bibnamefont {Glaudell}}, \bibinfo
  {author} {\bibfnamefont {T.~E.}\ \bibnamefont {Mates}}, \bibinfo {author}
  {\bibfnamefont {S.}~\bibnamefont {Stemmer}}, \ and\ \bibinfo {author}
  {\bibfnamefont {M.~L.}\ \bibnamefont {Chabinyc}},\ }\href {\doibase
  10.1063/1.4915269} {\bibfield  {journal} {\bibinfo  {journal} {Applied
  Physics Letters}\ }\textbf {\bibinfo {volume} {106}},\ \bibinfo {pages}
  {122102} (\bibinfo {year} {2015})},\ \Eprint
  {http://arxiv.org/abs/https://doi.org/10.1063/1.4915269}
  {https://doi.org/10.1063/1.4915269} \BibitemShut {NoStop}%
\bibitem [{\citenamefont {Ha}\ \emph {et~al.}(2013{\natexlab{b}})\citenamefont
  {Ha}, \citenamefont {Vetter}, \citenamefont {Shi},\ and\ \citenamefont
  {Ramanathan}}]{Ha:2013p183102}%
  \BibitemOpen
  \bibfield  {author} {\bibinfo {author} {\bibfnamefont {S.~D.}\ \bibnamefont
  {Ha}}, \bibinfo {author} {\bibfnamefont {U.}~\bibnamefont {Vetter}}, \bibinfo
  {author} {\bibfnamefont {J.}~\bibnamefont {Shi}}, \ and\ \bibinfo {author}
  {\bibfnamefont {S.}~\bibnamefont {Ramanathan}},\ }\href@noop {} {\bibfield
  {journal} {\bibinfo  {journal} {Applied Physics Letters}\ }\textbf {\bibinfo
  {volume} {102}},\ \bibinfo {pages} {183102} (\bibinfo {year}
  {2013}{\natexlab{b}})}\BibitemShut {NoStop}%
\bibitem [{\citenamefont {Rost}\ \emph {et~al.}(2015)\citenamefont {Rost},
  \citenamefont {Sachet}, \citenamefont {Borman}, \citenamefont {Moballegh},
  \citenamefont {Dickey}, \citenamefont {Hou}, \citenamefont {Jones},
  \citenamefont {Curtarolo},\ and\ \citenamefont {Maria}}]{Rost:2015p8485}%
  \BibitemOpen
  \bibfield  {author} {\bibinfo {author} {\bibfnamefont {C.~M.}\ \bibnamefont
  {Rost}}, \bibinfo {author} {\bibfnamefont {E.}~\bibnamefont {Sachet}},
  \bibinfo {author} {\bibfnamefont {T.}~\bibnamefont {Borman}}, \bibinfo
  {author} {\bibfnamefont {A.}~\bibnamefont {Moballegh}}, \bibinfo {author}
  {\bibfnamefont {E.~C.}\ \bibnamefont {Dickey}}, \bibinfo {author}
  {\bibfnamefont {D.}~\bibnamefont {Hou}}, \bibinfo {author} {\bibfnamefont
  {J.~L.}\ \bibnamefont {Jones}}, \bibinfo {author} {\bibfnamefont
  {S.}~\bibnamefont {Curtarolo}}, \ and\ \bibinfo {author} {\bibfnamefont
  {J.-P.}\ \bibnamefont {Maria}},\ }\href@noop {} {\bibfield  {journal}
  {\bibinfo  {journal} {Nature communications}\ }\textbf {\bibinfo {volume}
  {6}},\ \bibinfo {pages} {8485} (\bibinfo {year} {2015})}\BibitemShut
  {NoStop}%
\bibitem [{\citenamefont {Stoica}\ \emph {et~al.}(2020)\citenamefont {Stoica},
  \citenamefont {Puggioni}, \citenamefont {Zhang}, \citenamefont {Singla},
  \citenamefont {Dakovski}, \citenamefont {Coslovich}, \citenamefont {Seaberg},
  \citenamefont {Kareev}, \citenamefont {Middey}, \citenamefont {Kissin} \emph
  {et~al.}}]{Stoica:2020disentangling}%
  \BibitemOpen
  \bibfield  {author} {\bibinfo {author} {\bibfnamefont {V.}~\bibnamefont
  {Stoica}}, \bibinfo {author} {\bibfnamefont {D.}~\bibnamefont {Puggioni}},
  \bibinfo {author} {\bibfnamefont {J.}~\bibnamefont {Zhang}}, \bibinfo
  {author} {\bibfnamefont {R.}~\bibnamefont {Singla}}, \bibinfo {author}
  {\bibfnamefont {G.}~\bibnamefont {Dakovski}}, \bibinfo {author}
  {\bibfnamefont {G.}~\bibnamefont {Coslovich}}, \bibinfo {author}
  {\bibfnamefont {M.}~\bibnamefont {Seaberg}}, \bibinfo {author} {\bibfnamefont
  {M.}~\bibnamefont {Kareev}}, \bibinfo {author} {\bibfnamefont
  {S.}~\bibnamefont {Middey}}, \bibinfo {author} {\bibfnamefont
  {P.}~\bibnamefont {Kissin}},  \emph {et~al.},\ }\href@noop {} {\bibfield
  {journal} {\bibinfo  {journal} {arXiv preprint arXiv:2004.03694}\ } (\bibinfo
  {year} {2020})}\BibitemShut {NoStop}%
\bibitem [{\citenamefont {F{\"o}rst}\ \emph {et~al.}(2011)\citenamefont
  {F{\"o}rst}, \citenamefont {Manzoni}, \citenamefont {Kaiser}, \citenamefont
  {Tomioka}, \citenamefont {Tokura}, \citenamefont {Merlin},\ and\
  \citenamefont {Cavalleri}}]{Forst:2011p854}%
  \BibitemOpen
  \bibfield  {author} {\bibinfo {author} {\bibfnamefont {M.}~\bibnamefont
  {F{\"o}rst}}, \bibinfo {author} {\bibfnamefont {C.}~\bibnamefont {Manzoni}},
  \bibinfo {author} {\bibfnamefont {S.}~\bibnamefont {Kaiser}}, \bibinfo
  {author} {\bibfnamefont {Y.}~\bibnamefont {Tomioka}}, \bibinfo {author}
  {\bibfnamefont {Y.}~\bibnamefont {Tokura}}, \bibinfo {author} {\bibfnamefont
  {R.}~\bibnamefont {Merlin}}, \ and\ \bibinfo {author} {\bibfnamefont
  {A.}~\bibnamefont {Cavalleri}},\ }\href {\doibase 10.1038/nphys2055}
  {\bibfield  {journal} {\bibinfo  {journal} {Nature Physics}\ }\textbf
  {\bibinfo {volume} {7}},\ \bibinfo {pages} {854} (\bibinfo {year}
  {2011})}\BibitemShut {NoStop}%
\bibitem [{\citenamefont {Abreu}\ \emph {et~al.}(2020)\citenamefont {Abreu},
  \citenamefont {Meyers}, \citenamefont {Thorsm{\o}lle}, \citenamefont {Zhang},
  \citenamefont {Liu}, \citenamefont {Geng}, \citenamefont {Chakhalian},\ and\
  \citenamefont {Averitt}}]{Abreu:2020p7422}%
  \BibitemOpen
  \bibfield  {author} {\bibinfo {author} {\bibfnamefont {E.}~\bibnamefont
  {Abreu}}, \bibinfo {author} {\bibfnamefont {D.}~\bibnamefont {Meyers}},
  \bibinfo {author} {\bibfnamefont {V.~K.}\ \bibnamefont {Thorsm{\o}lle}},
  \bibinfo {author} {\bibfnamefont {J.}~\bibnamefont {Zhang}}, \bibinfo
  {author} {\bibfnamefont {X.}~\bibnamefont {Liu}}, \bibinfo {author}
  {\bibfnamefont {K.}~\bibnamefont {Geng}}, \bibinfo {author} {\bibfnamefont
  {J.}~\bibnamefont {Chakhalian}}, \ and\ \bibinfo {author} {\bibfnamefont
  {R.~D.}\ \bibnamefont {Averitt}},\ }\href@noop {} {\bibfield  {journal}
  {\bibinfo  {journal} {Nano letters}\ }\textbf {\bibinfo {volume} {20}},\
  \bibinfo {pages} {7422} (\bibinfo {year} {2020})}\BibitemShut {NoStop}%
\end{thebibliography}
%

\end{document}